\documentclass[fleqn,usenatbib]{mnras}

\usepackage[T1]{fontenc}

\DeclareRobustCommand{\VAN}[3]{#2}
\let\VANthebibliography\thebibliography
\def\thebibliography{\DeclareRobustCommand{\VAN}[3]{##3}\VANthebibliography}

\usepackage{graphicx}	
\usepackage{amsmath}	
\usepackage{amssymb}	
\usepackage{bm}
\usepackage{afterpage}
\usepackage{pdflscape}	
\usepackage{algpseudocode}
\usepackage{algorithmicx}
\usepackage{algorithm}
\usepackage{empheq}
\usepackage{cases}
\usepackage{caption}
\usepackage{subcaption}
\usepackage{standalone}
\usepackage{tikz}
\usepackage{pgfplots}

\algblock[Block]{Block}{EndBlock}
\algblockdefx[Block]{Block}{EndBlock}%
[1]{{#1}}%
[1]{{#1}}
\newcommand{\Given}[1]{\State{\bf given} {#1}}

\newcommand{\RepeatFor}[1]{\Repeat {\bf~for} {#1}}

\newcommand{\RR}{\ensuremath{\mathbb{R}}}
\newcommand{\NN}{\ensuremath{\mathbb{N}}}
\newcommand{\CC}{\ensuremath{\mathbb{C}}}

\usepackage{newtxtext,newtxmath}

\title[Interferometric imaging with learned denoisers]{Image reconstruction algorithms in radio interferometry: from handcrafted to learned regularization denoisers}

\author[M. Terris et al.]{
Matthieu Terris,$^{1}$\thanks{E-mail: m.terris@hw.ac.uk}
Arwa Dabbech,$^{1}$
Chao Tang,$^{1}$
Yves Wiaux$^{1}$
\\
$^{1}$Institute of Sensors, Signals and Systems, Heriot-Watt University, Edinburgh EH14 4AS, UK
}

\date{Accepted XXX. Received YYY; in original form ZZZ}

\pubyear{2021}

\begin{document}
\label{firstpage}
\pagerange{\pageref{firstpage}--\pageref{lastpage}}
\maketitle

\begin{abstract} 

We introduce a new class of iterative image reconstruction algorithms for radio interferometry, at the interface of convex optimization and deep learning, inspired by plug-and-play methods. The approach consists in learning a prior image model by training  a deep neural network (DNN) as a denoiser, and substituting it for the handcrafted proximal regularization operator of an optimization algorithm. The proposed AIRI (``AI for Regularization in radio-interferometric Imaging'') framework, for imaging complex intensity structure with diffuse and faint emission from visibility data, inherits the robustness and interpretability of optimization, and the learning power and speed of networks. Our approach relies on three steps. Firstly, we design a low dynamic range training database from optical intensity images. Secondly, we train a DNN denoiser at a noise level inferred from the signal-to-noise ratio of the data. We use training losses enhanced with a nonexpansiveness term ensuring algorithm convergence, and including on-the-fly database dynamic range enhancement via exponentiation. Thirdly, we plug the learned denoiser into the forward-backward optimization algorithm, resulting in a simple iterative structure alternating a denoising step with a gradient-descent data-fidelity step. We have validated AIRI against CLEAN, optimization algorithms of the SARA family, and a DNN trained to reconstruct the image directly from visibility data. Simulation results show that AIRI is competitive in imaging quality with SARA and its unconstrained forward-backward-based version uSARA, while providing significant acceleration. CLEAN remains faster but offers lower quality. The end-to-end DNN offers further acceleration, but with far lower quality than AIRI.
\end{abstract}




\section{Introduction}

Data acquisition by interferometry in radio astronomy relies on an array of antennas measuring an incomplete coverage of the spatial Fourier domain of the image of interest, yielding an ill-posed inverse problem towards image formation. Modern and upcoming radio telescopes are designed to bring unprecedented resolution and sensitivity. In this context, the algorithms to be deployed to solve the inverse imaging problem face an ever increasing requirement to jointly deliver precision (\emph{i.e.}~high resolution and dynamic range), robustness (\emph{i.e.}~endowed with calibration and uncertainty quantification functionalities), and scalability (\emph{i.e.}~the capability to process sheer data volumes).

Most RI imaging pipelines utilized by astronomers are based on the CLEAN algorithm, originally proposed by \citet{hogbom1974aperture}. Assuming a sparse sky model, this greedy algorithm iteratively removes the contribution of each point source in the dirty image (more generally known as the backprojected image, defined as the inverse Fourier transform of the data). Numerous extensions of CLEAN have been devised over the last fifty years \citep[\emph{e.g.}][]{schwarz1978mathematical, schwab1983global, bhatnagar2004scale, cornwell2008noncoplanar, thompson2017interferometry}. Albeit simple and computationally efficient, CLEAN-based methods often require extensive manual intervention to ensure their stability. Furthermore, CLEAN restored images  often exhibit sub-optimal imaging quality. On the one hand, their resolution is limited to that of the observations due to the convolution with a smoothing beam. On the other hand, their sensitivity is limited to the noise level, through the addition of the residual image.

Since their first inception by \citet{wiaux2009compressed}, convex and nonconvex optimization algorithms underpinned by sparsity priors have emerged in radio interferometric (RI) imaging \citep{li2011application,  dabbech2015moresane, garsden2015lofar, onose2016scalable, onose2017accelerated, repetti2017non, pratley2018robust, dabbech2018cygnus, repetti2018uncertainty, repetti2019scalable, birdi20, thouvenin2021parallel, thouvenin2021parallel2} and have led to a much higher image precision than that offered by advanced versions of CLEAN. In this framework, a so-called objective function is defined, typically as the sum of a data fidelity term and a regularization term injecting a prior image model to compensate for data incompleteness. The image estimate is defined as the minimizer of this objective and is reached via provably convergent algorithms. The obtained solution can also be understood in a Bayesian framework as a \emph{maximum a posteriori} (MAP) estimate with respect to a posterior distribution, the negative logarithm of which is the objective. Uncertainty quantification approaches fully powered by optimization algorithms have also been proposed \citep{repetti2018uncertainty, repetti2019scalable}. The versatility of the framework has progressively imposed optimization algorithms as a cornerstone of current state-of-the-art techniques in RI imaging. We note that they have also proven their worth for image reconstruction in the related fields of optical interferometry \citep{thiebaut2017principles} and very long baseline interferometry (VLBI) \citep{akiyama2019first, akiyama2022first}. However, albeit robust and interpretable, optimization algorithms come at the expense of a significant increase of the computational cost over CLEAN, ultimately affecting their scalability to large data volumes.

Bayesian inference approaches have also been proposed to address the RI image formation problem \citep{cai2018uncertainty, arras2019unified}, also for VLBI \citep{akiyama2022first}, naturally enabling uncertainty quantification, but remaining computationally very demanding for the data volumes expected by modern radio telescopes.

Recent works have contemplated the use of deep neural networks (DNNs) for end-to-end reconstruction or postprocessing in RI imaging \citep{terris2019deep, connor2021deep, gheller2021convolutional}. Despite promising results and tremendous scalability capabilities, these approaches have neither been thoroughly validated for high resolution high dynamic range imaging of complex structure involving diffuse and faint emission, nor compared with CLEAN or state-of-the-art optimization approaches. Moreover, end-to-end DNNs are known to be subject to robustness issues \citep{goodfellow2014explaining, nguyen2015deep, pang2018towards}. Their exploitation raises questions, not only with regards to the interpretability of the obtained solution, specifically in a Bayesian context, but also their generalizability, \emph{i.e.}~their ability to provide accurate reconstruction quality in the case of acquisition conditions unseen during training.

In this work, we propose a new RI image reconstruction approach based on the Plug-and-Play (PnP) framework, whereby a prior image model is learned by training a DNN as a denoiser, and substituted for the so-called proximal operator enforcing regularization at the heart of proximal optimization algorithms \citep{venkatakrishnan2013plug, chan2016plug, zhang2017beyond, zhang2019deep}. PnP algorithms were shown to deliver outstanding image reconstruction quality in applications such as image restoration \citep{zhang2020plug} and medical imaging \citep{ahmad2020plug}. Importantly, as the DNN is trained as a denoiser on an image database to serve as a simple regularization operator, it is by construction applicable for any sensing procedure, \emph{de facto} avoiding the generalizability issue affecting end-to-end approaches. Leveraging monotone operator theory, recent works have shown that these algorithms provably converge to well characterized solutions under nonexpansiveness conditions on the denoiser, yielding similar interpretations to that of MAP estimators \citep{cohen2021regularization, pesquet2020learning}. In summary, this versatile framework simultaneously inherits the robustness and interpretability of optimization approaches, and the learning power and scalability capabilities of DNNs.

Building on \citet{pesquet2020learning}, we propose a PnP framework dubbed AIRI, standing for ``AI for Regularization in radio-interferometric Imaging''. Focusing on monochromatic intensity imaging, our approach for developing a first AIRI algorithm capable of imaging complex structure with diffuse and faint emission can be summarized in three steps. The first step is to design, from publicly available optical images, a realistic but low dynamic range database of intensity images for supervised training. The second step is to train a DNN denoiser with basic architecture ensuring positivity of the reconstructed images, at the signal-to-noise ratio of the data. The training relies on a simple $\ell_2$ or $\ell_1$ loss enhanced with a firm nonexpansiveness term ensuring algorithm convergence, and on an on-the-fly exponentiation procedure for dynamic range enhancement. The third step resides in plugging the DNN into the FB optimization algorithm, resulting in a simple iterative structure alternating between a denoising step and a gradient-descent data-fidelity step.

We finally validate the resulting AIRI-$\ell_2$ and AIRI-$\ell_1$ algorithms, implemented in Matlab, on high resolution high dynamic range simulations utilizing intensity images containing diffuse and faint emission across the field of view. Our test images are 3c353, Hercules~A, Centaurus~A, and Cygnus~A, with size $512 \times 512$. Our benchmark algorithms are (i) the multi-scale CLEAN version implemented in the C++ WSClean software package \citep{offringa2017wsclean}, (ii) two optimization algorithms from the SARA family leveraging the handcrafted ``average sparsity'' proximal regularization operator: SARA itself \citep{onose2017accelerated} and its unconstrained FB-based version uSARA that we introduce here, and (iii) a UNet trained in an end-to-end fashion to reconstruct target images from dirty images. While SARA relies on an advanced primal-dual FB (PDFB) algorithm \citep{pesquet2014class} enabling non-differentiable data-fidelity constraints, uSARA only differs from AIRI-$\ell_2$ and AIRI-$\ell_1$ by the use of the average sparsity proximal operator in lieu of a learned denoiser for regularization. It in fact corresponds to the standalone version of the imaging module of the joint calibration and imaging approach proposed in \citet{repetti2017non} and \citet{dabbech2021cygnus}. Results show that AIRI-$\ell_2$ and AIRI-$\ell_1$ are already competitive with SARA and uSARA in terms of imaging quality, while providing a significant acceleration. In a nutshell, AIRI does indeed inherit the robustness and interpretability of optimization approaches, and the learning power and speed of DNNs. The WSClean code remains significantly faster but offers lower reconstruction quality. The UNet offers further acceleration, but with far lower quality than AIRI.

The remainder of the paper is organized as follows. In Section~\ref{sect:prox_algos}, we review imaging approaches from proximal optimization, recall SARA, and introduce its unconstrained version uSARA. In Section~\ref{sect:pnp_algos}, we review PnP algorithms through the prism of proximal algorithms and introduce AIRI. In Section~\ref{sect:experiments}, we study the performance of AIRI-$\ell_2$ and AIRI-$\ell_1$ through extensive simulation in comparison with the benchmark algorithms. In Section~\ref{sect:conclusion}, we draw our conclusions and discuss necessary future work.

\section{State-of-art optimization algorithms \& RI}
\label{sect:prox_algos}

\subsection{RI imaging problem}
\label{ssect:measmod}
Aperture synthesis in radio astronomy probes the sky by measuring an incomplete coverage of the spatial Fourier domain of the image of interest through an array of antennas. Focusing on monochromatic intensity imaging, assuming a narrow field of view, and in the absence of atmospheric and instrumental perturbations, each pair of antennas acquires a noisy Fourier component of the intensity image to be formed, called a visibility. The associated Fourier mode (also called $uv$-point) is given by the projection of the corresponding baseline, expressed in units of the observation wavelength, onto the plane perpendicular to the line of sight \citep{thompson2017interferometry}. A discrete formulation of the resulting linear RI image formation problem, aiming to restore a target intensity image $\overline{\bm{x}}\in \RR^{n}$ from the measured complex visibilities  $\bm{y} \in \CC^{m}$, reads \citep{onose2016scalable}
\begin{equation}
    \bm{y} = \bm{\Phi} \overline{\bm{x}}+\bm{e},
\label{eq:invpb}
\end{equation}
where $\bm{\Phi} = \bm{\mathrm{GFZ}}\in \CC^{m \times n}$ is the measurement operator, $\bm{\mathrm{G}} \in \CC^{m \times d}$ is a sparse interpolation matrix, encoding the non-uniform Fourier transform, $\bm{\mathrm{F}} \in \CC^{d\times d}$ is the 2D Discrete Fourier Transform, $\bm{\mathrm{Z}} \in \RR^{d\times n}$ is a zero-padding operator, incorporating the correction for the convolution performed through the operator $\bm{\mathrm{G}}$, and $\bm{e}\in \CC^{m}$ is a realization of some i.i.d. Gaussian random noise, with zero mean and standard deviation $\tau>0$. We refer to Appendix~\ref{appendix:ri_model} for considerations regarding more general noise distributions, and the inclusion of so-called direction-dependent effects (DDEs) in $\bm{\mathrm{G}}$.

We also note that, backprojecting problem \eqref{eq:invpb} into the image domain gives 
\begin{equation}
    \operatorname{Re}\{\bm{\Phi}^\dagger \bm{y}\}= \operatorname{Re}\{\bm{\Phi}^\dagger\bm{\Phi}\} \overline{\bm{x}}+\operatorname{Re}\{\bm{\Phi}^\dagger \bm{e}\},
\label{eq:invpbim}
\end{equation}
where $(\cdot)^\dagger$ denotes the complex conjugate transpose and $\operatorname{Re}\{\cdot\}$ denotes the real part. $\operatorname{Re}\{\bm{\Phi}^\dagger \bm{y}\}$ is known as the dirty image, $\operatorname{Re}\{\bm{\Phi}^\dagger\bm{\Phi}\}$ is the operator representing the convolution of $\overline{\bm{x}}$ by the point spread function, also known as the dirty beam, and $\operatorname{Re}\{\bm{\Phi}^\dagger \bm{e}\}$ is the noise backprojected in the image domain.

We finally emphasise that in the optical interferometry and VLBI contexts, imaging is often performed from closure quantities derived from the visibilities, bringing a nonlinear inverse problem affected by non-Gaussian noise \citep{akiyama2019first, akiyama2022first}. Imaging in the presence of DDEs or from closure quantities lies beyond the scope of the present work.

\subsection{Proximal algorithms \& FB}
\label{sect:prox_tools}

Leaving the nature of $\bm{\Phi}$ aside, problems of the likes of \eqref{eq:invpb} are ubiquitous in imaging sciences, and arise for instance in image restoration \citep{levin2009understanding, yang2010image, bredies2020higher}, hyperspectral imaging \citep{wang2015dual, xie2019multispectral} and medical imaging \citep{gupta2018cnn, zbontar2018fastmri, fessler2020optimization}, to name a few. A widespread approach to solve such a problem is to reformulate it as a convex minimization problem
\begin{equation}
    \underset{\bm{x}\in \RR^n}{\text{minimize}}\, f(\bm{x})+\lambda r(\bm{x}),
\label{eq:min_pb}
\end{equation}
where $f(\bm{x})+\lambda r(\bm{x})$ is called the objective function, $f\in \Gamma_0(\RR^n)$\footnote{$\Gamma_0(\RR^n)$ denotes the set of convex, proper and lower-semicontinuous functions from $\RR^n$ to $(-\infty, \infty]$.} is the term enforcing data-fidelity, and $r\in \Gamma_0(\RR^n)$ is the regularization term introduced to address the ill-posedness of \eqref{eq:invpb} by enforcing a prior image model. 

The theory of optimization offers a myriad of algorithms to solve such minimization problems, ranging from the simple Forward-Backward (FB) algorithm \citep{bauschke2017convex}, applicable if one of the two terms is differentiable, to ADMM \citep{boyd2011distributed}, or the Douglas-Rachford algorithm \citep{eckstein1992douglas}, applicable even if both $f$ and $r$ are non-differentiable, and more evolved structures, such as PDFB \citep{pesquet2014class}, that can handle multi-term objective functions with as many non-differentiable data-fidelity and regularization terms processed in parallel. All these algorithms are part of the same class of proximal algorithms, where differentiable and non-differentiable terms respectively translate in the algorithm into gradient and proximal operators.

Considering $r \in \Gamma_0(\RR^n)$, its proximal operator $\operatorname{prox}_{r}$ is defined as
\begin{equation}
(\forall \bm{z} \in \RR^n),\qquad \operatorname{prox}_{r}(\bm{z}) = \underset{\bm{u}\in \RR^n}{\text{argmin}} \frac{1}{2}\|\bm{z}-\bm{u}\|^2+r(\bm{u}),
\label{eq:prox_def}
\end{equation}
and can be interpreted as a generalization of a projection operator, or interestingly, as a denoising operator. Indeed, problem \eqref{eq:prox_def} can be interpreted as the minimization problem to be solved for a simple denoising problem of the form $\bm{z}=\bm{u}+\bm{w}$ where the data $\bm{z}$ result from adding some i.i.d. Gaussian noise $\bm{w}$ to the unknown $\bm{u}$. From this perspective, the first term of the objective function in \eqref{eq:prox_def} would be the standard data-fidelity term, given the Gaussian nature of the noise, and $r$ the regularization term. A proximal operator is a denoiser!

The choice of $f$ and $r$ in \eqref{eq:min_pb} is of paramount importance as it influences the final solution to the minimization task \citep{mallat1999wavelet, selesnick2005dual, bredies2020higher}. Interpreting the reconstructed image as a MAP estimate leads naturally to choosing $f$ as the negative log-likelihood associated with a given statistical model of the noise (typically a squared $\ell_2$ norm for i.i.d. additive Gaussian noise), and  $r$ as the negative logarithm associated with a statistical prior image model (typically, $\ell_1$ norm for a Laplace prior). However, optimization algorithms are not tied to this statistical interpretation. In the context of compressive sensing theory \citep{candes2006robust, donoho2006compressed, baraniuk2007compressive}, where sparsity is the postulated signal model, the choice of the $\ell_1$ norm as a regularization term simply emanates from it being the closest convex relaxation of the $\ell_0$ norm that is the natural sparsity measure. $\ell_1$ regularization was shown to yield state-of-the-art image reconstruction methods beyond the context of compressive sensing. For instance, sparsity of the gradient is enforced as   the $\ell_1$ norm of the magnitude of the image gradient, that is the so-called Total Variation semi-norm (TV) \citep{rudin1992nonlinear, bredies2010total, bredies2020higher}, whereas sparsity of the sought image in a sparsifying domain $\bm{\mathrm{A}}\in \RR^{n\times p}$ can be enforced by choosing $r(\bm{x}) = \|\bm{\mathrm{A}}^\dagger \bm{x}\|_1$ where $\bm{\mathrm{A}}$ can be a wavelet dictionary \citep{mallat1999wavelet}, an x-let transform \citep{candes2003curvelets, do2003contourlets}, or a learned dictionary \citep{mairal2009online}, to name a few. Generally, the choice of the data-fidelity term is often driven by the statistical nature of the measurement noise, while regularization terms are carefully handcrafted to meet the specificity of the applications, with sparsity models defining the state-of-the-art. In summary, the MAP interpretation has its limitation, and in particular, setting the regularization parameter $\lambda$ according to the proper normalization of the statistical models of which $f$ and $r$ would be the negative logarithms, is known to be suboptimal. 

The FB algorithm is a proximal algorithm designed to solve problems of the form \eqref{eq:min_pb}, where $f,r\in\Gamma_0(\RR^n)$ and $f$ is differentiable. It is defined by the iterative sequence 
\begin{equation}
(\forall k \in \NN),\qquad   \bm{x}_{k+1} = \operatorname{prox}_{\gamma \lambda r}(\bm{x}_k-\gamma \nabla f(\bm{x}_k)),
\label{eq:prox_fb}
\end{equation}
which is proven to converge towards a minimizer of the objective function \eqref{eq:min_pb}, provided that $0<\gamma<2/L$ where $L$ is the Lipschitz constant\footnote{The Lipschitz constant of an operator $h:\RR^n\to\RR^p$ is defined as $\text{sup}_{x\neq y}\|h(x)-h(y)\|/\|x-y\|$.} of $\nabla f$ \citep{bauschke2017convex}. Accelerated versions of FB, leveraging inertial terms, preconditioning, and stochastic approaches, also exist. We also emphasize that FB can be used to solve problems involving nonconvex regularization terms $r$ \citep{beck2009fast, attouch2013convergence, chouzenoux2014variable, combettes2015stochastic, repetti2021variable}.

\subsection{SARA and its unconstrained version}
\label{ssect:sarausaraRI}

\subsubsection{SARA}
\label{ssect:sara}

The SARA family represent state-of-the-art optimization algorithms for RI imaging \citep{carrillo2012sparsity, carrillo2014purify,Abdulaziz2016,onose2016scalable,onose2017accelerated,birdi2018sparse,pratley2018robust,dabbech2018cygnus,abdulaziz2019wideband, thouvenin2021parallel}. In its monochromatic intensity imaging version, SARA consists in solving the constrained minimization problem given by
\begin{equation}
    \underset{\bm{x}\in\RR^n}{\text{minimize}}\,\, \iota_{\mathcal{B}(\bm{y},\epsilon)}(\boldsymbol{\Phi} \bm{x}) + \lambda r(\bm{x}),
\label{eq:SARA}
\end{equation}
where data fidelity is imposed via a constraint encoded by the indicator function\footnote{The indicator function of a non-empty, closed, and convex set $\mathcal{S}$ is defined as $\iota_\mathcal{S}(\bm{x}) = 0$ if $\bm{x}\in\mathcal{S}$, else $\iota_\mathcal{S}(\bm{x}) = +\infty$.} $\iota_{\mathcal{B}(\bm{y},\epsilon)}$ of  the $\ell_2$-ball centred at the data $\bm{y}$ and of radius $\epsilon > 0$: $\mathcal{B}(\bm{y},\epsilon) = \left\{ \bm{u} \in \CC^m \, | \, \|\bm{y}-\bm{u}\| \leq \epsilon \right\}$. The $\ell_2$-bound  $\epsilon$ is derived from the assumed i.i.d. Gaussian random noise statistics as $\epsilon^2=(2m+4\sqrt{m})\tau^2/2$ \citep{carrillo2012sparsity}. The function $r$ defines the SARA prior,  also called the ``average sparsity'' prior, which consists of a positivity constraint and a log-sum prior promoting average sparsity in an over-complete dictionary $\bm{\Psi} \in\RR^{n\times bn}$. More specifically, the dictionary $\bm{\Psi}$ is defined as the concatenation of $b = 9$ orthogonal bases (the first eight Daubechies wavelets and the Dirac basis), re-normalized by $\sqrt{b}=3$ to ensure that $\bm{\Psi}\bm{\Psi}^\dagger = \mathbf{I}$, where $\mathbf{I}$ denotes the identity operator. The SARA prior explicitly reads 
\begin{equation}
r(\bm{x}) = \rho \sum_{j=1}^{bn} \log\left(\rho^{-1}\left|\left(\bm{\mathrm{\Psi}}^\dagger \bm{x}\right)_{j}\right|+1\right)+\iota_{\RR^n_+}(\bm{x}),
\label{eq:sara_prior}
\end{equation}
where $\left(.\right)_j$ denotes the $j^\text{th}$ coefficient of its argument vector, and $\RR^n_+$ denotes the $n$-dimensional real positive orthant. The parameter $\lambda>0$ is a regularization parameter, while $\rho>0$ is used to avoid reaching zero values in the argument of the logarithmic terms.  \citet{thouvenin2021parallel} suggest to set both parameters to an estimate of the standard deviation of the measurement noise in the wavelet domain.

Minimizing \eqref{eq:SARA} should typically involve the proximal operator $\operatorname{prox}_{r}$, dubbed the ``average sparsity'' proximal operator. However, if the log-sum prior enforces a stronger sparsity of the solution than a typical $\ell_1$ prior, it is also nonconvex. To address the resulting nonconvex minimization task, a reweighting procedure is adopted, where a sequence of convex surrogate minimization problems is solved iteratively \citep{candes2008enhancing,carrillo2012sparsity}, each involving a weighted-$\ell_1$ prior $g$, given by
\begin{equation}
\label{eq:weighted_l1_prior}
g(\bm{x}, \bm{\mathrm{W}}) =  \|\bm{\mathrm{W}}\bm{\mathrm{\Psi}}^\dagger \bm{x}\|_1+\iota_{\RR^n_+}(\bm{x}),
\end{equation}
where ${\bm{\mathrm{W}} \in \RR^{bn\times bn}}$ is a diagonal weighting matrix that needs to be updated after each resolution of the surrogate problem. The full SARA algorithm for solving \eqref{eq:SARA} can be summarized as
\begin{equation}
\label{eq:SARA-rw}
\begin{array}{l}
\text{for}\;i=0,1,\ldots\\
\left\lfloor
\begin{array}{ll}
\bm{\tilde x}_{i+1}&\hspace{-1em}=\underset{\bm{x}\in\RR^n}{\text{argmin}}\,\, \iota_{\mathcal{B}(\bm{y},\epsilon)}(\boldsymbol{\Phi} \bm{x}) + g(\bm{x},\bm{\mathrm W}_i)\\
\bm{\mathrm{W}}_{i+1}&\hspace{-1em}=\mathrm{Diag}\left( \rho\left/\left(\rho+\bm{\mathrm{\Psi}}^\dagger \bm{\tilde x}_{i+1}\right)\right.\right),\\
\end{array}
\right.\\
\end{array}
\end{equation}
with $\bm{\mathrm{W}}_{0}=\mathbf{I}$, where $\text{Diag}(\cdot)$ denotes the diagonal matrix containing its (vector) argument on the diagonal. Minimizing each of these problems will now involve the weighted-$\ell_1$ prior $g$ via its proximal operator $\operatorname{prox}_{g}$. One should acknowledge that the reweighting strategy also introduces significant complexity to the algorithm since a sequence of convex minimization problems needs to be solved.

Given the non-differentiability of the data-fidelity term in \eqref{eq:SARA}, the PDFB algorithm \citep{pesquet2014class} is leveraged to solve each minimization task of the form \eqref{eq:SARA-rw}. We note that because $\bm{\Psi}^\dagger$ is over-complete and due to the presence of the positivity constraint in \eqref{eq:weighted_l1_prior}, the proximal operator $\operatorname{prox}_{g}$ does not admit a closed-form solution, thus \emph{a priori} requiring a sub-iterative structure. However, the full splitting functionalities of PDFB enable the decomposition of $g$ into functions admitting simple proximal operators (a component-wise projection on the real positive orthant and a component-wise thresholding operation). These proximal operators are handled in parallel by PDFB without the need for sub-iterations \citep{onose2016scalable}. Interestingly, the same functionalities enable handling large data volumes via parallel processing of data blocks \citep{onose2016scalable}, and large image sizes via parallel processing of image facets \citep{thouvenin2021parallel}, providing significant scalability to SARA.

\subsubsection{Unconstrained SARA (uSARA)}
\label{sect:uSARA}
Building on the works of \citet{repetti2020forward, repetti2021variable}, we propose to focus on the unconstrained formulation of the SARA problem, that we dub unconstrained SARA (uSARA), again looked at from the prism of monochromatic intensity RI imaging only. More precisely, we consider the problem \eqref{eq:min_pb}, with a differentiable data-fidelity term $f$ chosen as the mean squared error loss, corresponding to the negative log-likelihood of the data under the i.i.d. Gaussian random noise assumption, given by
\begin{equation}
    f(\bm{x})=\frac{1}{2}\|\bm{\Phi} \bm{x}-\bm{y}\|^2.
\label{eq:def_fl2}
\end{equation}
Adopting the prior model $r$ described in \eqref{eq:sara_prior}, the resulting nonconvex minimization task writes
\begin{equation}
  \underset{\bm{x}\in\RR^n}{\text{minimize}}\, \frac{1}{2}\|\bm{\Phi} \bm{x}-\bm{y}\|^2+ \lambda r(\bm{x}),
\label{eq:uSARA}
\end{equation}
where $\lambda >0$ is a regularization parameter. The nonconvexity of the prior $r$ from \eqref{eq:sara_prior} is handled through a similar reweighting procedure as in \eqref{eq:SARA-rw}:
\begin{equation}
\label{eq:uSARA-rw}
\begin{array}{l}
\text{for}\;i=0,1,\ldots\\
\left\lfloor
\begin{array}{ll}
\bm{\tilde x}_{i+1}&\hspace{-1em}=\underset{\bm{x}\in\RR^n}{\text{argmin}}\,\, \frac{1}{2}\|\boldsymbol{\Phi} \bm{x}-\bm{y}\|^2 + \lambda g(\bm{x},\bm{\mathrm W}_i)\\
\bm{\mathrm{W}}_{i+1}&\hspace{-1em}=\mathrm{Diag}\left( \rho\left/\left(\rho+\bm{\mathrm{\Psi}}^\dagger \bm{\tilde x}_{i+1}\right)\right.\right),\\
\end{array}
\right.\\
\end{array}
\end{equation}
with $\bm{\mathrm{W}}_{0}=\mathbf{I}$. At each reweighting iteration, problem \eqref{eq:uSARA-rw} can be solved with a standard FB algorithm. The full reweighting procedure for solving the uSARA problem \eqref{eq:uSARA} is summarized in Algorithm~\ref{algo:sara_fb}, where $\xi_1>0$ is a relative variation convergence criterion, and where the gradient of $f$ at $\bm{x}$ reads as
\begin{equation}
\label{eq:gradf}
\nabla f(\bm{x}) = \text{Re}\{\boldsymbol{\Phi}^\dagger\boldsymbol{\Phi}\}\bm{x}-\text{Re}\{\boldsymbol{\Phi}^\dagger\bm{y}\},
\end{equation}
and its Lipschitz constant is the spectral norm\footnote{The spectral norm of a linear operator is its maximum singular value, which can be computed via the power method~(see \emph{e.g.}~\citet{golub2013matrix}).} of $\text{Re}\{\boldsymbol{\Phi}^\dagger\boldsymbol{\Phi}\}$, \emph{i.e.}
\begin{equation}
\label{eq:lipgradf}
L=\|\text{Re}\{\boldsymbol{\Phi}^\dagger\boldsymbol{\Phi}\}\|_{\rm{S}}.
\end{equation}
We recall that the stepsize $\gamma$ in \eqref{eq:prox_fb} is to be upper-bounded by $2/L$.

As appears in Algorithm~\ref{algo:sara_fb} solving \eqref{eq:uSARA-rw} (steps \ref{step:begin-fb}-\ref{step:end-fb}), the weighted-$\ell_1$ prior $g$ is involved via $\operatorname{prox}_{\gamma \lambda g}$. The algorithmic structure does not exhibit the same full splitting functionalities as PDFB, and $\operatorname{prox}_{\gamma \lambda g}$ must be computed iteratively. This can typically be achieved via a so-called dual FB algorithm \citep{combettes2011proximal}, detailed in Algorithm~\ref{algo:dfb}, where $\xi_2>0$ is a relative variation convergence criterion. This algorithm alternates between a projection on the real positive orthant (Step~\ref{step:proj_orthant}), denoted by $\Pi_{\mathbb{R}^n_+}$, and the proximal operator of the dual of the weighted-$\ell_1$ prior, which involves a component-wise soft-thresholding operator $\operatorname{prox}_{\gamma\lambda\|\bm{\mathrm{W}}\cdot\|_1}$\footnote{This operator, applied to $\bm{z}$ boils down component-wise to $(\operatorname{prox}_{\gamma\lambda\|\bm{\mathrm{W}}\cdot\|_1}(\bm{z}))_{j}=\operatorname{prox}_{\eta_{j} |\cdot|}(z_j)=\operatorname{sign}(z_j)\operatorname{max}\{|z_j|-\eta_j, 0\}$ where $\eta_{j} = \gamma\lambda \mathrm{W}_{j,j}$ is called the soft-thresholding parameter. All values below the threshold are set to $0$ while those above are reduced by the value of the threshold (in absolute value).} (Step~\ref{step:prox_dual}).

Formally, Algorithms~\ref{algo:sara_fb} and \ref{algo:dfb} lead to a triply sub-iterative structure: a proximal operator loop (Step \ref{step:prox_loop}), inside a weighted-$\ell_1$ loop (Step~\ref{step:weighted_loop}), inside a reweighting loop (Step~\ref{step:reweight_loop}). This suggests a very computationally expensive structure. Nonetheless, firstly, \citet{repetti2021variable} have shown that only a fixed number of FB iterations $K$ are required (Step \ref{step:prox_loop}), corresponding to the approximate minimization of \eqref{eq:uSARA-rw}, while preserving the convergence of the overall algorithm to a minimizer of the nonconvex objective \eqref{eq:uSARA}. This contrasts with SARA, where the sub-problems \eqref{eq:SARA-rw} need to be solved to convergence. The value of $K$ can in fact be optimized to provide significant acceleration of Algorithm~\ref{algo:sara_fb}, effectively removing one iteration layer. Secondly, the number of iterations in Algorithm~\ref{algo:dfb} is moderate in practice when using an appropriate initialization strategy for the dual variable $\bm{v}_0$, leading to mild computational cost of $\operatorname{prox}_{\gamma \lambda g}$.

We conclude this section by underlining that uSARA corresponds to the imaging module of the joint calibration imaging approach described in \citet{repetti2017non} and \citet{dabbech2021cygnus}.
\begin{algorithm}[t]
\caption{Re-weighted FB algorithm for uSARA}
\begin{algorithmic}[1]
\small
\Given{$0<\gamma<2/L$, $\lambda>0$, $\rho>0$, $\bm{\tilde x}_0\in\RR^{n}$, $\xi_1>0$}
\State {Set $\bm{\mathrm{W}}_0=\mathbf{I}$}
\RepeatFor{$i=0,1,\ldots$}\label{step:reweight_loop}
\State {$\bm{x}_0=\bm{\tilde x}_i$}
\State{\textbf{repeat for }{$k=0,\ldots,K$}}\label{step:weighted_loop} \label{step:begin-fb}
\quad\quad
\State {
\quad\quad
$
\bm{x}_{k+1} = \operatorname{prox}_{\gamma  \lambda g(\cdot, \bm{\mathrm{W}}_i)}(\bm{x}_k-\gamma \nabla f(\bm{x}_k))
$} \label{step:fb}
\State{\textbf{end for}
}  \label{step:end-fb}
 \State {${\bm{\tilde x}}_{i+1} = \bm{x}_{K}$}
\State {$\bm{\mathrm{W}}_{i+1} =\mathrm{Diag}\left( \rho\left/\left(\rho+\bm{\mathrm{\Psi}}^\dagger \bm{\tilde x}_{i+1}\right)\right.\right)$}
\Until {{{$\|x_{k+1}-x_{k}\|/\|x_{k+1}\|<\xi_1$}}} \label{step:dfb:conv_crit}
\State \Return {${\bm{\tilde x}}_{i+1}$}
\end{algorithmic}
\label{algo:sara_fb}
\end{algorithm}

\begin{algorithm}[t]
\caption{Dual FB algorithm for computing $\operatorname{prox}_{\gamma \lambda g(\cdot, \bm{\mathrm{W}})}(\bm{z})$ in Algorithm~\ref{algo:sara_fb}}
\begin{algorithmic}[1]
\small
\Given{$\gamma>0$, $\lambda>0$, $\bm{\mathrm{W}}$, $\bm{z}\in\RR^{n}$, $\bm{v}_0\in \text{Span}(\bm{\mathrm{\Psi}}^\dagger$), $\xi_2>0$}
\RepeatFor{$l=0,1,\ldots$}\label{step:prox_loop}
\State {$
\bm{x}_{l+1} = \Pi_{\RR_+^n}(\bm{z}-\bm{\mathrm{\Psi}} \bm{v}_l)
$} \label{step:proj_orthant}
\State {$
\bm{v}_{l+1} = \left(\mathbf{I}-\operatorname{prox}_{\gamma \lambda \|\bm{\mathrm{W}}\cdot\|_1}\right)\left(\bm{v}_l+\bm{\mathrm{\Psi}}^\dagger \bm{x}_{l+1}\right)
$} \label{step:prox_dual}
\Until {{{$\|x_{k+1}-x_{k}\|/\|x_{k+1}\|<\xi_2$}}}
\State \Return {$\bm{x}_{l+1}$}
\end{algorithmic}
\label{algo:dfb}
\end{algorithm}

\section{AIRI: AI for Regularization in RI Imaging}
\label{sect:pnp_algos}

\subsection{PnP-FB}
\label{sect:pnp_fb}

When solving an inverse imaging problem from an optimization theory viewpoint, one defines an objective function, of which the sought image would be a minimizer and which can be obtained via an iterative algorithm. The PnP approach \citep{venkatakrishnan2013plug} looks at the problem directly through the lens of the algorithm. As discussed in Section~\ref{sect:prox_tools}, it follows from the definition of a proximal operator that it can be interpreted as a denoiser. Borrowing a proximal optimization algorithm, the PnP approach proposes to replace the proximal regularization operator with a more general denoiser. We here note that this procedure can be applied to a wide class of algorithms, ranging from FB to PDFB. Specifically, assuming a minimization problem of the form \eqref{eq:min_pb} with the differentiable data fidelity term $f$, recall that the FB algorithm reads as in \eqref{eq:prox_fb}, with the proximal operator \eqref{eq:prox_def}. Its PnP counterpart (PnP-FB) then simply follows as 
\begin{equation}
(\forall k \in \NN),\qquad   \bm{x}_{k+1} = \operatorname{D}(\bm{x}_k-\gamma \nabla f(\bm{x}_k)),
\label{eq:pnp_fb}
\end{equation}
where $\operatorname{D}$ is a denoising operator, \emph{i.e.} an operator specifically designed to remove i.i.d. Gaussian random noise from an image. 

Denoisers are ubiquitous in imaging sciences, and a large variety of denoisers have been defined and studied in the signal processing literature, including Gaussian filters, BM3D \citep{dabov2007image}, non local means (NLM) \citep{buades2011non}, with, more recently, deep neural networks (DNNs) leading the state of the art \citep{zhang2017beyond}. It is therefore very tempting to plug powerful denoisers in lieu of proximal operators... and play. In this context, we aim at taking advantage of the learning capabilities of DNNs in order to learn an appropriate regularization denoiser. Just as handcrafting the regularization term $r$ (from which the proximal regularization operator $\operatorname{prox}_{r}$ results) is crucial to the reconstruction quality for a pure optimization approach, learning a powerful denoiser is cornerstone for the reconstruction quality of PnP algorithms. Recent advances in DNNs for image denoising tasks have shown new state of the art results over traditional denoisers \citep{zhang2017beyond, wang2018esrgan, zhou2020awgn}. Simultaneously, PnP algorithms with DNNs as denoisers have become the new state-of-the-art in image reconstruction \citep{zhang2020plug}, significantly improving over their traditional optimization counterparts.

Yet, replacing the proximal operator with an off-the-shelf denoising operator is not inconsequential. In particular, a general non-proximal denoiser can \emph{a priori} not be related with a regularization term in some overarching objective function. As a consequence, the characterization of the PnP solution (\emph{e.g.} as a MAP estimator), and worse, the convergence properties of the iterative structure, are not ensured anymore, thus questioning the robustness and interpretability of PnP solutions. To overcome that issue, we follow the approach of \citet{pesquet2020learning} allowing to enforce the so-called ``firm nonexpansiveness constraint'' on the denoiser $\operatorname{D}$ during training. This constraint ensures that $\operatorname{D}$ ``contracts distances'' and yields both the convergence of Algorithm~\ref{algo:pnp} and the characterization of the limit point (see Appendix~\ref{sect:mon_tech} for more details). The PnP-FB algorithm with established convergence guarantees is summarized in Algorithm~\ref{algo:pnp}, where $\xi_3>0$ is a relative variation convergence criterion. In the RI imaging case of interest, the PnP counterpart to uSARA follows from defining $f$ as in \eqref{eq:def_fl2}, with the measurement operator $\bm{\Phi}$ from \eqref{eq:invpb}. The gradient of $f$ at $\bm{x}$ therefore reads as in \eqref{eq:gradf}, with Lipschitz constant given in \eqref{eq:lipgradf}.

\begin{algorithm}[t]
\caption{AIRI (PnP-FB) algorithm}
\begin{algorithmic}[1]
\small
\Given{$0<\gamma<2/L$, denoiser $\operatorname{D}$, $\bm{x}_0\in\RR^n$, $\xi_3>0$}
\RepeatFor{$k=0,1,\ldots$}
\State {$
\bm{x}_{k+1} = \operatorname{D}(\bm{x}_k-\gamma \nabla f(\bm{x}_k))
$}
\Until{{{$\|x_{k+1}-x_{k}\|/\|x_{k+1}\|<\xi_3$}}}
\State \Return {$x_{k+1}$}
\end{algorithmic}
\label{algo:pnp}
\end{algorithm}

\subsection{Training noise level vs. uSARA regularization parameter}
\label{sect:heuristic}

\subsubsection{Denoising i.i.d.~Gaussian noise of specific standard deviation}
\label{sssect:singlesigmadenoiser}

Building from \citet{pesquet2020learning}, our AIRI denoiser $\operatorname{D}$ will be trained in a supervised manner to remove i.i.d.~Gaussian random noise of specific standard deviation $\sigma$. In other words, $\operatorname{D}$ is set to tackle the denoising problem
\begin{equation}
\label{eq:Dproxdenoisingprob}
\bm{z} = \bm{u}+\sigma \bm{w},
\end{equation}
where $\bm{z}$ denotes the noisy data input to the DNN, $\bm{u}$ is the unknown image corresponding to the target output of the DNN, and $\bm{w}\sim \mathcal{N}(0,\mathbf{I})$. As discussed in the next section, the input signal-to-noise ratio of the RI data provides a reliable handle on $\sigma$.

\subsubsection{Equating training \& target dynamic ranges}
\label{ssect:heuristic}

As already emphasized, AIRI will emerge from PnP-FB with $f$ from \eqref{eq:def_fl2} and $\bm{\Phi}$ from \eqref{eq:invpb}. We propose to set the training noise level $\sigma$ for $\operatorname{D}$ to an appropriate estimate of the standard deviation of some effective image-domain noise induced by the original i.i.d. Gaussian random noise of standard deviation $\tau$ on the Fourier data in \eqref{eq:invpb}. We propose below a procedure to estimate the standard deviation of this image-domain noise, which, for simplicity, we also approximate to be i.i.d. Gaussian random noise.

Firstly, we resort to the image-domain formulation \eqref{eq:invpbim} of \eqref{eq:invpb},  normalized by the Lipschitz constant $L$ in \eqref{eq:lipgradf}: $\operatorname{Re}\{\bm{\Phi}^\dagger \bm{y}\}/L = \operatorname{Re}\{\bm{\Phi}^\dagger \bm{\Phi}\}\overline{\bm{x}}/L+\operatorname{Re}\{\bm{\Phi}^\dagger \bm{e}\}/L$. By construction, the chosen normalization ensures that the convolution operator $\operatorname{Re}\{\bm{\Phi}^\dagger \bm{\Phi}\}/L$ has unit spectral norm, so that the dirty image is at the same scale as the original image. The covariance matrix of the image-domain noise $\operatorname{Re}\{\bm{\Phi}^\dagger \bm{e}\}/L$ thus writes $\tau^2\operatorname{Re}\{\bm{\Phi}^\dagger\bm{\Phi}\}/2L^2$. Secondly, as proposed in \citet{thouvenin2021parallel}, we discard the correlation structure of the image-domain noise via the approximation $\operatorname{Re}\{\bm{\Phi}^\dagger \bm{\Phi}\}\simeq L\mathbf{I}$, leading to a noise covariance matrix approximation $\tau^2\mathbf{I}/2L$. As a result the effective image-domain noise is assumed i.i.d. Gaussian, and the original standard deviation $\tau$ in the measurement domain is rescaled to $\tau/\sqrt{2L}$ in the transfer to the image domain. Equating the standard deviations of the training noise for $\operatorname{D}$ and the image-domain noise leads to the following heuristic:
\begin{equation}
\label{eq:heuristic}
    \sigma = \frac{\tau}{\sqrt{2L}}.
\end{equation}
We note that this heuristic is algorithm agnostic, \emph{i.e.~a priori} independent of the algorithmic structure in which the denoiser is plugged. In this sense, while similar, it is simpler and more general than the one proposed in \citet{pesquet2020learning}, which was derived specifically by estimating $\sigma$ around the fixed point of \eqref{eq:pnp_fb}. As will be shown through simulations in Section~\ref{sect:experiments}, the proposed heuristic provides a very precise estimate of the optimal training noise level.

Assuming the validity of this heuristic, one must acknowledge a dependency of the training noise level $\sigma$ on the measurement model, via $\tau$ and $L$, suggesting that a denoiser should be trained independently for each target reconstruction. Also, we consider a normalized database of images whose maximum intensity values are lower or equal to 1 (see Section~\ref{sect:dataset}), so that the denoising performance cannot be ascertained on non-normalized images. These two generalizability questions can be solved by a simple rescaling of the original inverse problem \eqref{eq:invpb} to simultaneously fit the normalization constraint (an upper bound on the peak intensity of the image being accessible from the dirty image) and rescale the noise level in the data to the training noise level of some already available denoiser. \citet{dabbech2022first} and \citet{wilber2022first} build from these considerations and propose both a ``single denoiser'' approach and a more flexible ``denoiser shelf'' approach, where denoisers are respectively trained at a single predefined noise level or on a limited number of such noise levels, to which the target inverse problems can be rescaled at will. It also appears from the above rescaling considerations that the dimension of $\sigma$ is that of an inverse input image-domain peak signal-to-noise ratio rather than an absolute noise level. As the AIRI denoiser consistently categorizes $\sigma$-level values as noise to be removed, the input image-domain signal-to-noise ratio also naturally sets the target output peak signal-to-noise ratio or dynamic range of the reconstruction. In summary, our heuristic simply stipulates that the training dynamic range should be set equal to the target reconstruction dynamic range.

\subsubsection{uSARA regularization parameter}

We further note the very similar roles of the noise level $\sigma$ of the regularization denoiser in PnP-FB and the regularization parameter $\lambda$ in the FB algorithm \eqref{eq:prox_fb} solving \eqref{eq:min_pb}. At the algorithmic level, on the one hand, $\operatorname{D}$ is a denoiser and we have proposed to set $\sigma$ to the estimate \eqref{eq:heuristic} of the effective image-domain noise level, or more generally to the inverse target reconstruction dynamic range value. On the other hand, for the traditional FB algorithm, $\lambda$ is involved in \eqref{eq:prox_fb} via $\operatorname{prox}_{\gamma \lambda r}$. Focusing on the specific uSARA instance described in Section~\ref{sect:uSARA}, and summarized on Algorithms~\ref{algo:sara_fb} and \ref{algo:dfb}, $\gamma \lambda$ typically acts as a soft-thresholding parameter, thresholding out small values in the wavelet domain defined by the average sparsity dictionary. Setting its value to the standard deviation of the effective wavelet-domain noise, which according to \citet{thouvenin2021parallel} is the same as in the image domain given the unit spectral norm of the average sparsity dictionary $\bm{\Psi}$, yields the heuristic:
\begin{equation}
\label{eq:heuristic_uSARA}
    \gamma \lambda = \frac{\tau}{\sqrt{2L}}.
\end{equation}
We readily note that, in contrast with the reasoning leading to \eqref{eq:heuristic}, the above reasoning for $\lambda$ is specific to the FB structure underpinning uSARA. We also recall the parameter $\rho>0$ in the definition of the average sparsity prior $r$ in \eqref{eq:sara_prior}, used to avoid reaching zero values in the argument of the logarithmic terms. This parameter represents a floor level for the wavelet coefficients \citep{carrillo2012sparsity, thouvenin2021parallel}, and should naturally be set to the same effective wavelet-domain noise level estimate as $\lambda$: $ \rho = \lambda$. We further note that a variation of the proposed heuristic for the wavelet-domain noise level arises from a slightly different approximation in the transfer of the noise level from the image domain to the wavelet domain. Instead of assuming noise level conservation due to the normalization of $\bm{\Psi}$, one can analyze the noise in each of the $9$ bases of the dictionary separately, discarding the noise correlation structure between each pair of bases. In this case, the orthonormality of each basis and its normalization by $3$ suggests a correction factor $1/3$ in the heuristic values: $\rho = \gamma \lambda \simeq \tau/ 3\sqrt{2L}$.

As will be shown in Section~\ref{sect:experiments}, relation \eqref{eq:heuristic_uSARA} provides a useful reference to set $\lambda$ in uSARA. Even though the $1/3$ correction will be shown to bring a closer-to-optimal value, it is still not as precise an estimate as \eqref{eq:heuristic} is for the training noise level of AIRI denoisers. We emphasize that a significant body of work has been dedicated to the question of setting the regularization parameter in problems of the form \eqref{eq:min_pb}, with no simple and accurate solution to our knowledge \citep{donoho1995adapting, luisier2007new, vidal2020maximum}. In this context, the fact that \eqref{eq:heuristic_uSARA} provides an appropriate reference value for $\lambda$ in uSARA is an interesting result. However, the fact that the PnP-FB admits a very accurate heuristic to set the training noise level, avoiding the requirement for fine-tuning an arbitrary regularization parameter, represents a major advantage of AIRI over uSARA.

\subsection{Training database with adaptive dynamic range}
\label{sect:dataset}

\subsubsection{State-of-the-art and proposed approach}

Supervised training of DNNs for imaging requires the definition of a database of images, which remains a main difficulty in translating deep learning methods to RI imaging due to the absence of groundtruths. The few works that have leveraged deep learning techniques for RI imaging so far resorted to various approaches. In \citet{terris2019deep}, we used a high dynamic range simulated radio map from \citet{bonaldi2018square} of size $32000\times32000$ containing mainly elliptical Gaussian sources, broken down into a database of 1100 images of size $512 \times  512$.  \citet{gheller2021convolutional} relied on 1000 radio images of size $2000\times 2000$ simulated with ENZO \citep{bryan2014enzo} and augmented by random rotations, resulting in the appearance of realistic extended emission in the field of view. \citet{connor2021deep} generated a database of 900 synthetic images of size $2000\times 2000$ containing elliptical Gaussian sources.

In this work, we are targeting high dynamic range imaging of complex structure with diffuse and faint emission across the field of view. The database to be designed for the training of the regularization denoiser thus needs to reflect this target. We propose (i) to source publicly available optical astronomy images containing a rich combination of compact and diffuse emission, (ii) to preprocess these images, aiming to remove existing noise and artefacts via a preliminary denoising procedure, and (iii) to exponentiate their intensity values to create high dynamic range images. The procedure results in the creation of a rich and clean training database of $2235$ training images of size $512\times 512$. More technical details are provided below.

\subsubsection{Building a rich low dynamic range database}
\label{sssect:databaselow}

Firstly, we constituted a set of 32 large grayscale images $(\bm{u}^{\text{raw}}_k)_{1 \leq k\leq 32}$ from optical astronomical intensity images available online, containing complex structure with both compact and diffuse emission. Each image is normalized with maximum intensity value equal to $1$, and when the image contained colour channels, those were averaged. The typical size of each of these images is $3500\times3500$. The sample standard deviation of the residual noise in image regions with no signal, averaged over the 32 images reads $\sigma_0\simeq1/64$, leading to a typical dynamic range of $\sigma_0^{-1}\simeq64$, orders of magnitude below the values of interest in modern RI observations.

Secondly, the images are preprocessed with the aim of removing the residual noise, which includes various compression artefacts, and would otherwise be learned as part of the image model during training. To that effect, for all $1\leq k\leq 32$, a simple denoising problem is formulated for the recovery of a clean image $\overline{\bm{u}}_k$ from $\bm{u}_k^{\text{raw}}$: $\bm{u}_k^{\text{raw}}=\overline{\bm{u}}_k+\bm{e}_k$, assuming i.i.d. Gaussian random noise $\bm{e}_k$ with standard deviation $\widehat{\sigma}_k$. The denoising procedure applied consists in solving the optimization problem underpinning SARA in \eqref{eq:SARA}, with $\bm{\Phi}=\mathbf{I}$ and $\bm{y}=\bm{u}_k^{\text{raw}}$, resulting in a denoised database $(\bm{u}^{\text{low}}_k)_{1 \leq k\leq 32}$. We note that this procedure does not affect the (low) dynamic range of the images. In a nutshell, noise and artefacts are removed, but the floor signal level and maximum intensity remain essentially unchanged.

Finally, the cleaned images are split to create a rich and clean database $(\bm{u}^{\text{low}}_s)_{1 \leq s \leq S}$ with $S=2235$ images of size $512\times 512$ containing both compact and diffuse emission. When the image dimensions are not multiples of $512$, a symmetric padding is used before splitting. The resulting images exhibit a whole distribution of maximum intensity values bounded by $1$, and a corresponding distribution of dynamic range values, with $\sigma_0^{-1}\simeq64$ representing the ``nominal'' dynamic range value of the database.

\subsubsection{Enhancing database dynamic range to target}
\label{ssect:dr_enhancement}

In order to simulate the high dynamic range of interest in RI imaging, we propose to exponentiate pixel-wise the intensity value of all groundtruth images of the low dynamic range database. The exponentiation parameter will be set to achieve a final nominal dynamic range of the same order as the training dynamic range, which according to \eqref{eq:heuristic} is also equivalent to the target dynamic range of the reconstruction. Technically, the low dynamic range images $\bm{u}^{\text{low}}_s$ are exponentiated through pixel-wise as:
\begin{equation}
    \bm{u}_s = h(\bm{u}^{\text{low}}_s,a)
    = \frac{a^{\bm{u}^{\text{low}}_s}-1}{a},
\label{eq:exponentiation}
\end{equation}
for some parameter $a\gg1$. Recalling that the maximum span of the pixel values in any $\bm{u}^{\text{low}}_s$ is $[\sigma_0,1]$, the resulting maximum span of the exponentiated images is approximately $[a^{-1}(a^{\sigma_0}-1),1]$. The exponentiation thus preserves the database normalization with maximum intensity value across images bounded by $1$, while providing control on the final nominal dynamic range $a(a^{\sigma_0}-1)^{-1}$. Imposing this value to be equal to the training dynamic range leads to the following equation for $a$:
\begin{equation}
a=\left(1+a\sigma\right)^{\sigma_0^{-1}}.
\label{dynamicrangeheuristic}
\end{equation}
Parametrizing $a$ as $a=b\sigma^{-1}$ leads to the following equation for the scaling factor $b$: $ b = \sigma \left(1 + b \right)^{\sigma_0^{-1}}$. This equation can be solved numerically to estimate the exponentiation factor $a$ as a function of $\sigma_0$ and $\sigma$. In summary, the high dynamic range groundtruth database $(\bm{u}_s)_{1 \leq s \leq S}$ results from the low dynamic range groundtruth database $(\bm{u}^{\text{low}}_s)_{1 \leq s \leq S}$ via \eqref{eq:exponentiation}, with exponentiation parameter $a$ set in \eqref{dynamicrangeheuristic}.

\begin{figure*}
    \centering
         \includegraphics[trim={0 0em 0 0em}, width=0.99\textwidth, clip]{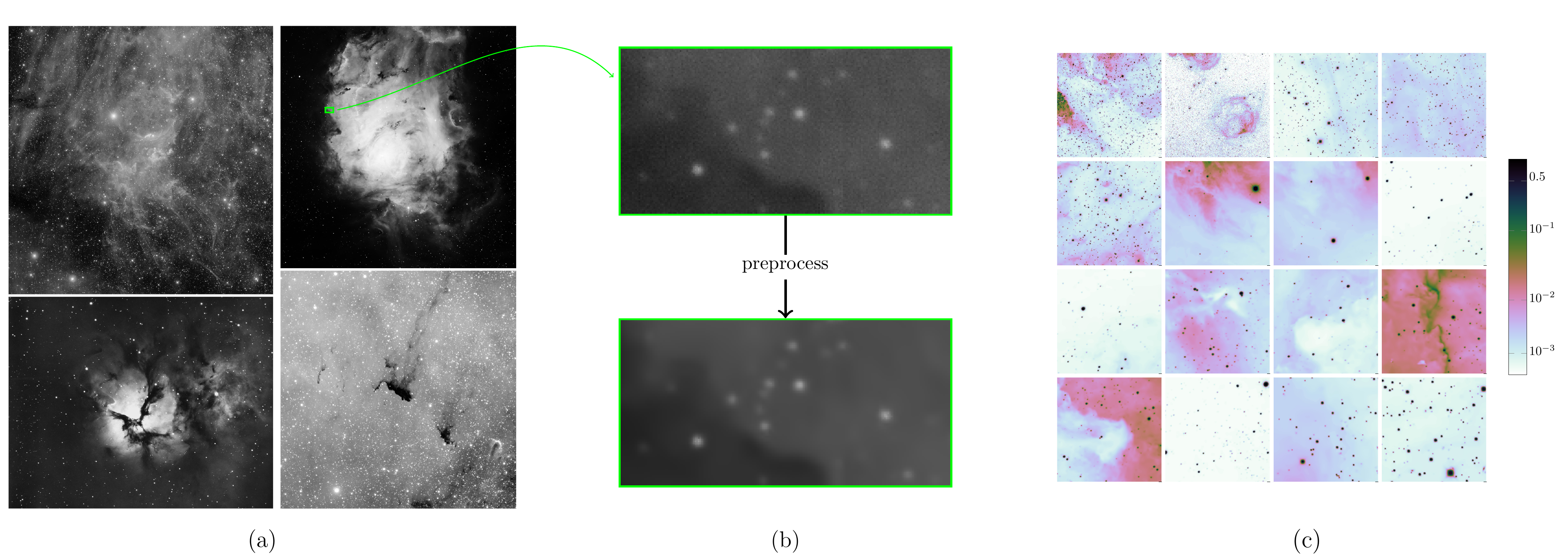}%
    \hfill
\vspace{-1em}
    \caption{Summary of the proposed strategy for building a training database of postprocessed radio groundtruths. We first gather a set of 32 astronomical (optical) images of high quality but low dynamic range: (a) shows a subset of four images. (b) illustrates the effect of preprocessing on a small part of the image from (a). Eventually, we split the images into $2235$ images of size $512\times 512$ to create the low dynamic range training database. Samples of our training database, after application of the exponentiation \eqref{eq:exponentiation} with $a=10^3$, are shown in (c) in logarithmic scale.}
    \label{fig:database_summary}
\end{figure*}

Figure~\ref{fig:database_summary} summarizes the preprocessing pipeline for the creation of the low and high dynamic range databases. Notice the diversity of images in the training samples displayed in Figure~\ref{fig:database_summary}~(c).

\begin{figure}
     \centering
     \begin{subfigure}[b]{0.51\linewidth}
         \centering
         \includegraphics[height=0.13\textheight, clip]{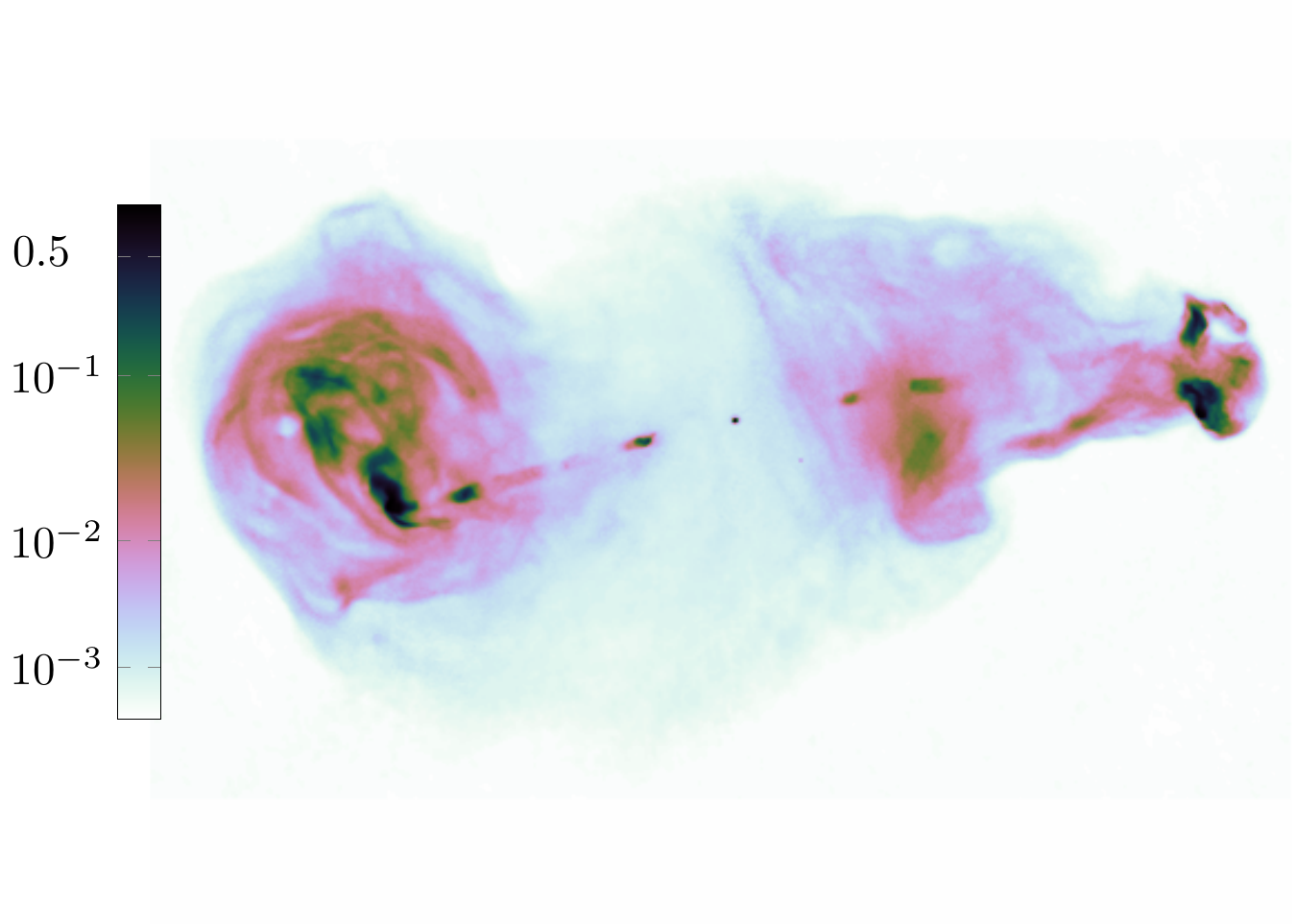}
         \vspace{-2em}
         \caption{3c353}
         \label{fig:gd:3c353}
     \end{subfigure}
     \hfill
     \begin{subfigure}[b]{0.48\linewidth}
         \centering
         \includegraphics[height=0.13\textheight, clip]{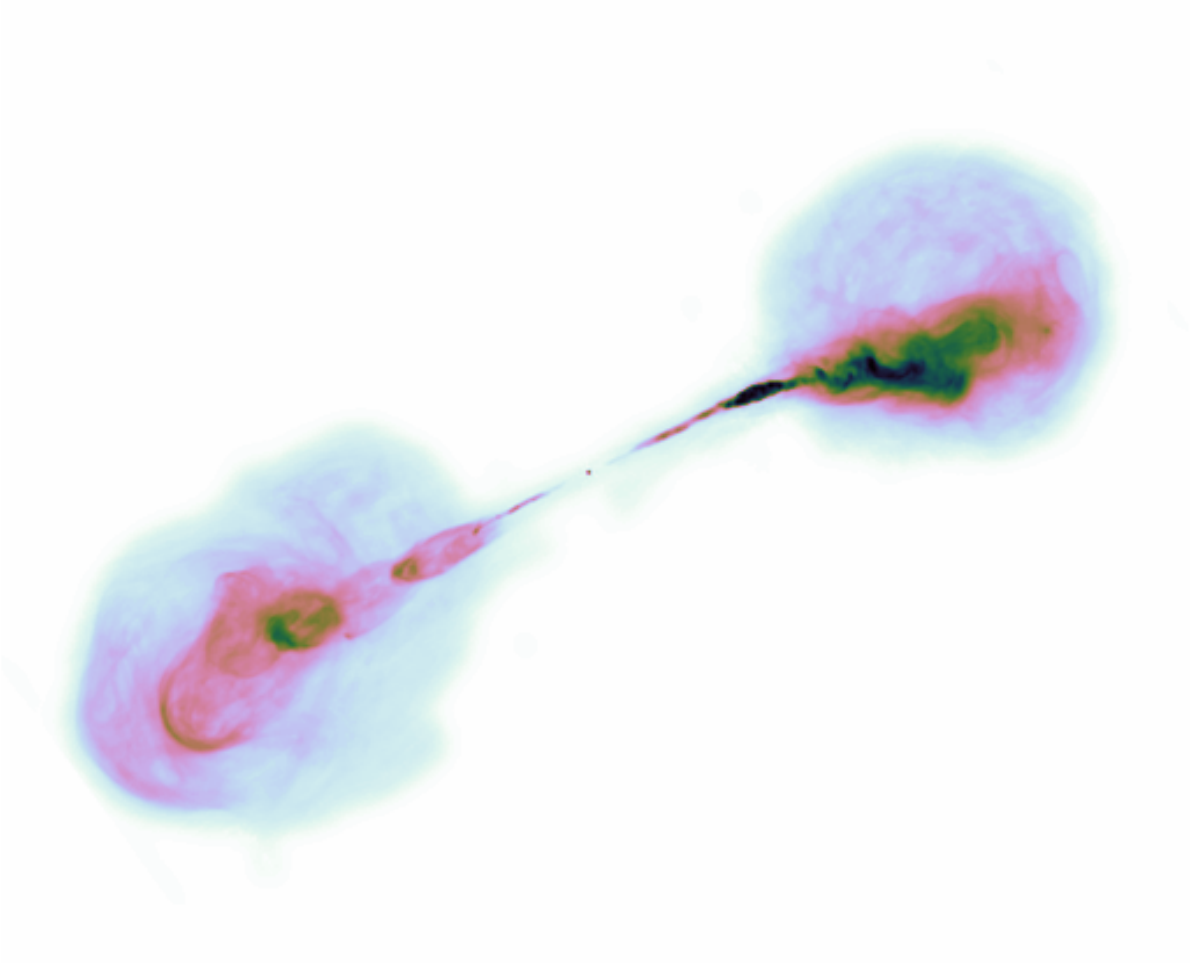}
         \vspace{-2em}
         \caption{Hercules~A}
         \label{fig:gd:hercA}
     \end{subfigure}
     
      \begin{subfigure}[b]{0.51\linewidth}
         \centering
         \includegraphics[height=0.13\textheight, clip]{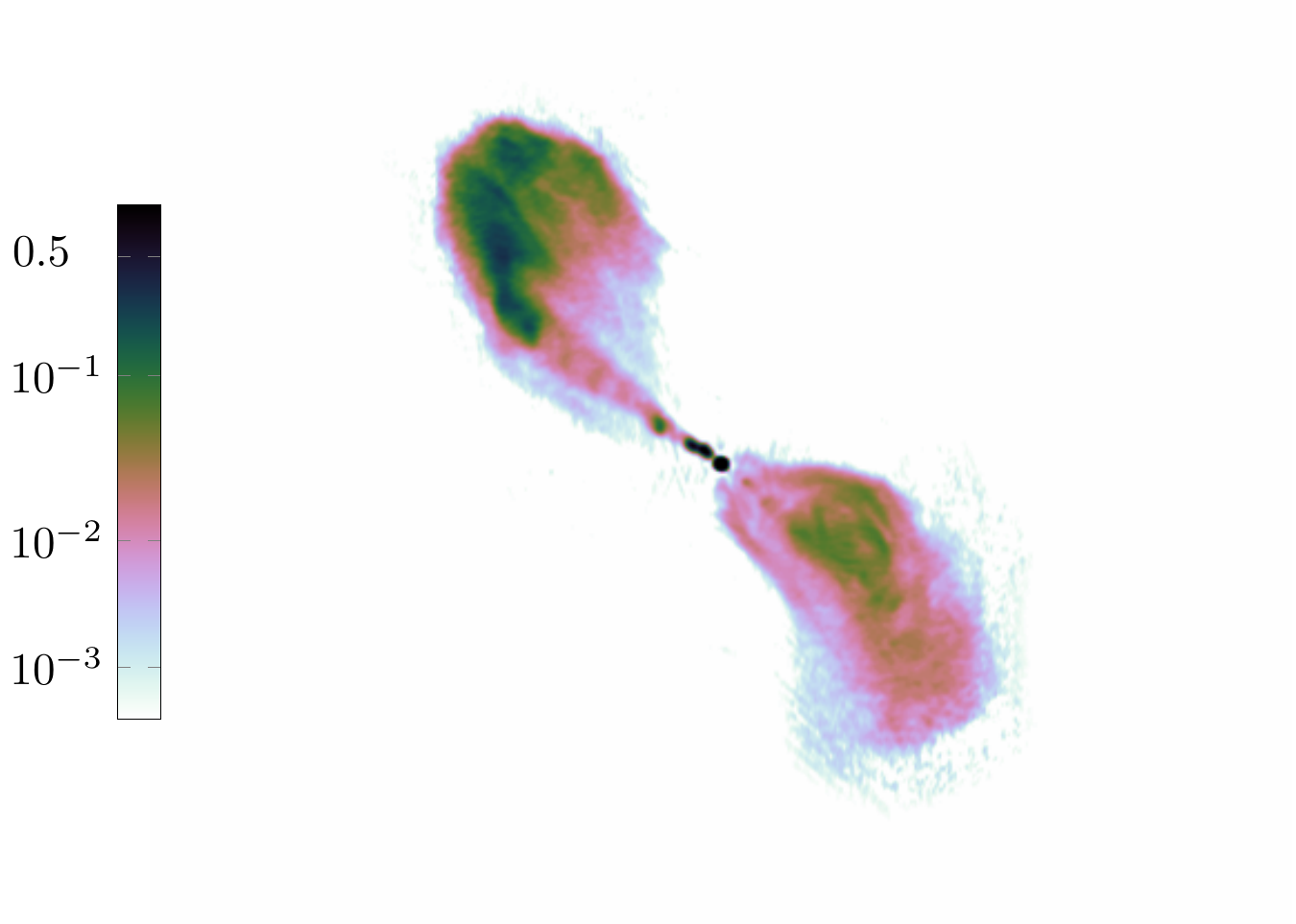}
         \vspace{-2em}
         \caption{Centaurus~A}
         \label{fig:gd:cenA}
     \end{subfigure}
     \hfill
     \begin{subfigure}[b]{0.48\linewidth}
         \centering
         \includegraphics[height=0.13\textheight, clip]{{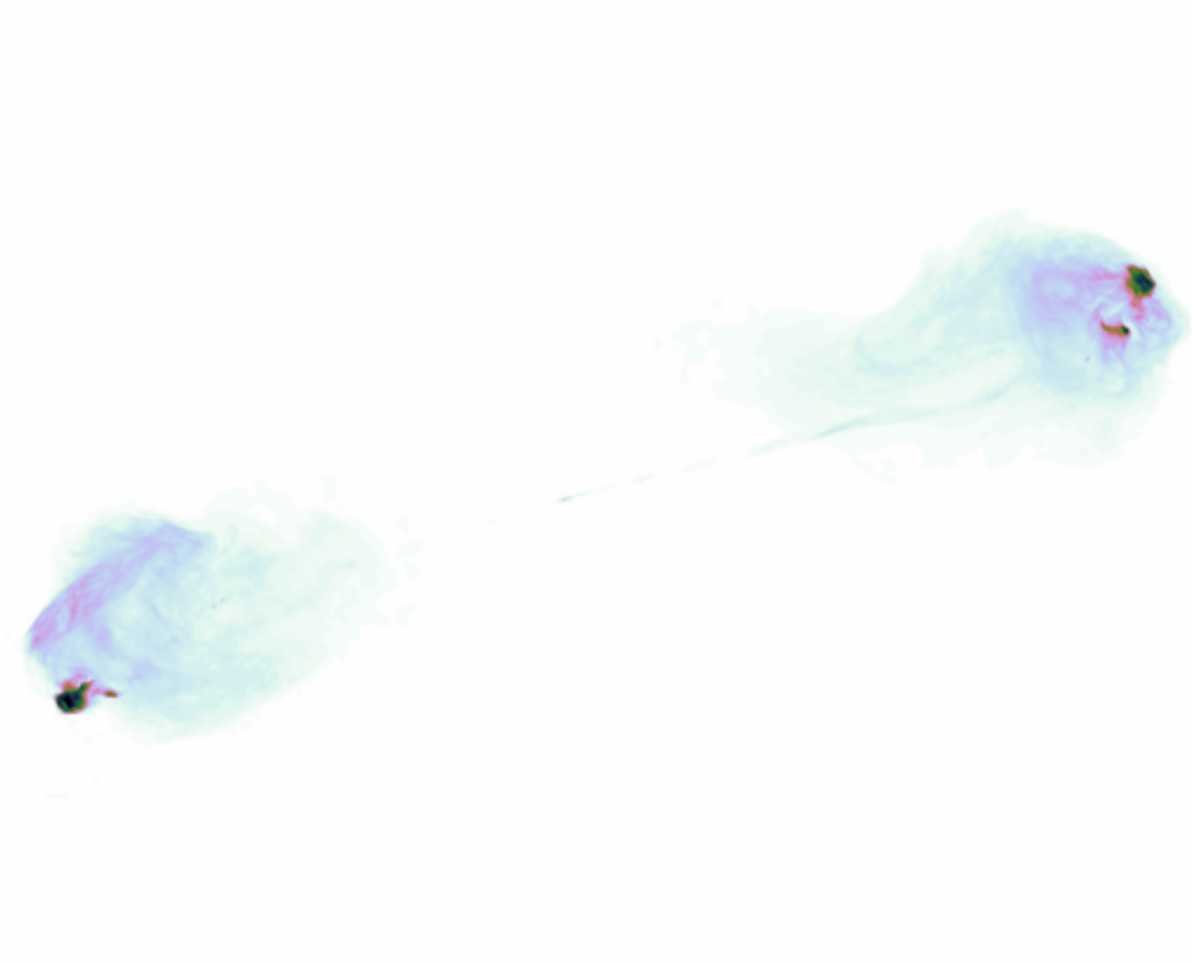}}
         \vspace{-2em}
         \caption{Cygnus~A}
         \label{fig:gd:cygA}
     \end{subfigure}
    \caption{The four test RI images, displayed in logarithmic scale.}
    \label{fig:gdths}
\vspace{-1.5em}
\end{figure}

\subsubsection{Generating noisy \& groundtruth pairs on-the-fly}
\label{ssect:databasepairs}

The supervised training approach considered relies on image pairs: the groundtruth images $(\bm{u}_s)_{1 \leq s \leq S}$ corresponding to the target outputs of the DNN, and the noisy images $(\bm{z}_s)_{1 \leq s \leq S}$ corresponding to the DNN input. Given our choice, described in Section~\ref{sssect:singlesigmadenoiser}, to build a denoiser tailored to tackle i.i.d. Gaussian random noise of specific standard deviation $\sigma$, the noisy high dynamic range images read
\begin{equation}
\label{eq:hdrnoisy}
\bm{z}_s = \bm{u}_s+\sigma \bm{w}_s,
\end{equation}
with $\bm{w}_s\sim \mathcal{N}(0,\bm{\mathrm{I}})$ and $\sigma$ given by \eqref{eq:heuristic}. 

Importantly, as the high dynamic range database of noisy and groundtruth pairs depends on the target dynamic range through \eqref{eq:exponentiation} and \eqref{eq:hdrnoisy}, one version of the database would need to be computed and stored for each dynamic range of interest, which is a highly impractical and unnecessary approach. Instead, $\bm{z}_s$ and $\bm{u}_s$ can simply be seen as functions of $\bm{u}^{\text{low}}_s$ and computed on the fly during training.

\subsection{Training loss function}
\label{ssect:methodology}

In this section, we introduce the training loss for the supervised training of the denoiser $\operatorname{D}$. Building on \citet{pesquet2020learning} and following technical arguments detailed in Appendix~\ref{sect:mon_tech}, we regularize the training loss with an appropriate Jacobian term to enforce the firm nonexpansiveness of $\operatorname{D}$. By denoting $\operatorname{Q}=2\operatorname{D}-\operatorname{I}$ where $\operatorname{I}$ denotes the identity operator, and denoting $\bm{\theta} \in \RR^c$ the learnable parameters of $\operatorname{D}$, the resulting training loss reads: 

\begin{equation}
\begin{aligned}
    \underset{\bm{\theta} \in \RR^c}{\text{minimize}}\, \frac{1}{S}\sum_{s=1}^S \bigg(\mathcal{L}(\operatorname{D}_{\bm{\theta}}(\bm{z}_s)-\bm{u}_s)+\kappa\operatorname{max}\{ \| \boldsymbol{\nabla} \operatorname{Q}_{\bm{\theta}}(\bm{z}_s)\|_{\rm{S}}, 1-\varepsilon\}\bigg),
\end{aligned}
\label{eq:training_loss}
\end{equation}
with $\bm{z}_s$ and $\bm{u}_s$ given in \eqref{eq:exponentiation} and \eqref{eq:hdrnoisy} respectively, as functions of $\bm{u}^{\text{low}}_s$ computed on the fly during training.
The second term in \eqref{eq:training_loss} is the firm nonexpansiveness regularization term, with $\kappa>0$ the associated regularization parameter, and $\varepsilon>0$ a safety margin parameter. The first term $\mathcal{L}$ is the standard part of the loss function, for which practical choices involve the $\ell_2$ loss \emph{i.e.} $\mathcal{L}(\cdot)=1/2\|\cdot\|^2$, and the $\ell_1$ loss, \emph{i.e.} $\mathcal{L}(\cdot)=\|\cdot\|_1$; both will be studied in Section~\ref{sect:experiments}. 

\begin{figure}
     \centering
     \includegraphics[width=0.4\textwidth, clip]{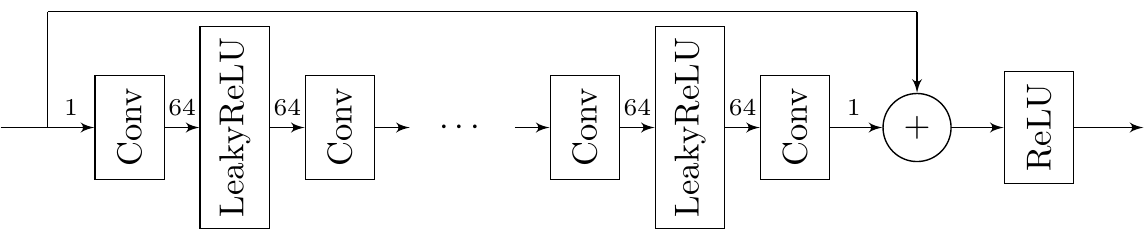}
    \caption{Architecture of the proposed denoising DNN, inspired from \citet{zhang2017beyond} and \citet{pesquet2020learning}, containing 20 convolutional layers in total. ``Conv'' denotes convolutional layers. The number of channels is given above the arrows. Notice the $\text{ReLU}$ layer at the output, ensuring that our reconstructed RI images are positive.}
    \label{fig:arch}
\end{figure}

\section{Simulations and results}
\label{sect:experiments}

\subsection{Overall simulation set up}
\label{ssect:testimmetrics}

\subsubsection{Test images}

Our test image database consists of four radio images, namely 3c353, Hercules~A, Centaurus~A and Cygnus~A, all resized to a size of $n=512 \times 512$, shown in Figure~\ref{fig:gdths}. These images all have a peak value equal to 1 and dynamic ranges above $10^{4}$. The simulations involving these images will target reconstruction dynamic ranges of the order of $10^{4}$. We emphasise that this is orders of magnitude higher than typical values in the PnP literature \citep{zhang2017beyond, zhang2020plug, pesquet2020learning}, and therefore significantly more challenging. 

\subsubsection{Benchmark methods}
\label{sssec:bench}

We benchmark AIRI with (i) uSARA  (Algorithm~\ref{algo:sara_fb}) solving the unconstrained problem \eqref{eq:uSARA}, (ii) SARA \citep{onose2017accelerated, thouvenin2021parallel}, solving the constrained problem \eqref{eq:SARA}, (iii) the multiscale variant of CLEAN implemented in the WSClean software \citep{offringa2017wsclean}, and (iv) a UNet \citep{jin2017deep} trained end-to-end on a large database to recover groundtruths from a database of dirty images.

All simulations involving uSARA rely on a value for $\rho$ using \eqref{eq:heuristic_uSARA} including the $1/3$ correction factor, while the values of $\lambda$ are fine-tuned around \eqref{eq:heuristic_uSARA} to optimize the reconstruction quality. As for SARA, we rely on the prescriptions of  \citet{thouvenin2021parallel} for setting its hyperparameters. While all other methods use natural weighting of the data, CLEAN makes use of uniform weighting, which relies on the density of the Fourier sampling for enhanced resolution (see Section~\ref{ssect:measmod} and Appendix~\ref{appendix:ri_model})\footnote{The specific command used is: \texttt{wsclean -multiscale -niter 30000 -weight uniform -gain 0.1 -mgain 0.6 -threshold 0.001 -no-reorder -minuvw-m 0.001 -auto-mask 0.001 -mem 1}.}. We underline that the CLEAN restored image is in units of Jansky per beam area, in contrast with SARA-based methods, AIRI, and the UNet, where the associated estimated images are in the units of Jansky per pixel. Therefore, in what follows, CLEAN images are rescaled by the area of the CLEAN beam, for visualization purposes. Finally, we note that we always show the dirty image computed with a uniform weighting scheme, and rescaled in the $[0,1]$ range for visualization purposes.

Regarding the end-to-end deep learning approach, we use a Deep Residual UNet from \citet{zhang2020plug} and train it to reconstruct a target image in \eqref{eq:invpb} from its dirty version in \eqref{eq:invpbim}. We generate a database $(\bm{u}_s, \operatorname{Re}\{\bm{\Phi}_s^\dagger \bm{y}_s\})_{1\leq s \leq S}$, where for every $s$, $\bm{y}_s$ represents the visibility vector computed from the groundtruth $\bm{u}_s$ and for a measurement operator $\bm{\Phi}_s = \bm{\mathrm{G}}_s\bm{\mathrm{FZ}}$ generated from a random pointing direction in the sky. The randomization ensures that the UNet training database contains a rich and diverse set of sampling patterns. We borrow the procedure detailed in Section~\ref{ssect:dr_enhancement} to enhance the nominal dynamic range of the groundtruth database $(\bm{u}_s)_{1 \leq s \leq S}$, and equate it to the target dynamic range of reconstruction. The UNet is then trained to minimize the $\ell_1$ loss between the groundtruth $\bm{u}_s$ and the reconstruction $\operatorname{UNet}(\operatorname{Re}\{\bm{\Phi}_s^\dagger \bm{y}_s\})$. The UNet is trained for 800 epochs with the Adam algorithm.

\subsubsection{Evaluation metrics}

For AIRI-$\ell_2$, AIRI-$\ell_1$, uSARA, SARA, and the UNet, the image estimates $\widehat{\bm{x}}$ are the reconstructed model images, while the image estimate considered for CLEAN is the restored image. The performance of all algorithms is assessed qualitatively via inspection of $\widehat{\bm{x}}$, alongside the residual images, defined as $\widehat{\bm{r}}=\beta \bm{\Phi}^\dagger (\bm{y}-\bm{\Phi} \widehat{\bm{x}})$, where $\beta$ is a normalization factor\footnote{The normalization factor is given by $\beta=1/\max_i(\bm{\Phi}^\dagger\bm{\Phi}{\bm{\delta}})_{i} $, where $\bm\delta$ is the image with value 1 at the phase centre and 0 otherwise. Hence, the dirty beam given by $\beta\bm{\Phi}^\dagger\bm{\Phi}{\bm{\delta}}$, has a peak value equal to 1.}. Given the high dynamic ranges of interest, the estimated images are shown in logarithmic scale, using the mapping \mbox{$\operatorname{rlog} = \bm{x} \mapsto \operatorname{log}_{10}(10^3\bm{x}+1)/3$}. The residual images are displayed in linear scale.

We also consider two quantitative evaluation metrics, namely the SNR and logSNR of the estimated images $\widehat{\bm{x}}$ with respect to the groundtruth $\overline{\bm{x}}$. The standard SNR metric (in dB) is defined as $\text{SNR}(\widehat{\bm{x}},\overline{\bm{x}}) = 20\operatorname{log}_{10}(\|\overline{\bm{x}}\|/\|\overline{\bm{x}}-\widehat{\bm{x}}\|)$. It simply corresponds to a logarithmic version of the norm of the error image $\overline{\bm{x}}-\widehat{\bm{x}}$. The logSNR (in dB) is the SNR evaluated on a logarithmic version of the intensity images. This metric is introduced given the high dynamic ranges of interest, and assigns more relative weight to low intensities than the SNR. Our definition is $\text{logSNR}(\widehat{\bm{x}},\overline{\bm{x}}) = \text{SNR}(\operatorname{rlog}(\widehat{\bm{x}}),\operatorname{rlog}(\overline{\bm{x}}))$. 

\subsubsection{DNN architecture \& training particulars}

Firstly, as highlighted already, the denoising losses that we use for $\mathcal{L}$ in our training loss \eqref{eq:training_loss} are chosen to be either the $\ell_2$ loss or the $\ell_1$ loss, giving rise to AIRI-$\ell_2$ and AIRI-$\ell_1$.

Secondly, we choose a basic DNN architecture for the denoiser $\operatorname{D}$, in the form of a modified DnCNN \citep{zhang2017beyond} where the batch normalization layers have been removed, as in \citet{pesquet2020learning}. But in this work, a ReLU has been added as the last layer. This is a simple way to enforce positivity at the output of the denoiser, and therefore of the final reconstruction, much necessary as the developed AIRI algorithm targets intensity (versus polarization) imaging (see Figure~\ref{fig:arch}).

Thirdly, the DNN is trained for $10^5$ epochs with the Adam algorithm \citep{kingma2014adam} and learning rate $10^{-4}$, the learning rate being divided by 2 every $2\times 10^4$ epochs. $(\bm{z}_s,\bm{u}_s)$ in \eqref{eq:training_loss} are of size $46\times 46$, generated according to \eqref{eq:hdrnoisy} and \eqref{eq:exponentiation} from patches randomly extracted from the low dynamic range database of $512\times 512$ images $(\bm{u}^{\text{low}}_s)_{1 \leq s \leq S}$, itself augmented with random rotations and zooms. The Jacobian regularization in \eqref{eq:training_loss} is computed with a power method with a number of iterations increased to 10 in the later stages of the training to improve the precision of the computation. As observed in \citet{pesquet2020learning}, starting the optimization of the network from a pretrained state  with $\kappa=0$ in \eqref{eq:training_loss} enhances the results significantly.

Finally, we have observed that performing, at each iteration of the PnP algorithm, random flips and $90$ degrees rotations of the image before applying the DNN, and inverting the transform after the denoising step, can significantly boost the reconstruction quality. We therefore adopt this strategy, inspired by \citet{zhang2020plug}.

\subsection{Experiment 1: validating the denoiser for denoising}

\subsubsection{Simulation setup}

The PnP approach suggests that the best denoisers should provide the best regularizers. In this first experiment, we compare the denoising capabilities of the AIRI-$\ell_2$ and AIRI-$\ell_1$ denoisers on one side, and the average sparsity proximal operator propelling uSARA and SARA on the other side. To this end, we focus on a denoising problem 
\begin{equation}
\label{eq:denoising}
    \bm{y} = \overline{\bm{x}}+ \bm{e},
\end{equation}
where $\bm{y}\in\RR^{n}$ is the data,  $\overline{\bm{x}}\in \RR^{n}$ is the groundtruth image, and $\bm{e}\in \RR^{n}$ is the realization of an i.i.d. Gaussian random noise with zero mean and standard deviation $\tau$. In other words, \eqref{eq:denoising} is a degenerate case of problem \eqref{eq:invpb} with $\bm{\Phi}=\textbf{I}$. We consider the 3c353 image as a groundtruth $\overline{\bm{x}}$. As the following experiments for the full validation of AIRI-$\ell_2$ and AIRI-$\ell_1$ operate at a target dynamic range around $10^{4}$, we design this preliminary pure denoising experiment with a similarly challenging target dynamic range, and set the observation noise level to $\tau=10^{-4}$.

Firstly, problem \eqref{eq:denoising} can be solved using AIRI via Algorithm \ref{algo:pnp}. As the image peaks at $1$, no rescaling of the inverse problem in \eqref{eq:invpb} is needed (see Section~\ref{sect:heuristic}). In this simple case, the gradient in \eqref{eq:gradf} reads $\nabla f(\bm{x})=\bm{x}-\bm{y}$ with Lipschitz constant in \eqref{eq:lipgradf} $L=1$. Taking $\gamma=1/L=1$, Algorithm \ref{algo:pnp} trivially boils down to a simple non-iterative application of $\operatorname{D}$ to the data, leading to the denoised image 
\begin{equation}
\label{eq:exp1Ddenoising}
\widehat{\bm{x}}=\operatorname{D}(\bm{y}).
\end{equation}
The AIRI-$\ell_2$ and AIRI-$\ell_1$ denoisers were trained with a noise level $\sigma=\tau=10^{-4}$ as per \eqref{eq:heuristic}, on the proposed synthetic training database, and following the procedure explained in Section~\ref{ssect:methodology}. Following the procedure detailed in Section~\ref{ssect:dr_enhancement}, the chosen exponentiation parameter for on-the-fly dynamic range enhancement of the database is $a=10^3$. Furthermore, we set the regularization parameter in the training loss \eqref{eq:training_loss} to $\kappa = 10^{-9}$  and $\kappa = 10^{-5}$ for AIRI-$\ell_2$ and AIRI-$\ell_1$ denoisers respectively. The safety margin parameter is set to $\varepsilon = 5\times 10^{-2}$. These values are chosen here as they are the ones to ensure the stability of Algorithm~\ref{algo:pnp} in our next experiments. 

Secondly, problem \eqref{eq:denoising} can also be solved in a proximal approach with uSARA, setting $\bm{\Phi}=\textbf{I}$. Interestingly, in this case, the uSARA objective boils down to the objective defining the average sparsity proximal operator (see definition \eqref{eq:prox_def}) of $\lambda r$, for the average sparsity prior $r$ in \eqref{eq:sara_prior}, and $\lambda=\tau$ following the reasoning around equation \eqref{eq:heuristic_uSARA}. Therefore, applying Algorithm~\ref{algo:sara_fb} results in computing the denoised image by simply applying the average sparsity proximal operator with appropriate $\lambda$:
\begin{equation}
\label{eq:exp1proxdenoising}
\widehat{\bm{x}}=\operatorname{prox}_{\lambda r}(\bm{y}).
\end{equation}
We note, given $\nabla f(\bm{x})=\bm{x}-\bm{y}$ and $L=1$, and choosing $K=1$ and $\gamma=1/L=1$, that Algorithm~\ref{algo:sara_fb} simplifies to the sequential application of $\operatorname{prox}_{\gamma  \lambda g}(\bm{y}, \bm{\mathrm{W}}_i)$ with only the weights being updated at each iteration. The algorithm was run with convergence criteria to $\xi_1=6\times 10^{-6}$ Algorithm~\ref{algo:sara_fb}, and $\xi_2 = 10^{-4}$ in Algorithm~\ref{algo:dfb}.

\subsubsection{Experimental results}

\begin{figure}
\captionsetup[subfigure]{justification=centering}%
     \begin{subfigure}[t]{0.49\linewidth}
         \centering%
         \includegraphics[trim={0 5em 0 10em}, height=0.112\textheight, clip]{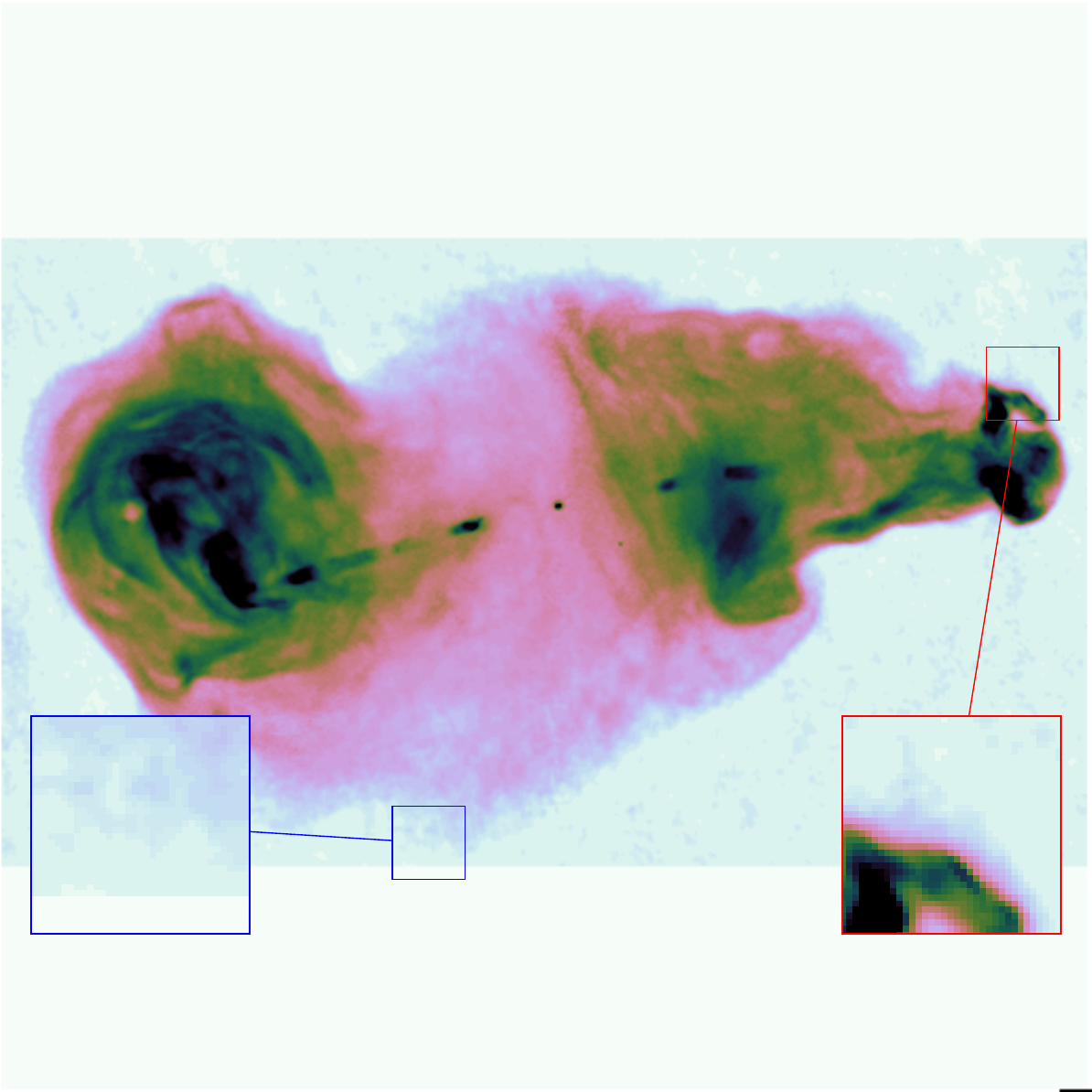}
\vspace{-0.5em}
         \caption{Groundtruth}
         \label{fig:dn:true}
     \end{subfigure}
     \hspace{-0.5em}
     \begin{subfigure}[t]{0.49\linewidth}
              \includegraphics[height=0.65\textwidth]{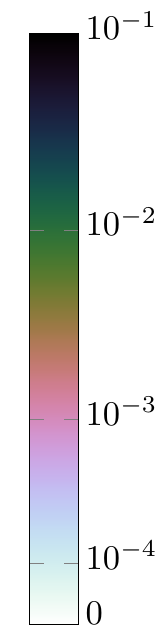}
     \end{subfigure}
     
     \begin{subfigure}[b]{0.49\linewidth}
         \centering
         \includegraphics[trim={0 5em 0 10em}, height=0.112\textheight, clip]{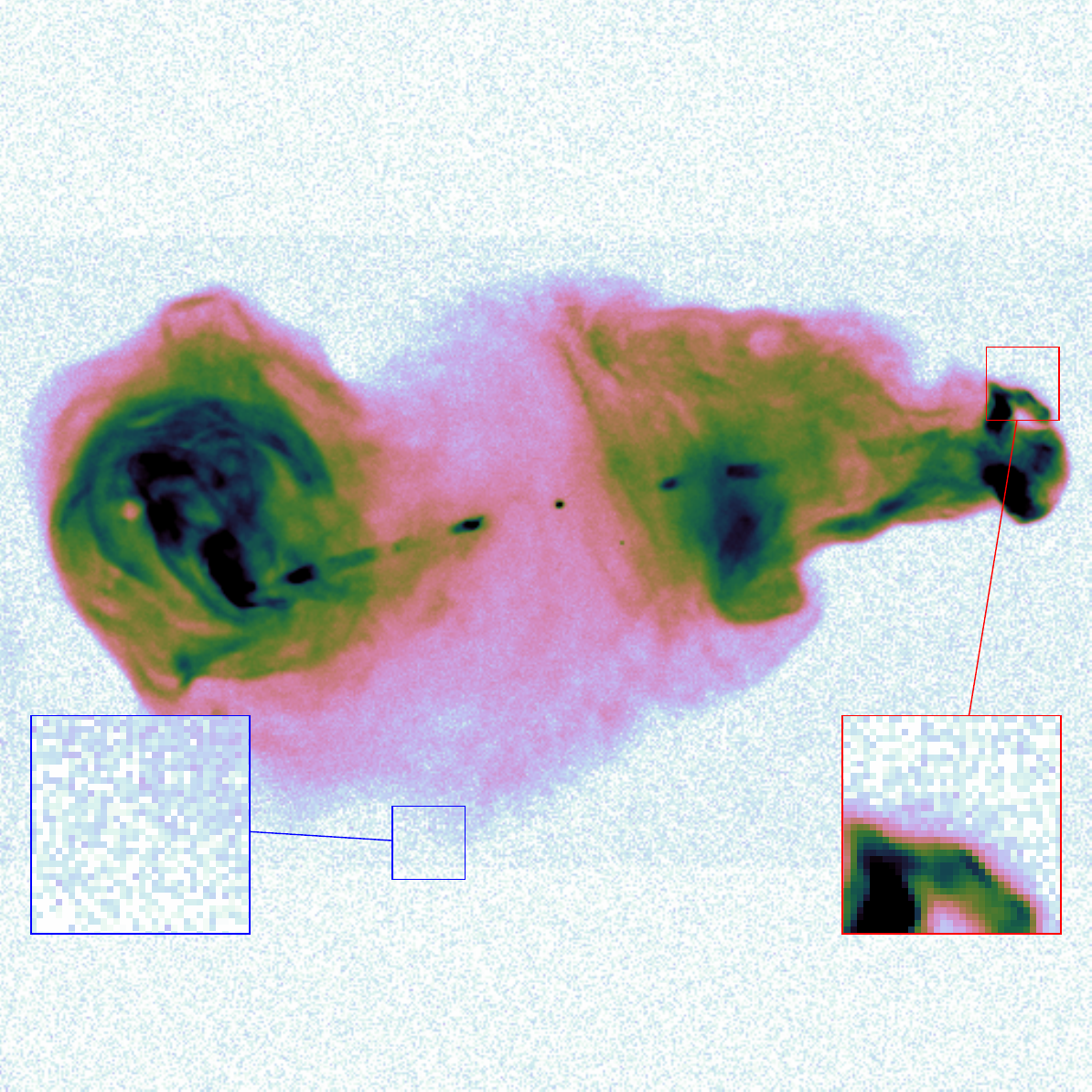}
\vspace{-0.5em}
         \caption{Noisy image \\ $(47.6\, \text{dB},  24.1\, \text{dB})$}
         \label{fig:dn:noisy}
     \end{subfigure}
     \hfill
     \begin{subfigure}[b]{0.49\linewidth}
         \centering
         \includegraphics[trim={0 5em 0 10em}, height=0.112\textheight, clip]{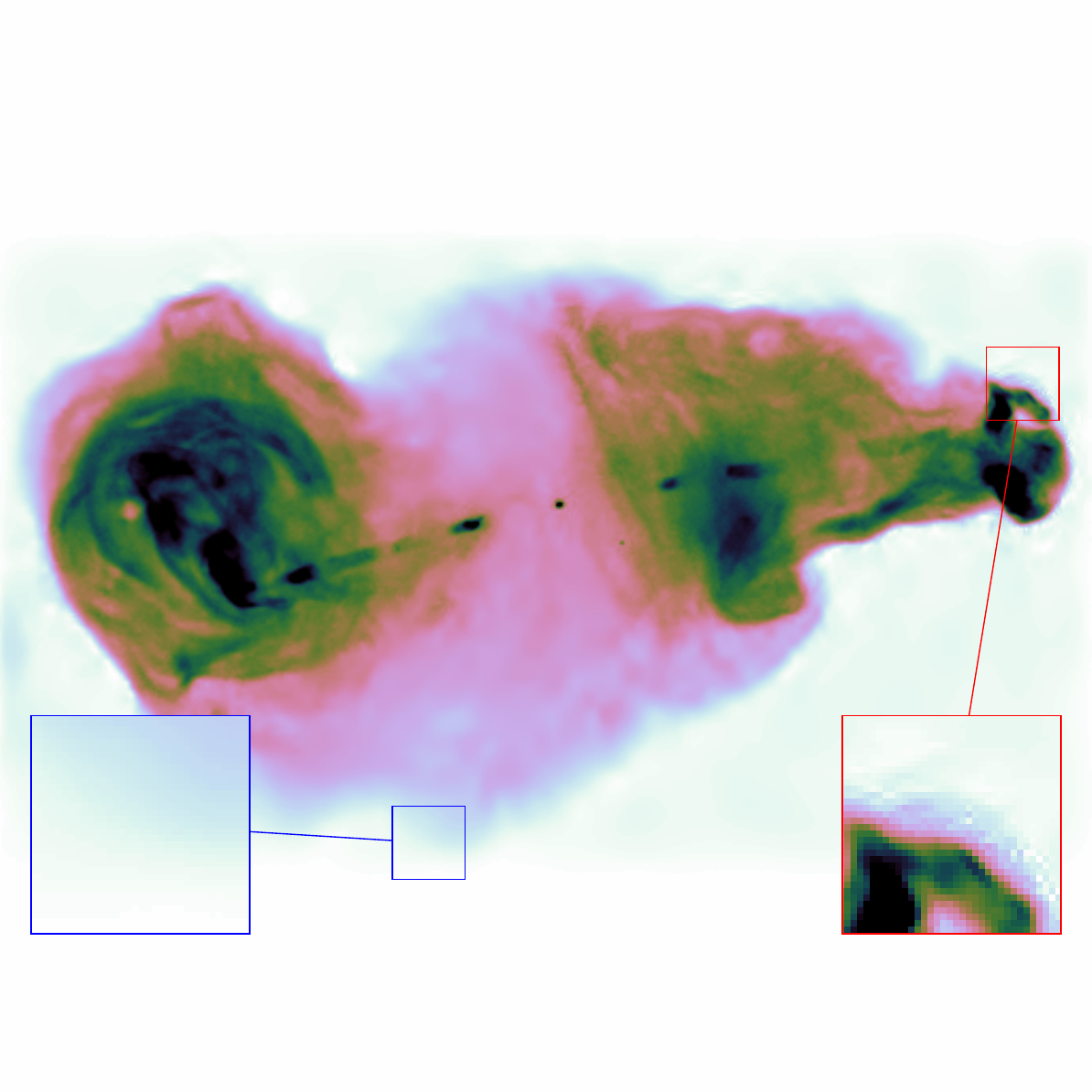}
\vspace{-0.5em}
         \caption{Average sparsity prox.~operator \\ $(53.8\, \text{dB}, 33.1\, \text{dB})$}
         \label{fig:dn:sara}
     \end{subfigure}
     
      \begin{subfigure}[b]{0.48\linewidth}
         \centering
         \includegraphics[trim={0 5em 0 10em}, height=0.112\textheight, clip]{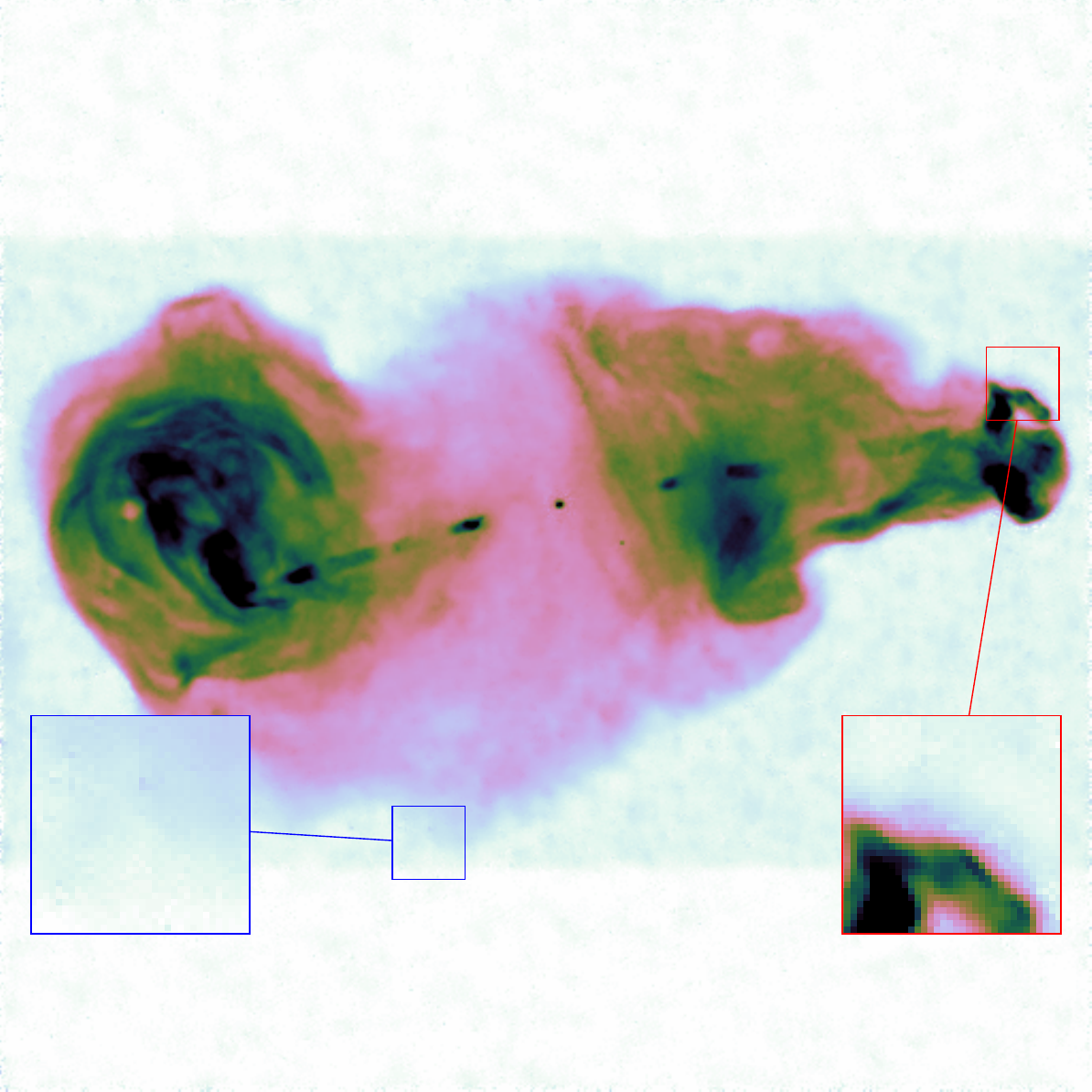}
\vspace{-2em}
         \caption{AIRI-$\ell_2$ denoiser \\ $(53.7\, \text{dB}, 35.1\, \text{dB})$}
         \label{fig:dn:pnpl2}
     \end{subfigure}
     \hfill
      \begin{subfigure}[b]{0.48\linewidth}
         \centering
         \includegraphics[trim={0 5em 0 10em}, height=0.112\textheight, clip]{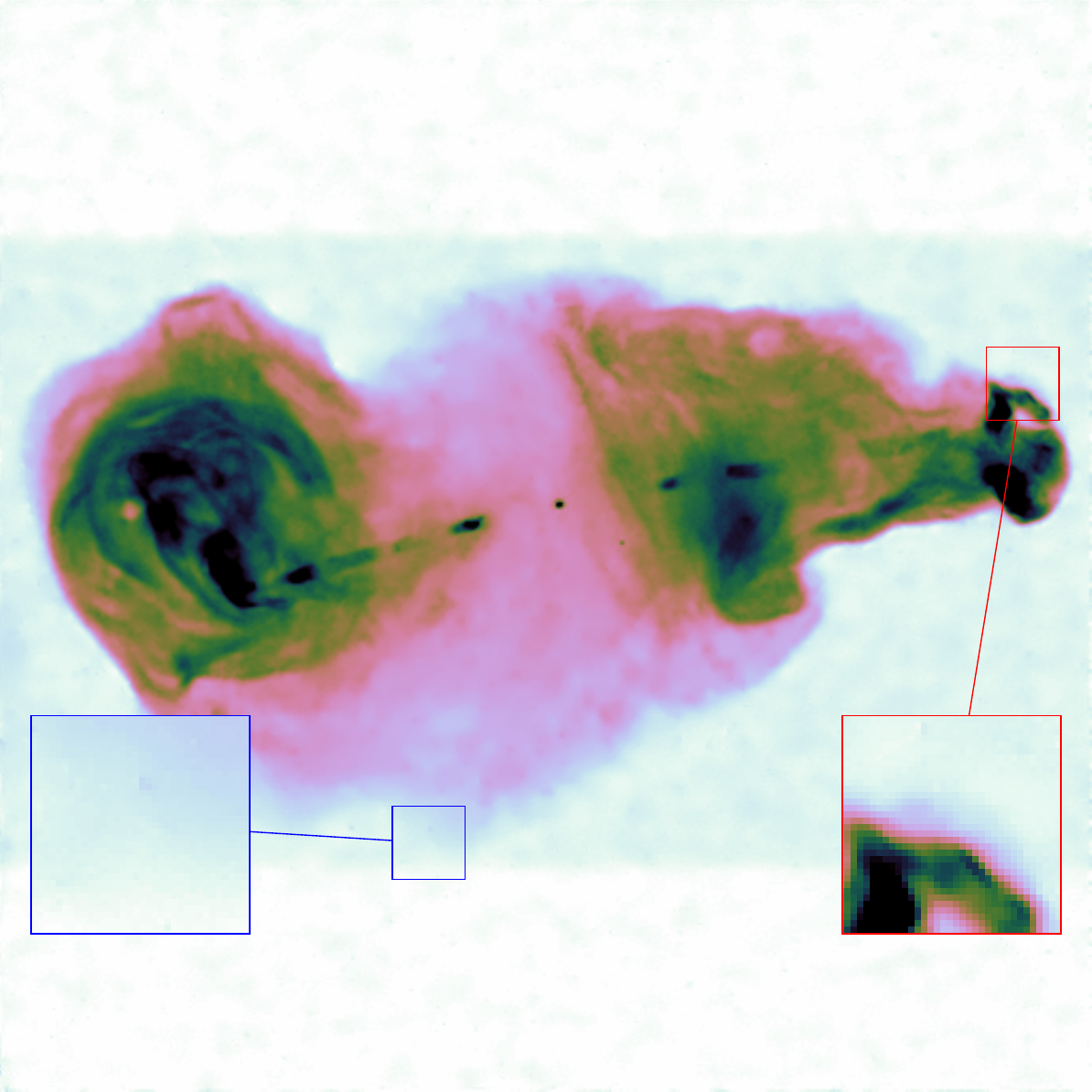}
\vspace{-2em}
         \caption{AIRI-$\ell_1$ denoiser \\ $(53.9\, \text{dB}, 35.8\, \text{dB})$ }
         \label{fig:dn:pnpl1}
     \end{subfigure}
\vspace{-1em}
\caption{Experiment 1 results: Denoising results  obtained with the proposed AIRI-$\ell_2$ and AIRI-$\ell_1$ denoisers in comparison with the averaged sparsity proximal operator. The respective groundtruth and noisy images are displayed in \ref{fig:dn:true} and \ref{fig:dn:noisy}. Denoising results obtained with the averaged sparsity proximal operator are shown in \ref{fig:dn:sara},  with AIRI-$\ell_2$ in \ref{fig:dn:pnpl2} and with AIRI-$\ell_1$ in \ref{fig:dn:pnpl1}. All images are displayed in logarithmic scale. The colourbar is saturated at $10^{-1}$ for clarity. Values of the obtained  evaluation metrics (SNR, logSNR) are indicated below their associated images.}
\label{fig:denoising}
\end{figure}

Denoising results are displayed in Figure~\ref{fig:denoising}. All three denoisers appear to be highly effective, both in terms of SNR and logSNR and visual denoising quality. The SNR values are very comparable. In terms of logSNR the AIRI-$\ell_1$ denoiser achieves better performance than the AIRI-$\ell_2$ denoiser, itself superior to the average sparsity proximal operator. Visually, the AIRI-$\ell_1$ denoiser also achieves the reconstruction with the best resolution and least amount of artefacts (see zooms).

These results suggests that the AIRI-$\ell_1$ and AIRI-$\ell_2$ denoisers do encapsulate a superior prior model to the average sparsity model encapsulated in the corresponding proximal operator.

\subsection{Experiment 2: validating the training noise level heuristic}

\subsubsection{Simulation setup}
\label{ssect:simul_setup}
Using our four $512\times512$ test images (see Figure~\ref{fig:gdths}), we study the impact of the noise level $\sigma$ involved in the training of the AIRI-$\ell_2$ and AIRI-$\ell_1$ denoisers according to the procedure set in Section~\ref{sect:pnp_algos}, with the aim to validate the heuristic \eqref{eq:heuristic}. We also study the optimal value of the regularization parameter $\lambda$ of uSARA around the heuristic \eqref{eq:heuristic_uSARA}. The performance of AIRI-$\ell_2$ and AIRI-$\ell_1$ in terms of reconstruction quality is compared to that of uSARA.

\begin{figure}
     \centering
     \begin{subfigure}[b]{0.49\linewidth}
         \centering
         \includegraphics[width=\textwidth]{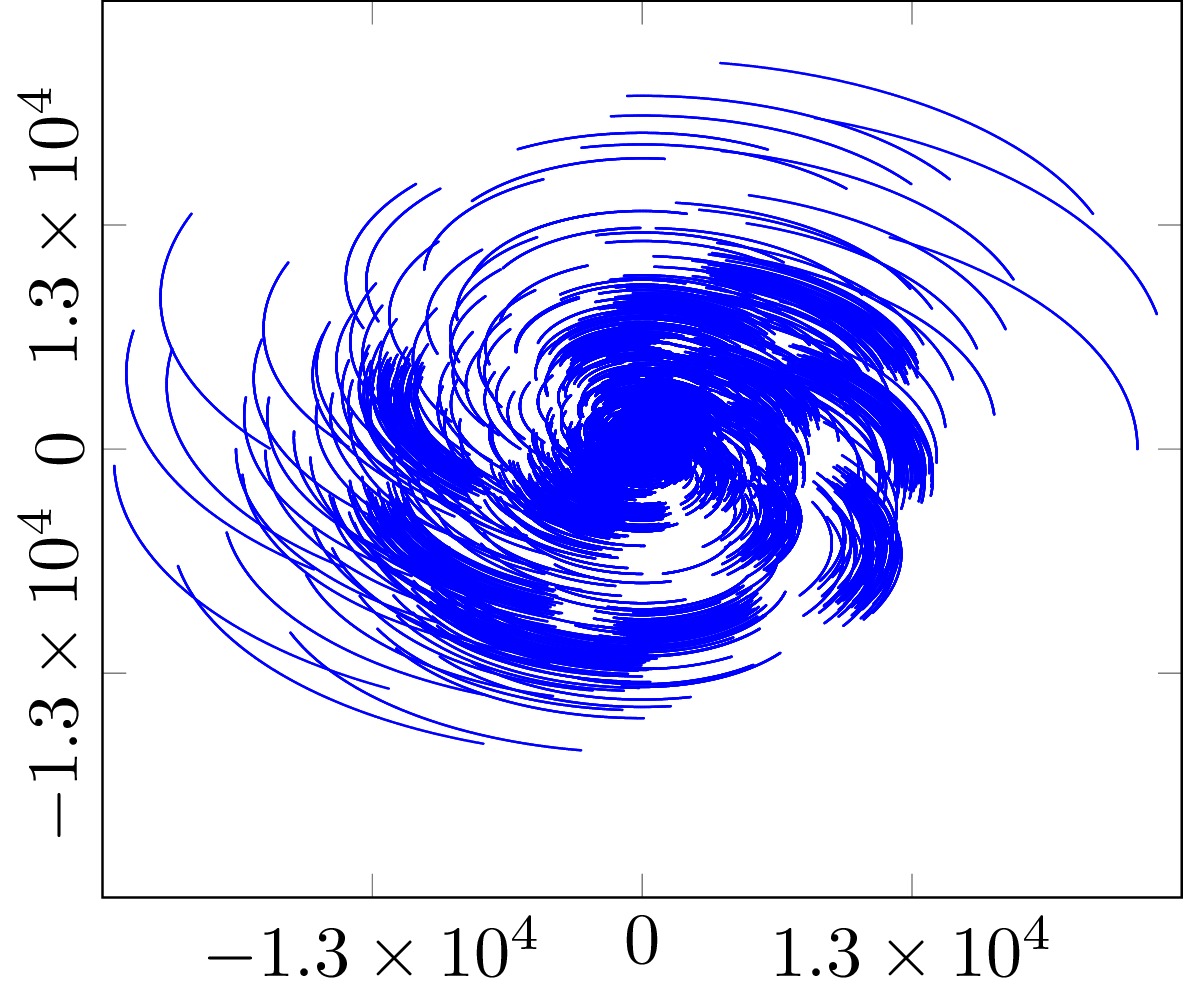}
\vspace{-2em}
         \caption{}
         \label{fig:samp:seed0}
     \end{subfigure}
     \hfill
     \begin{subfigure}[b]{0.49\linewidth}
         \centering
         \includegraphics[width=\textwidth]{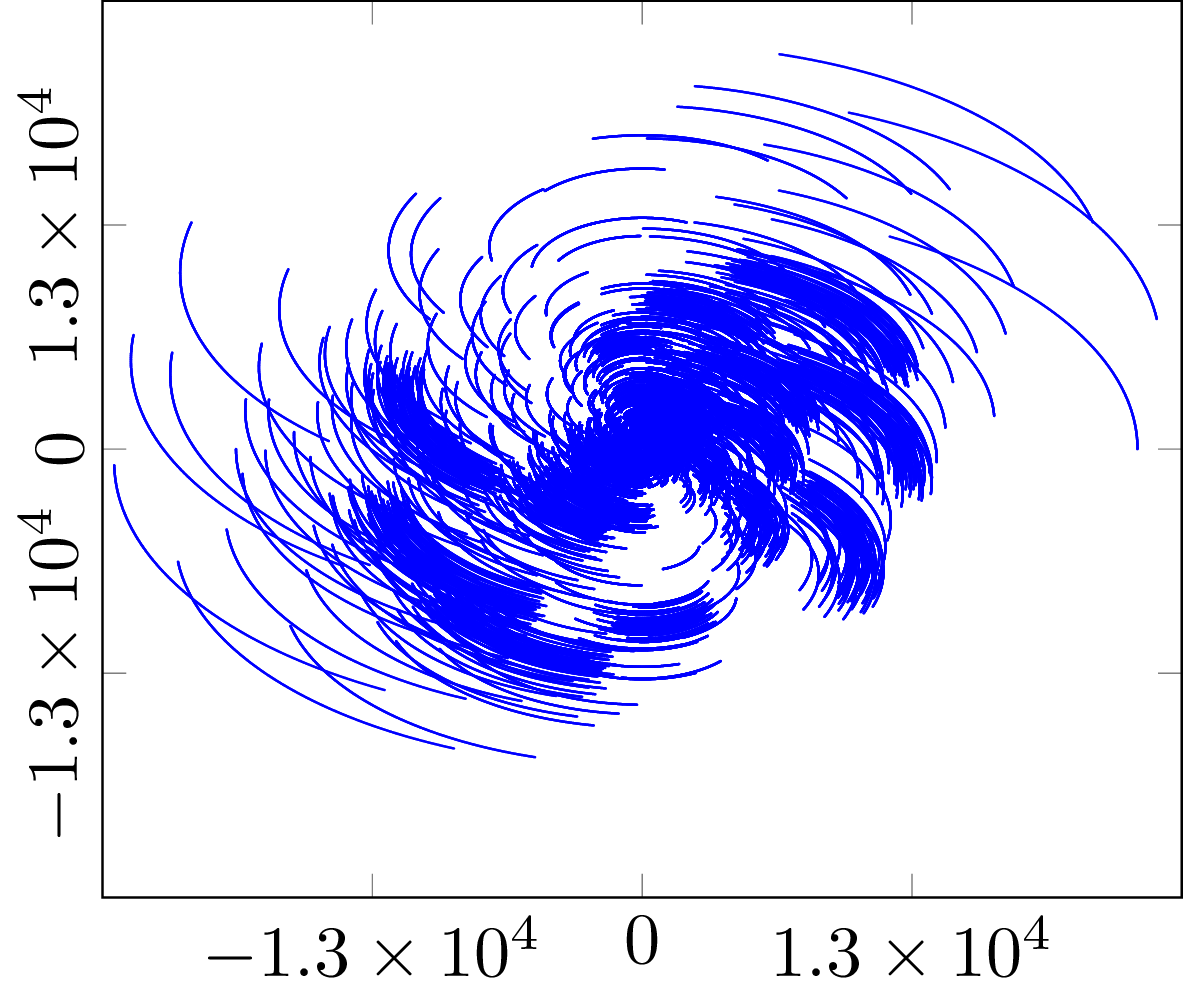}
\vspace{-2em}
         \caption{}
         \label{fig:samp:seed1}
     \end{subfigure}
     
     \hfill
     \begin{subfigure}[b]{0.49\linewidth}
         \centering
         \includegraphics[width=\textwidth]{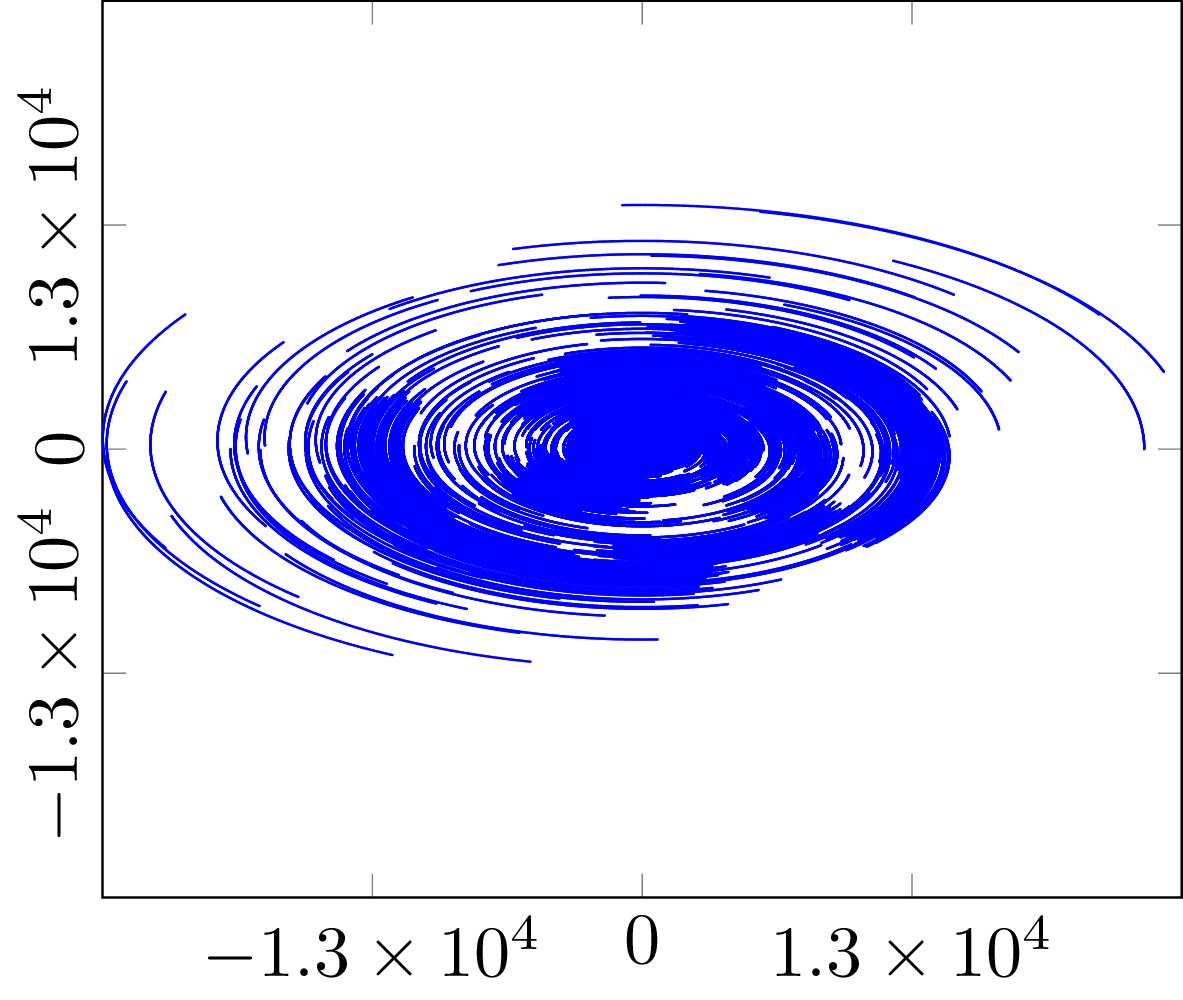}
\vspace{-2em}
         \caption{}
         \label{fig:samp:seed4}
     \end{subfigure}
     \hfill \begin{subfigure}[b]{0.49\linewidth}
         \centering
         \includegraphics[width=\textwidth]{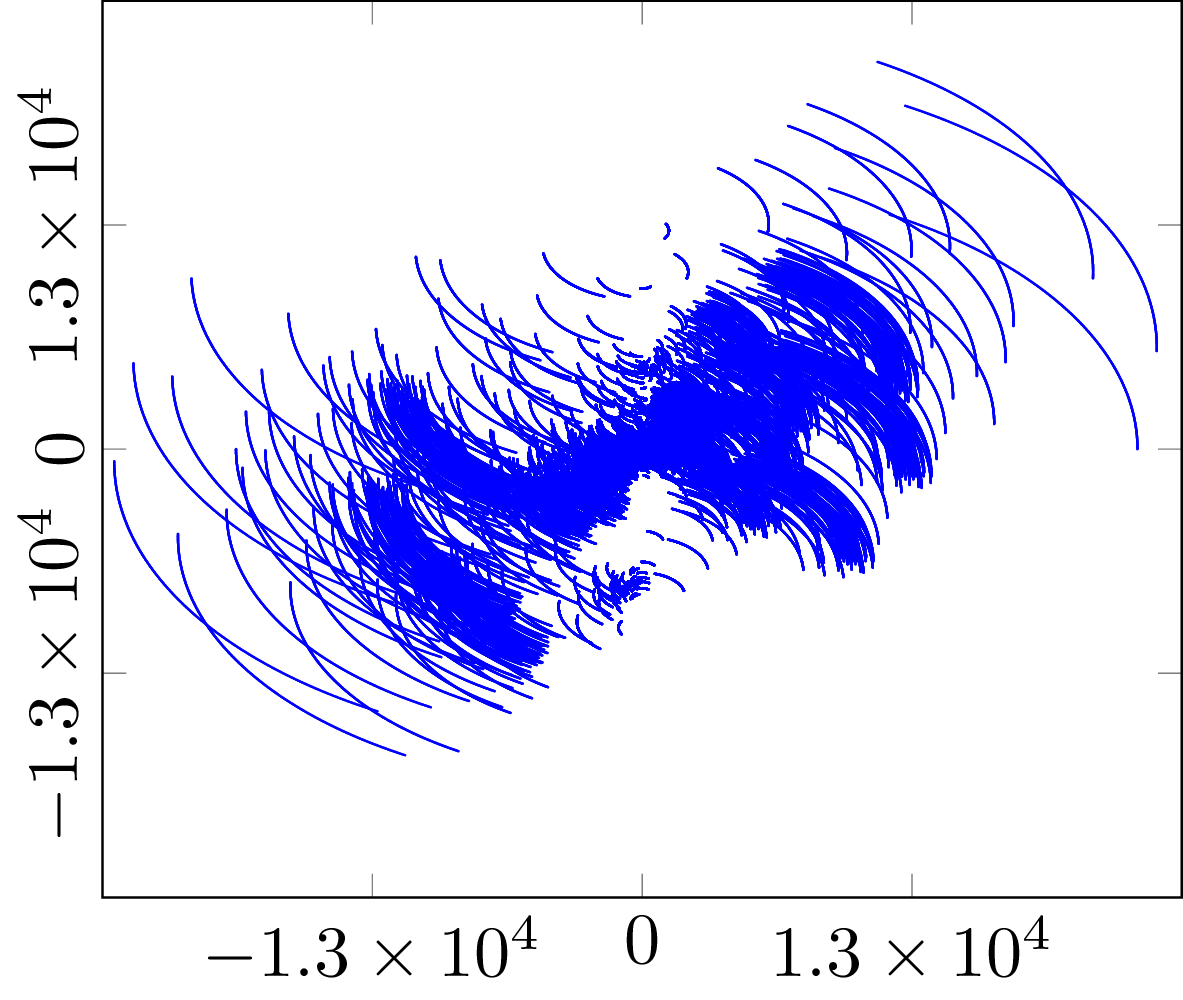}
\vspace{-2em}
         \caption{}
         \label{fig:samp:seed5}
     \end{subfigure}\hfill
     \hfill
\vspace{-1em}
\caption{Experiments 1-3: MeerKAT $uv$-coverages simulated for four different telescope pointing directions and fixed total observation duration $\Delta T = 4\,\text{h}$. The components are measured in units of the wavelength of the observation. Each $uv$-pattern contains $m=806400$ points.}
\label{fig:samp}
\vspace{-1em}
\end{figure}

\begin{figure}
     \centering
     \begin{subfigure}[b]{0.49\linewidth}
         \centering
         \includegraphics[width=\textwidth]{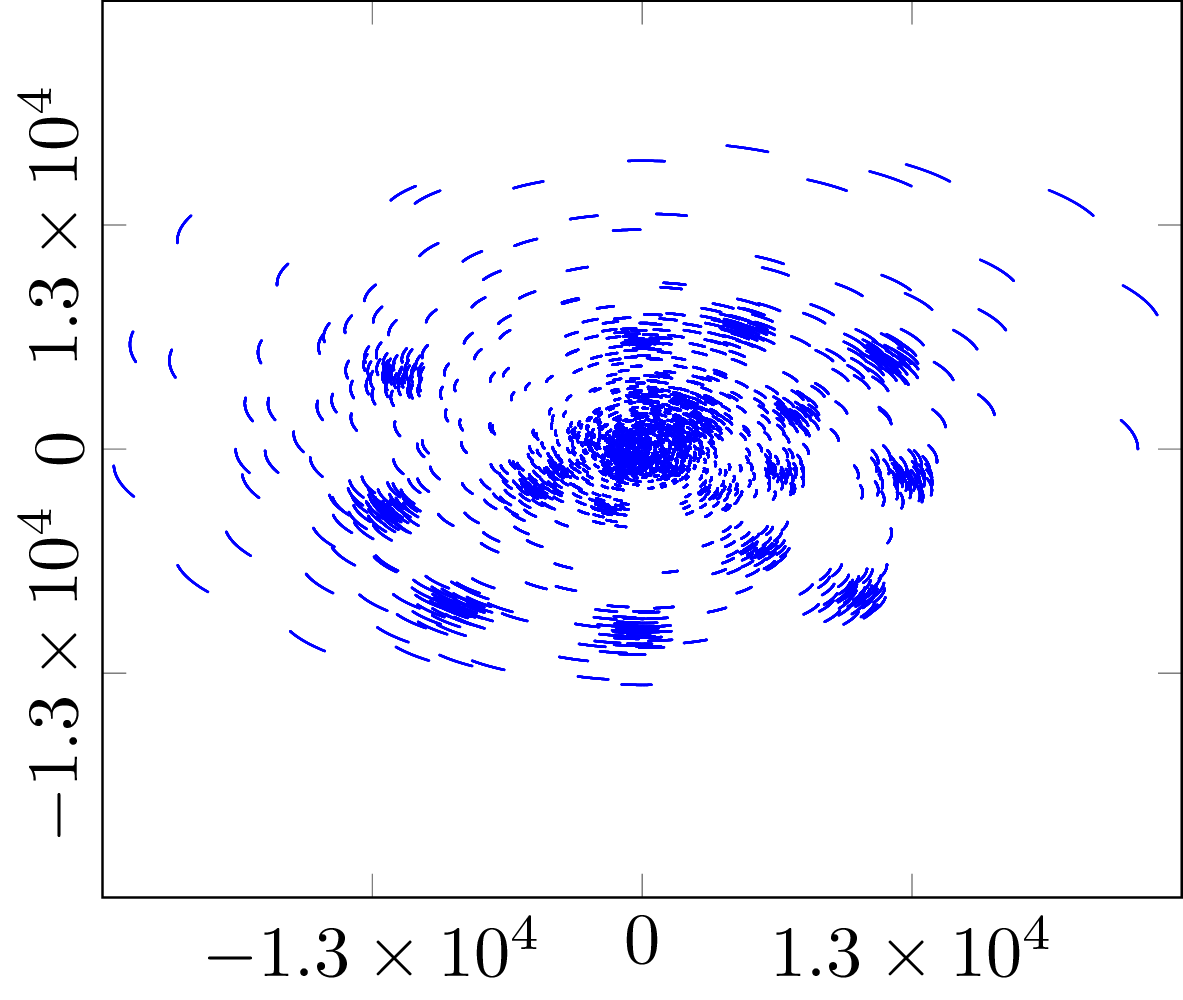}
\vspace{-2em}
         \caption{$\Delta T = 1\,\text{h}$}
         \label{fig:samp:seed2dt1}
     \end{subfigure}
     \hfill
     \begin{subfigure}[b]{0.49\linewidth}
         \centering
         \includegraphics[width=\textwidth]{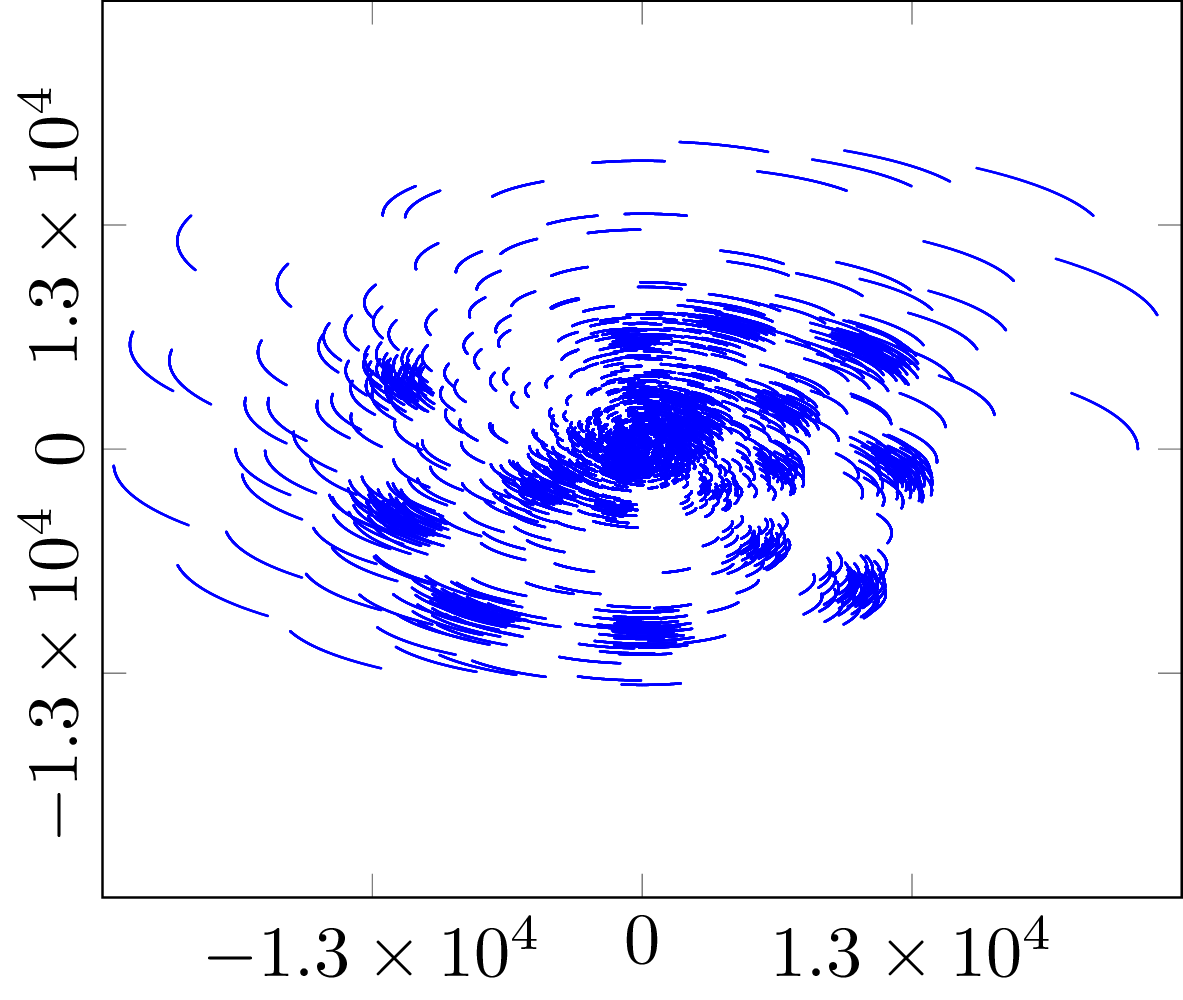}
\vspace{-2em}
         \caption{$\Delta T = 2\,\text{h}$}
         \label{fig:samp:seed2dt2}
     \end{subfigure}
     
      \begin{subfigure}[b]{0.49\linewidth}
         \centering
         \includegraphics[width=\textwidth]{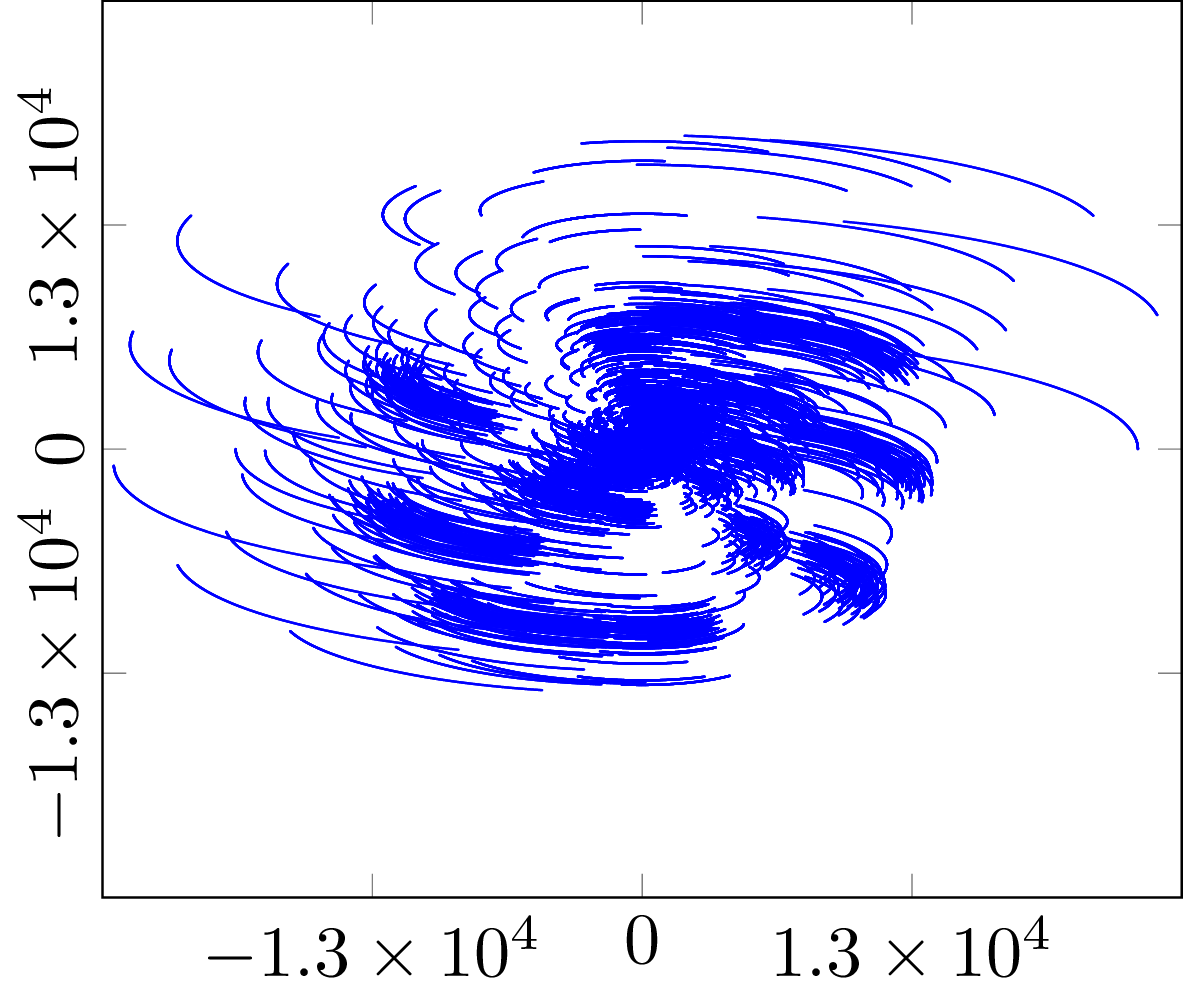}
\vspace{-2em}
         \caption{$\Delta T = 4\,\text{h}$}
         \label{fig:samp:seed2dt4}
     \end{subfigure}
     \hfill
      \begin{subfigure}[b]{0.49\linewidth}
         \centering
         \includegraphics[width=\textwidth]{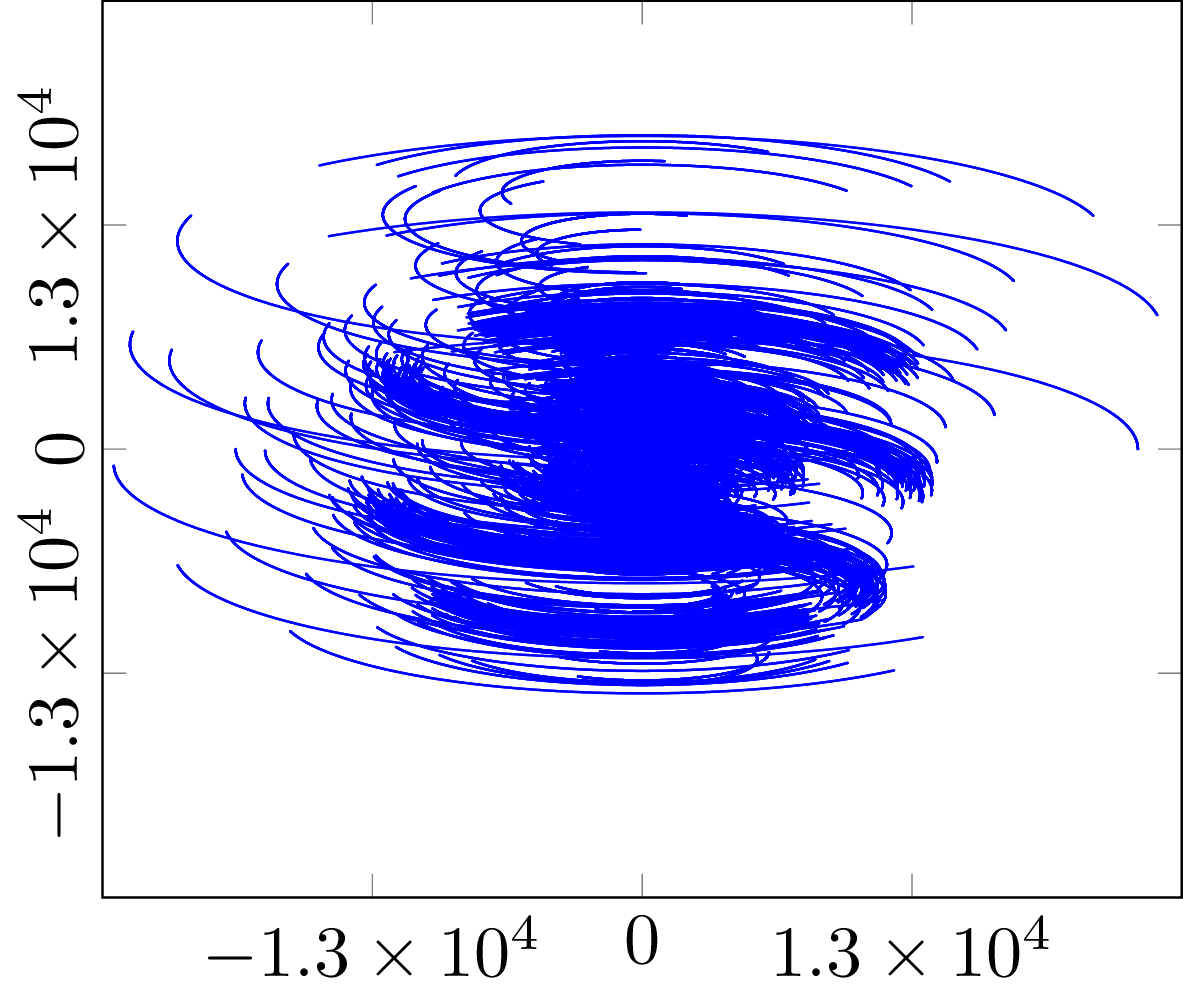}
\vspace{-2em}
         \caption{$\Delta T = 8\,\text{h}$}
         \label{fig:samp:seed2dt8}
     \end{subfigure}
     
\vspace{-1em}
    \caption{Experiments 1-3: MeerKAT $uv$-coverages associated with the four observation durations and fixed telescope pointing direction (complementary the four shown in Figure \ref{fig:samp}). The numbers of points in the $uv$-patterns are (a) $\Delta T = 1 \text{h}: 201600$, (b) $\Delta T = 2 \text{h}: 403200$, (c) $\Delta T = 4\,\text{h}: 806400$, and (d) $\Delta T = 8\,\text{h}: 1612800$.}
    \label{fig:samplings_seed2}
\vspace{-1em}
\end{figure}

We simulate RI observations utilizing simulated $uv$-coverages of the radio telescope MeerKAT. The $uv$-patterns correspond to five randomly selected pointing directions displayed in Figures~\ref{fig:samp} and \ref{fig:samp:seed2dt4}, and two of the four total observation durations displayed in Figure~\ref{fig:samplings_seed2}: $\Delta T \in\{4\,\text{h}, 8\,\text{h}\}$. The observed wavelength is $0.3\text{m}$ and the simulated sampling rate of the telescope is fixed to 100 points per hour. The data size thus increases linearly with $\Delta T$. Data are simulated following the model described in \eqref{eq:invpb}, with input SNR defined as $\text{iSNR}=20 \log_{10}\left( \|\bm{\Phi}\overline{\mathbf{x}}\|/ \tau \right)=30~\text{dB}$. We finally note that the $uv$-patterns extend to the edge of the considered Fourier domain as shown in Figures~\ref{fig:samp} and \ref{fig:samplings_seed2}. In other words, we target mild, if any, super-resolution with respect to the nominal resolution of the observation, as set by the largest baseline.

Since our groundtruth images have peak values at $1$, no rescaling of the inverse problem in \eqref{eq:invpb} is needed for AIRI, and a common heuristic value is reached for $\sigma$ in \eqref{eq:heuristic} and $\gamma\lambda$ in \eqref{eq:heuristic_uSARA}. Obviously, $\tau$ and $L$ are functions of $\bm{\Phi}$, and therefore of both the pointing direction and total duration of observation. However, the heuristic values are within a $25\%$ variation range across pointing directions for each $\Delta T$. For simplicity, a single heuristic value is considered for all five $uv$-patterns associated with the same $\Delta T$, defined as $\nu_{\Delta T}$, taken to be the average of the values resulting for each pointing direction. The two resulting values are as announced associated with a target dynamic range around $10^4$: $\nu_4=1.4\times 10^{-4}$ for $\Delta T=4\,\text{h}$, and $\nu_8=9.3\times 10^{-5}$ for $\Delta T=8\,\text{h}$. The values probed for $\gamma\lambda$ and $\sigma$ are sampled a factor $\sqrt{2}$ around the heuristic and by steps of $2$ thereafter.

Following the study by \citet{pesquet2020learning}, we choose the value of $\gamma=1.98/L$ in Algorithms~\ref{algo:sara_fb} and \ref{algo:pnp}, which saturates the theoretical bound. The convergence criteria of Algorithms~\ref{algo:sara_fb} and \ref{algo:pnp} are set to $\xi_1 = \xi_3 = 5\times 10^{-6}$, with a maximum number of iterations of $6\times 10^3$, and we set $\xi_2 = 10^{-5}$ in Algorithm~\ref{algo:dfb}.

We train our AIRI-$\ell_2$ and AIRI-$\ell_1$ denoisers with optimized values $\kappa = 10^{-9}$ and $\kappa = 10^{-5}$ respectively in the training loss \eqref{eq:training_loss}, with $\varepsilon = 5\times 10^{-2}$, and observe that this ensures the stability of Algorithm~\ref{algo:pnp} despite a spectral norm of the Jacobian slightly above $1$. Following the procedure detailed in Section~\ref{ssect:dr_enhancement}, the exponentiation parameter for on-the-fly dynamic range enhancement of the database is set to $a=10^3$.

\subsubsection{Experimental results}
\begin{figure*}%
\captionsetup[subfigure]{justification=centering}%
    \centering%
    \begin{subfigure}[t]{0.249\linewidth}%
         \centering%
         \includegraphics[width=0.9\textwidth]{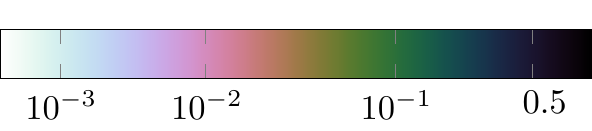}%
     \end{subfigure}%
    \hfill%
     \begin{subfigure}[t]{0.249\linewidth}%
         \centering%
         \phantom{\includegraphics[trim={0 5em 0 10em}, width=0.5\textwidth, clip]{./pictures/colorbar_log_bottom.pdf}}%
     \end{subfigure}%
     \hfill%
     \begin{subfigure}[t]{0.249\linewidth}%
         \centering
          \phantom{\includegraphics[trim={0 5em 0 10em}, width=0.5\textwidth, clip]{./pictures/colorbar_log_bottom.pdf}}
     \end{subfigure}
     \hfill%
     \begin{subfigure}[t]{0.249\linewidth}%
         \centering
          \phantom{\includegraphics[trim={0 5em 0 10em}, width=0.5\textwidth, clip]{./pictures/colorbar_log_bottom.pdf}}
     \end{subfigure}
     
     \begin{subfigure}[t]{0.249\linewidth}%
         \centering%
         \includegraphics[trim={0 5em 0 10em}, width=\textwidth, clip]{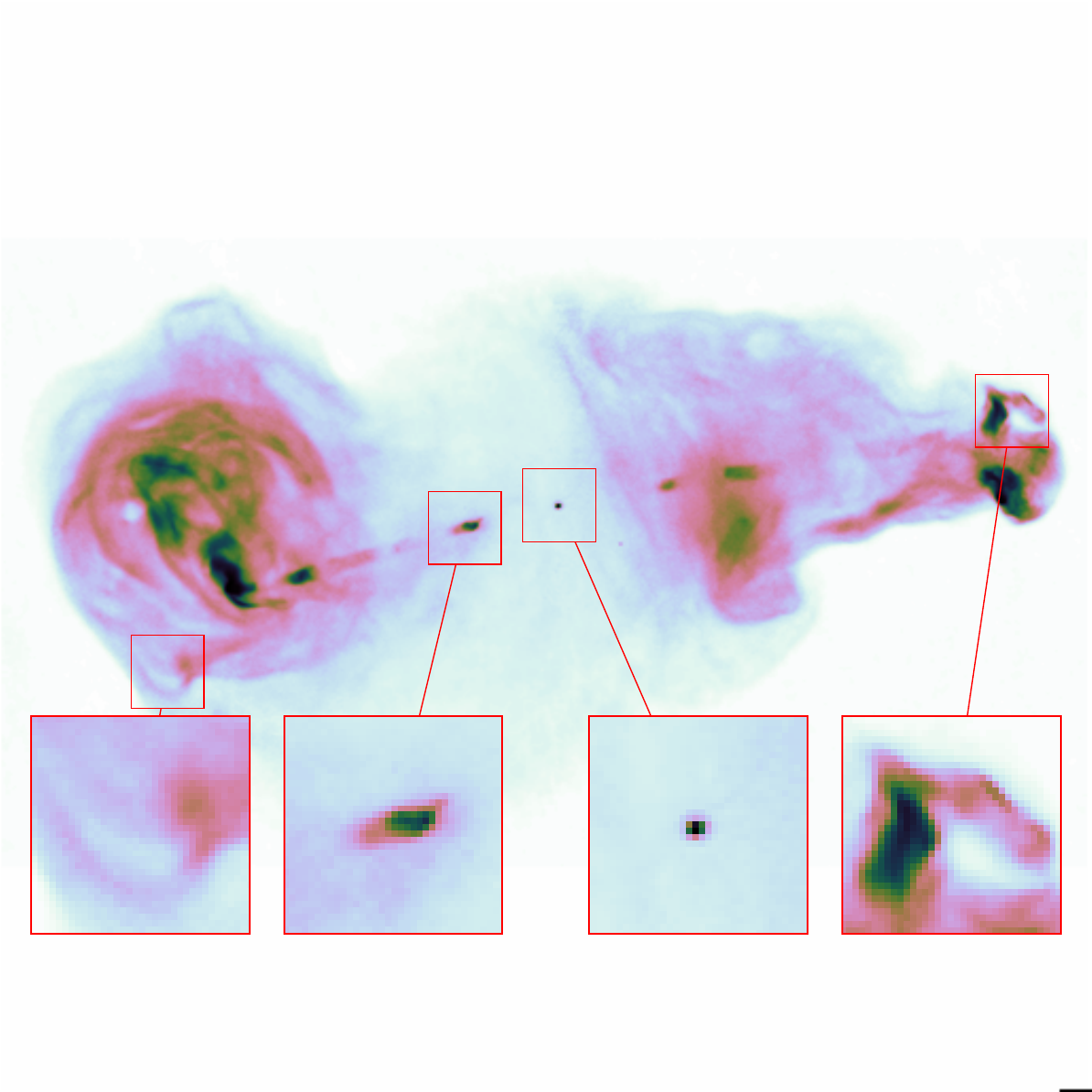}%
         \caption{Groundtruth}%
         \label{fig:3c353all:1}%
     \end{subfigure}%
     \hfill%
     \begin{subfigure}[t]{0.249\linewidth}%
         \centering%
         \includegraphics[trim={0 5em 0 10em}, width=\textwidth, clip]{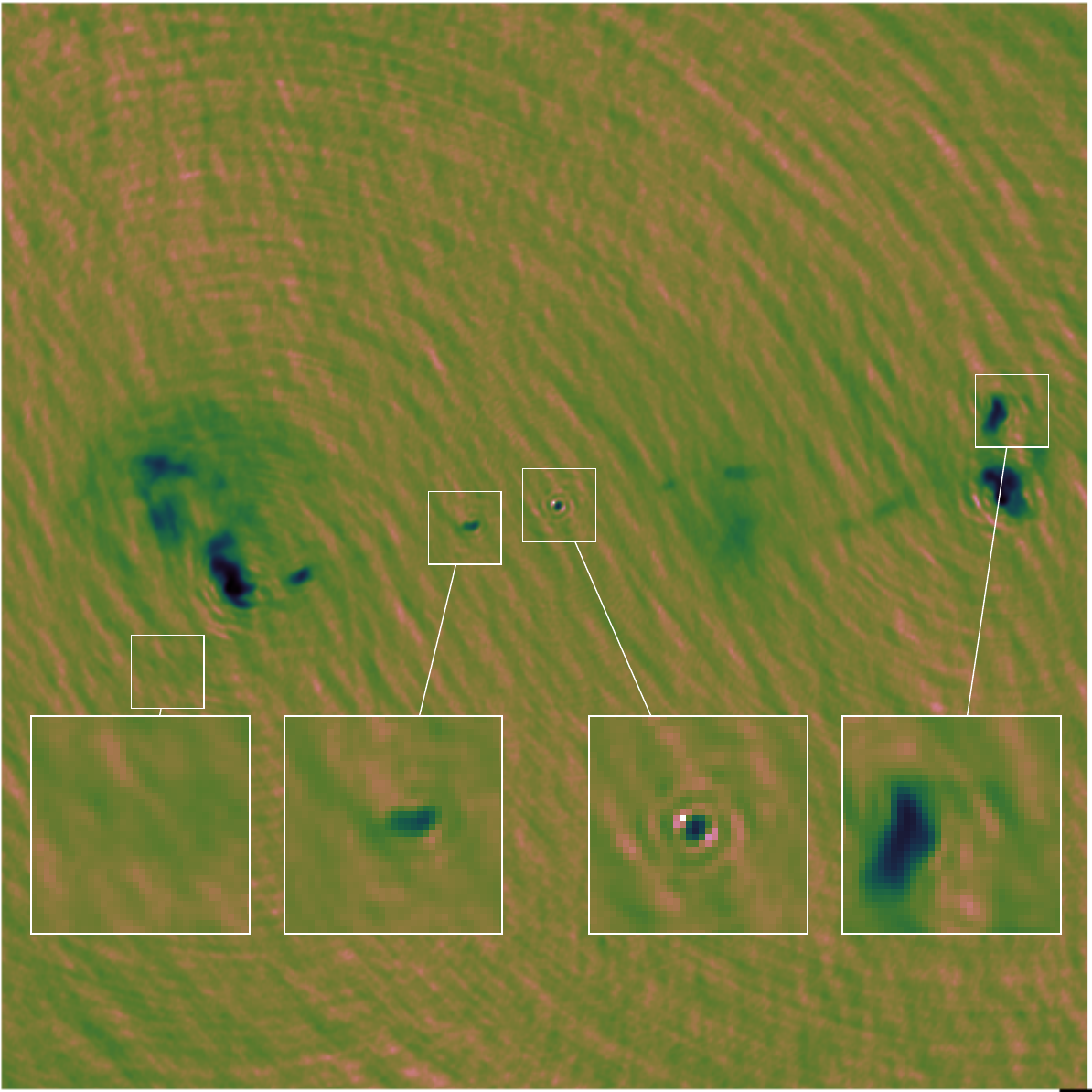}%
         \caption{Dirty image}%
         \label{fig:3c353all:2}%
     \end{subfigure}%
     \hfill%
     \begin{subfigure}[t]{0.249\linewidth}%
         \centering%
         \phantom{\includegraphics[trim={0 5em 0 10em}, width=\textwidth, clip]{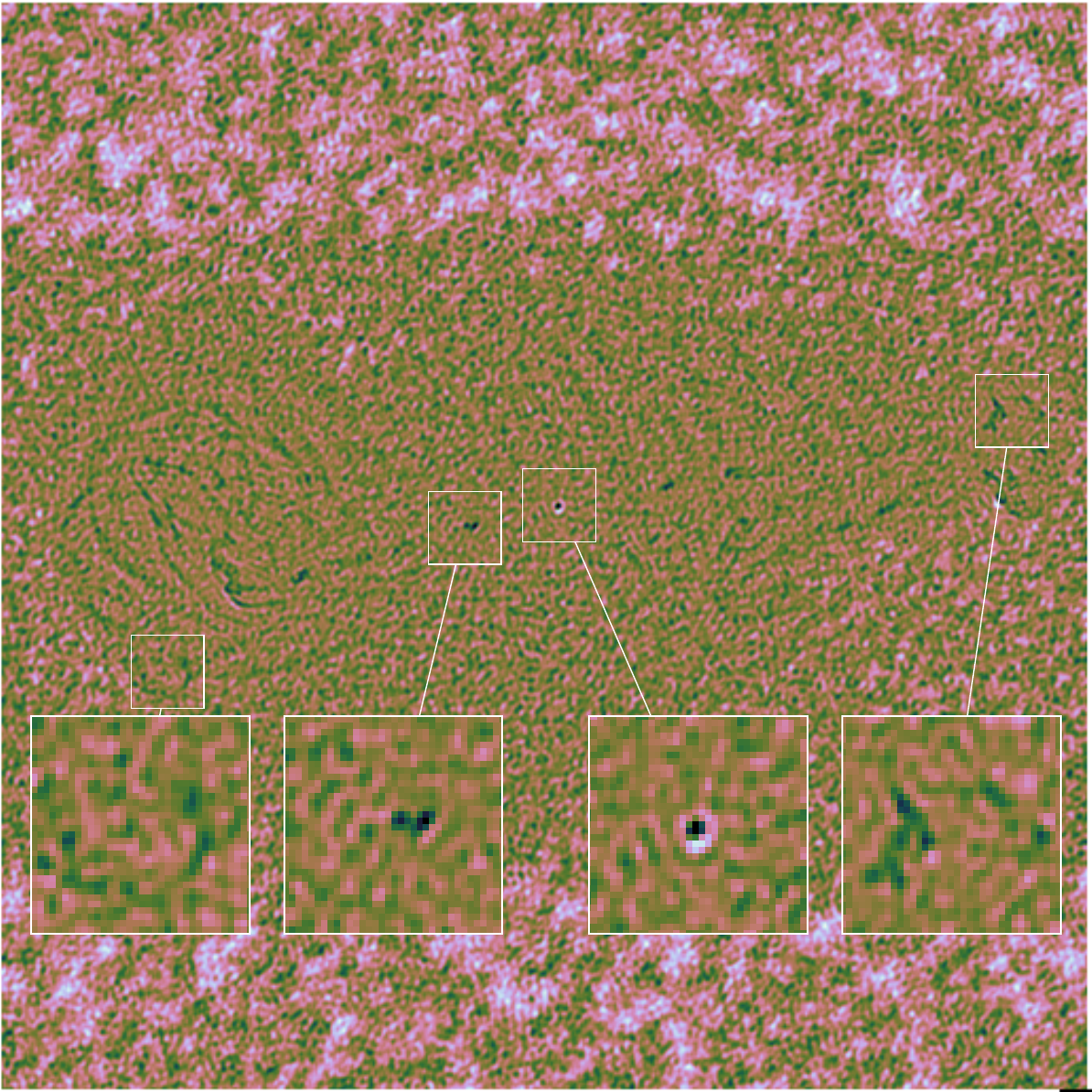}}%
     \end{subfigure}%
     \hfill%
     \begin{subfigure}[t]{0.249\linewidth}%
         \centering
          \phantom{\includegraphics[trim={0 5em 0 10em}, width=\textwidth,clip]{./pictures/3c353_residuals/zoom_sara_fb_residual_2.5e-05.pdf}}
     \end{subfigure}
     
     \begin{subfigure}[t]{0.249\linewidth}%
         \centering%
         \includegraphics[trim={0 5em 0 10em}, width=\textwidth, clip]{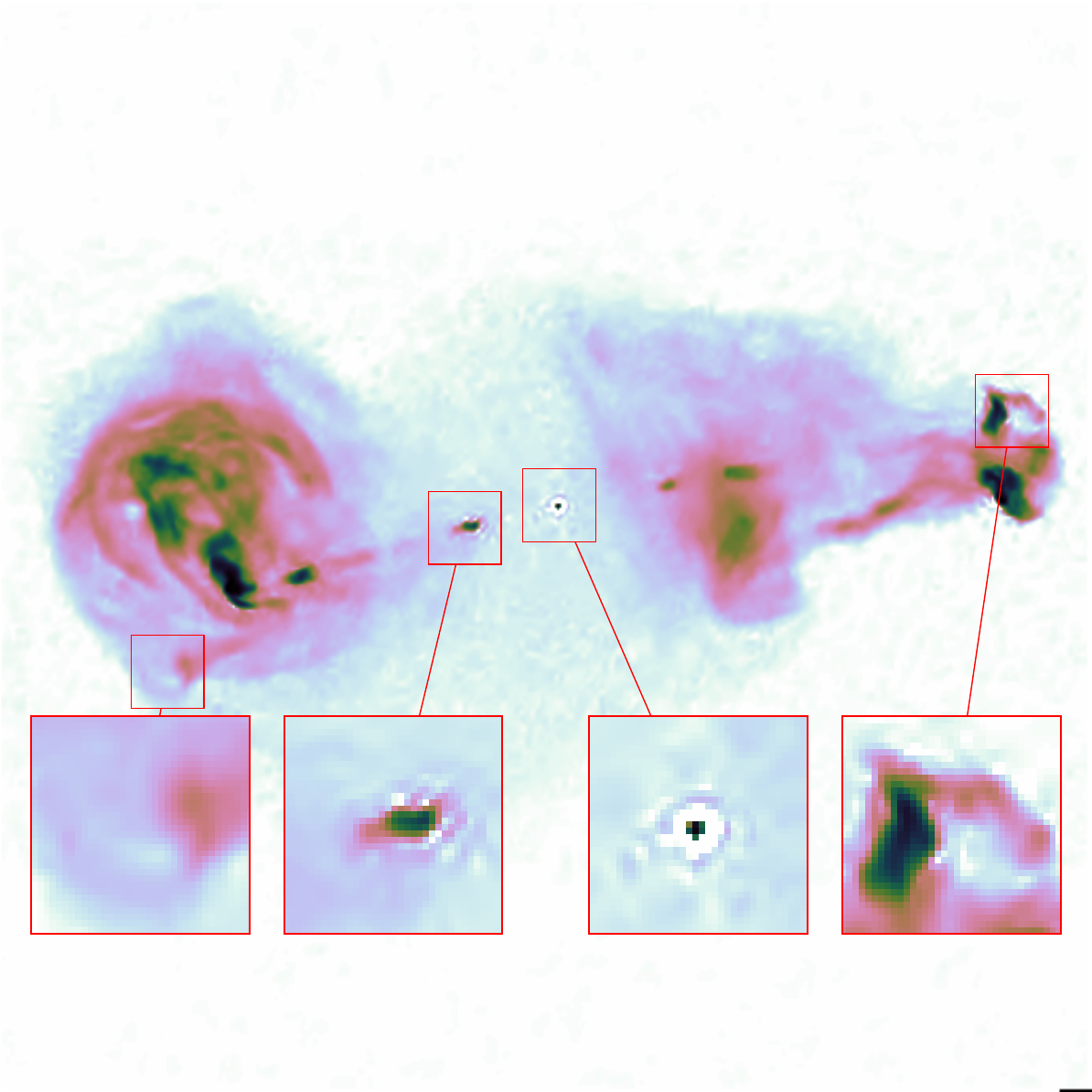}%
         \caption{uSARA, $\gamma\lambda=\nu_4/4\sqrt{2}$ \\ $(25.71\, \text{dB}, 23.47\, \text{dB})$}%
         \label{fig:3c353all:3}%
     \end{subfigure}%
     \hfill%
          \begin{subfigure}[t]{0.249\linewidth}%
         \centering%
         \includegraphics[trim={0 5em 0 10em}, width=\textwidth, clip]{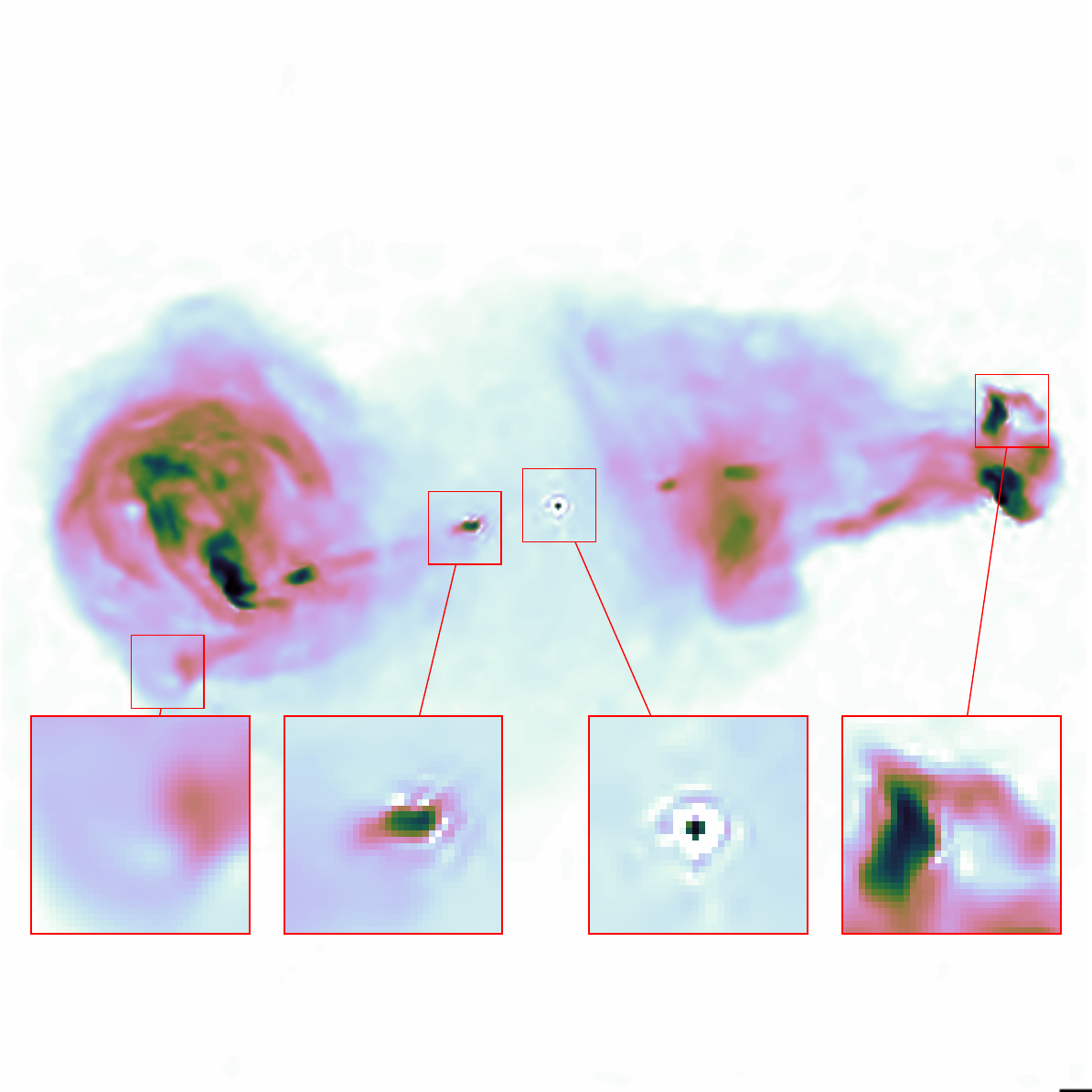}%
         \caption{uSARA, $\gamma\lambda = \nu_4/2\sqrt{2}$ \\ $(24.31\, \text{dB}, 25.12\, \text{dB})$}%
         \label{fig:3c353all:7}%
     \end{subfigure}%
     \hfill%
         \begin{subfigure}[t]{0.249\linewidth}%
         \centering%
         \includegraphics[trim={0 5em 0 10em}, width=\textwidth, clip]{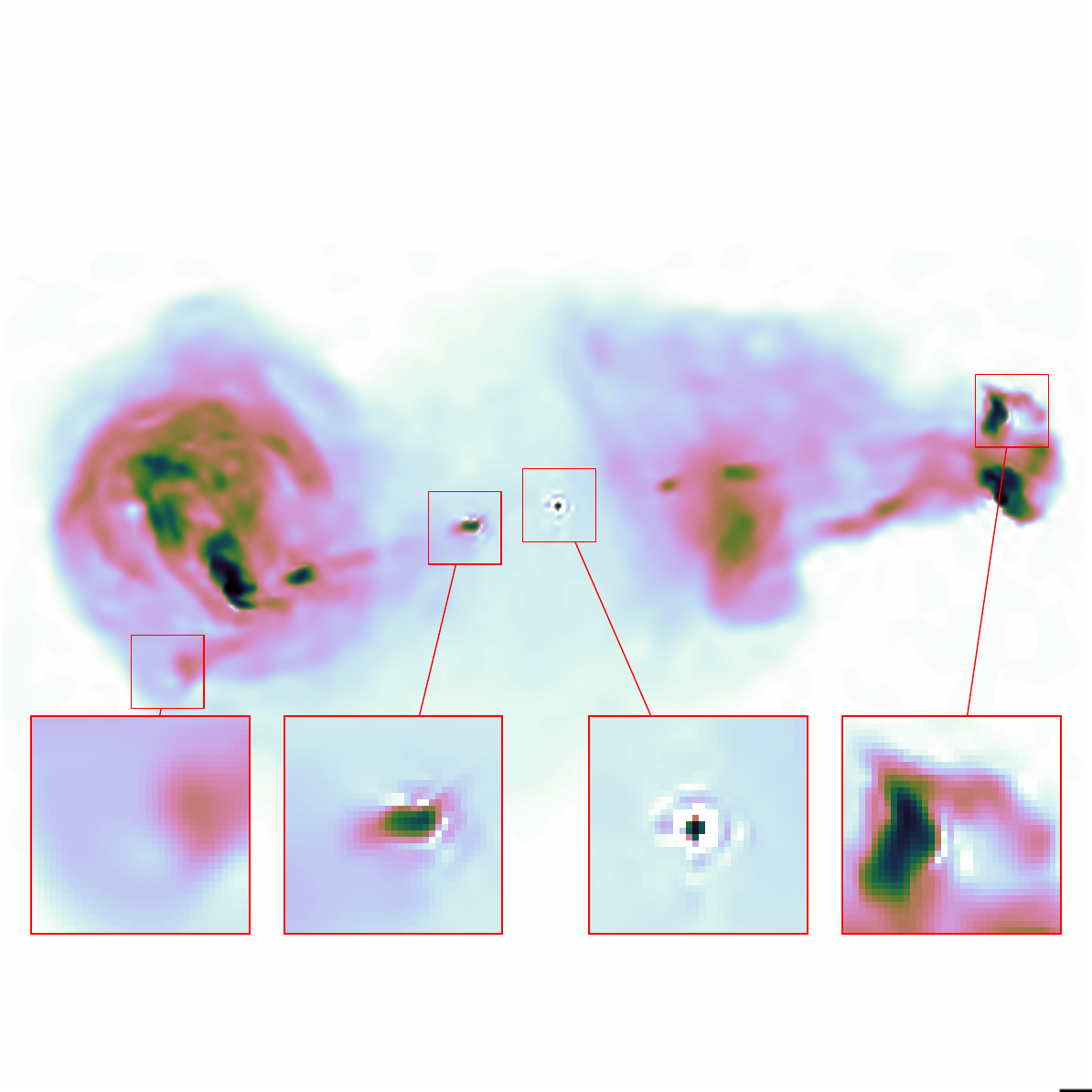}%
         \caption{uSARA, $\gamma\lambda=\nu_4/\sqrt{2}$ \\ $(22.77\, \text{dB}, 24.48\, \text{dB})$}%
         \label{fig:3c353all:11}%
     \end{subfigure}%
     \hfill%
    \begin{subfigure}[t]{0.249\linewidth}%
         \centering%
         \includegraphics[trim={0 5em 0 10em}, width=\textwidth, clip]{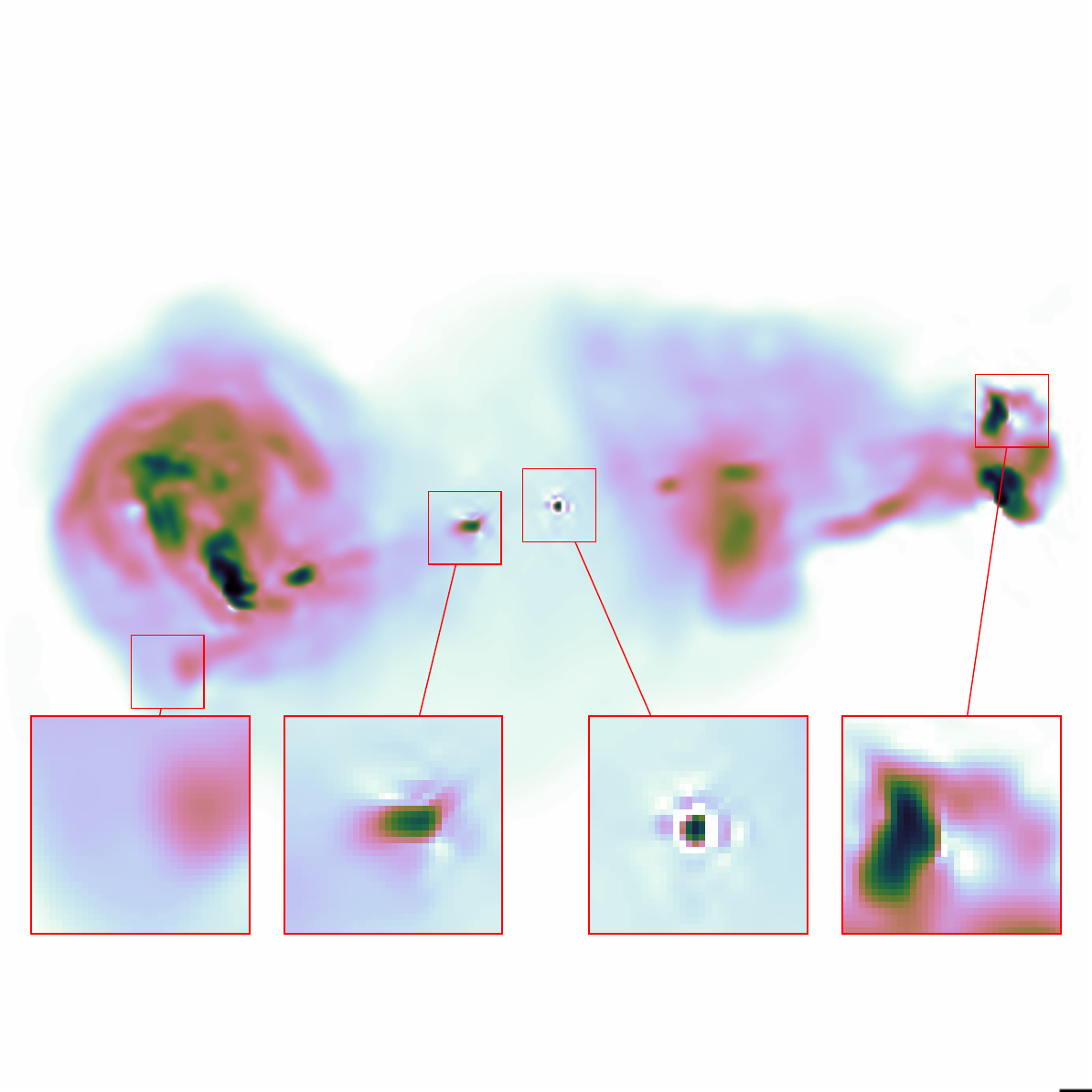}%
         \caption{uSARA, $\gamma\lambda = 2\sqrt{2}\nu_4$ \\ $(18.90\, \text{dB}, 21.54\, \text{dB})$}%
         \label{fig:3c353all:19}%
     \end{subfigure}%
     
    \begin{subfigure}[t]{0.249\linewidth}%
         \centering%
         \includegraphics[trim={0 5em 0 10em},width=\textwidth, clip]{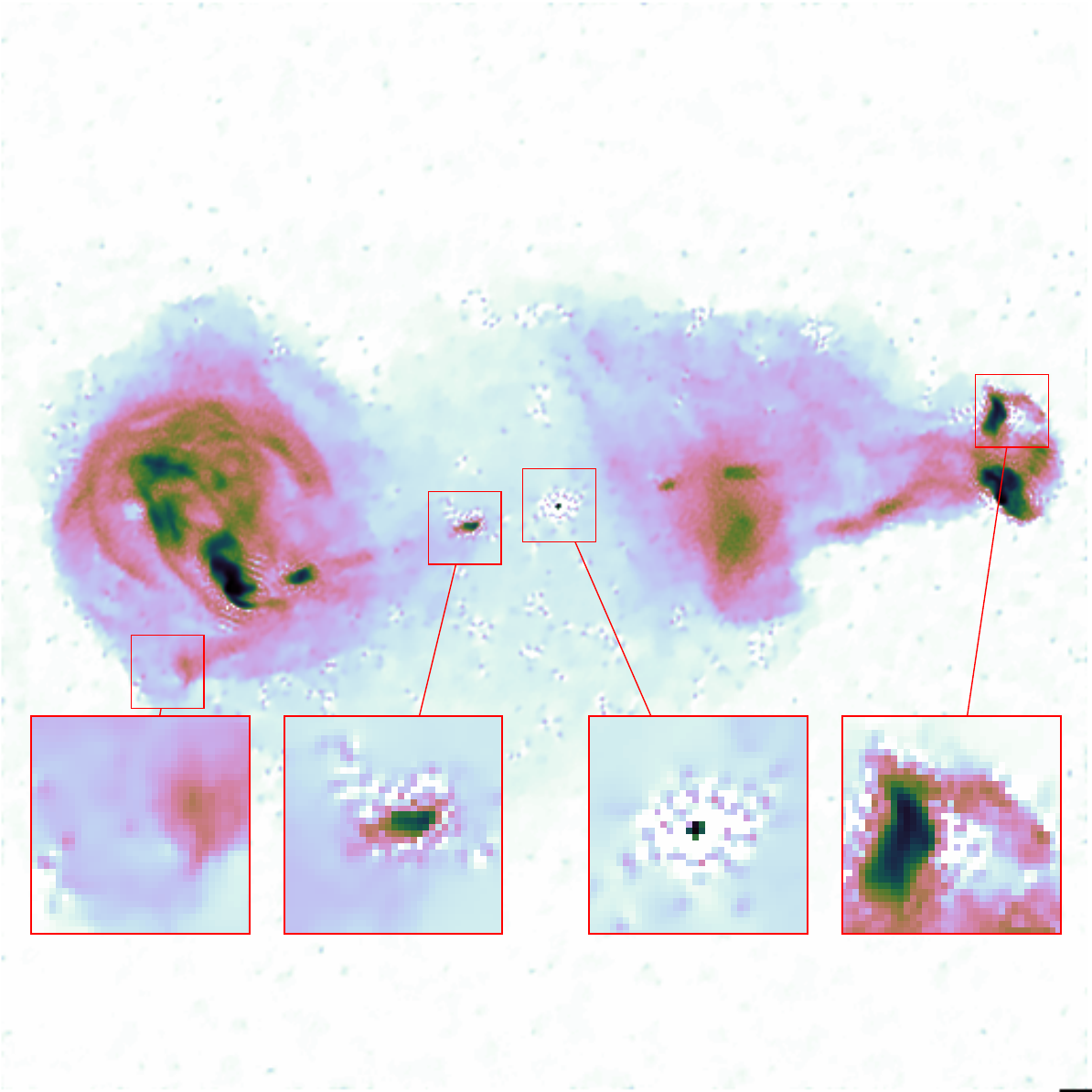}%
         \caption{AIRI-$\ell_2$, $\sigma=\nu_4/4\sqrt{2}$ \\ $(25.14\, \text{dB}, 19.03\, \text{dB})$}%
         \label{fig:3c353all:4}%
     \end{subfigure}%
     \hfill%
     \begin{subfigure}[t]{0.249\linewidth}%
         \centering%
         \includegraphics[trim={0 5em 0 10em},width=\textwidth, clip]{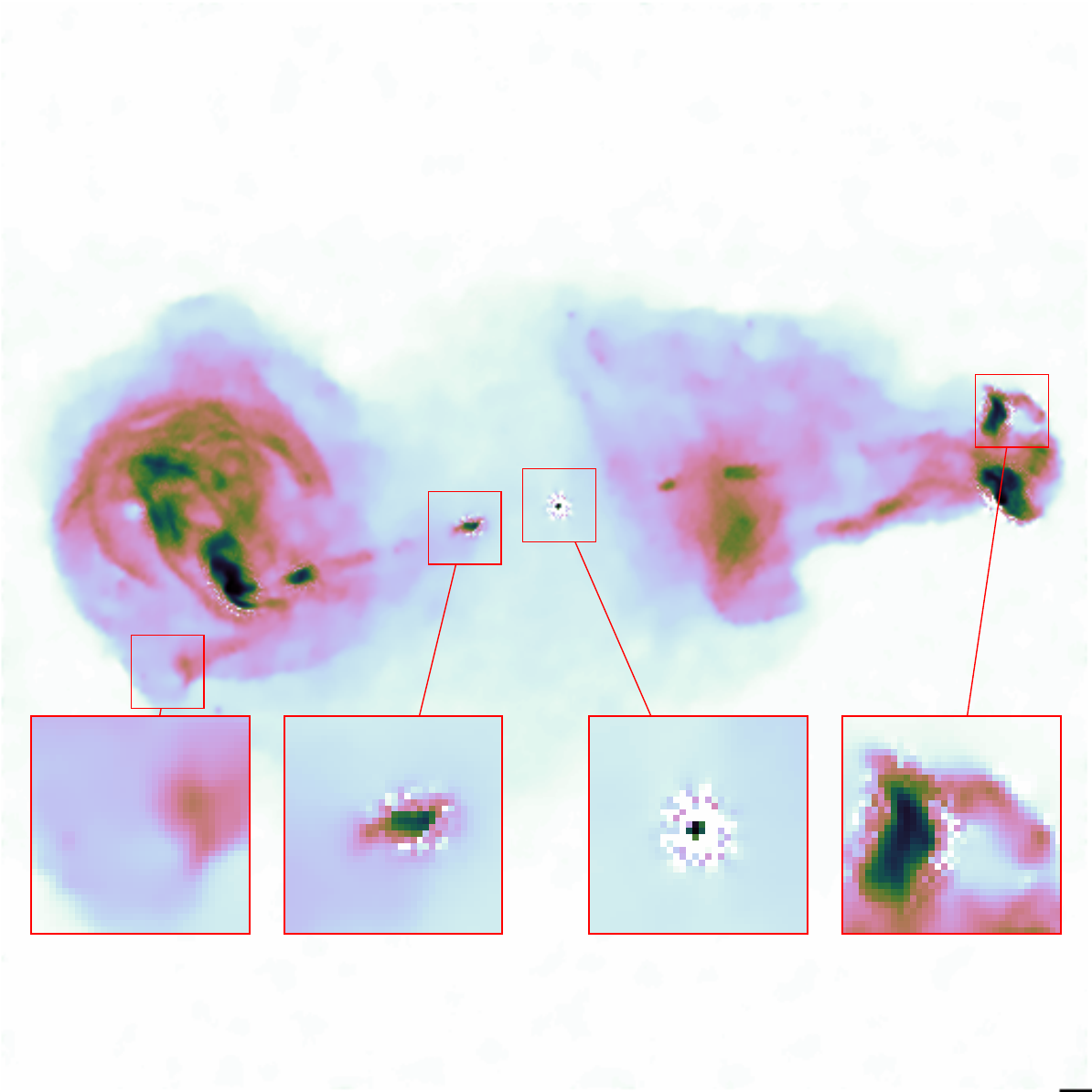}%
         \caption{AIRI-$\ell_2$, $\sigma = \nu_4/2\sqrt{2}$ \\ $(26.05\, \text{dB}, 24.03\, \text{dB})$}%
         \label{fig:3c353all:8}%
     \end{subfigure}%
     \hfill%
     \begin{subfigure}[t]{0.249\linewidth}%
         \centering%
         \includegraphics[trim={0 5em 0 10em},width=\textwidth, clip]{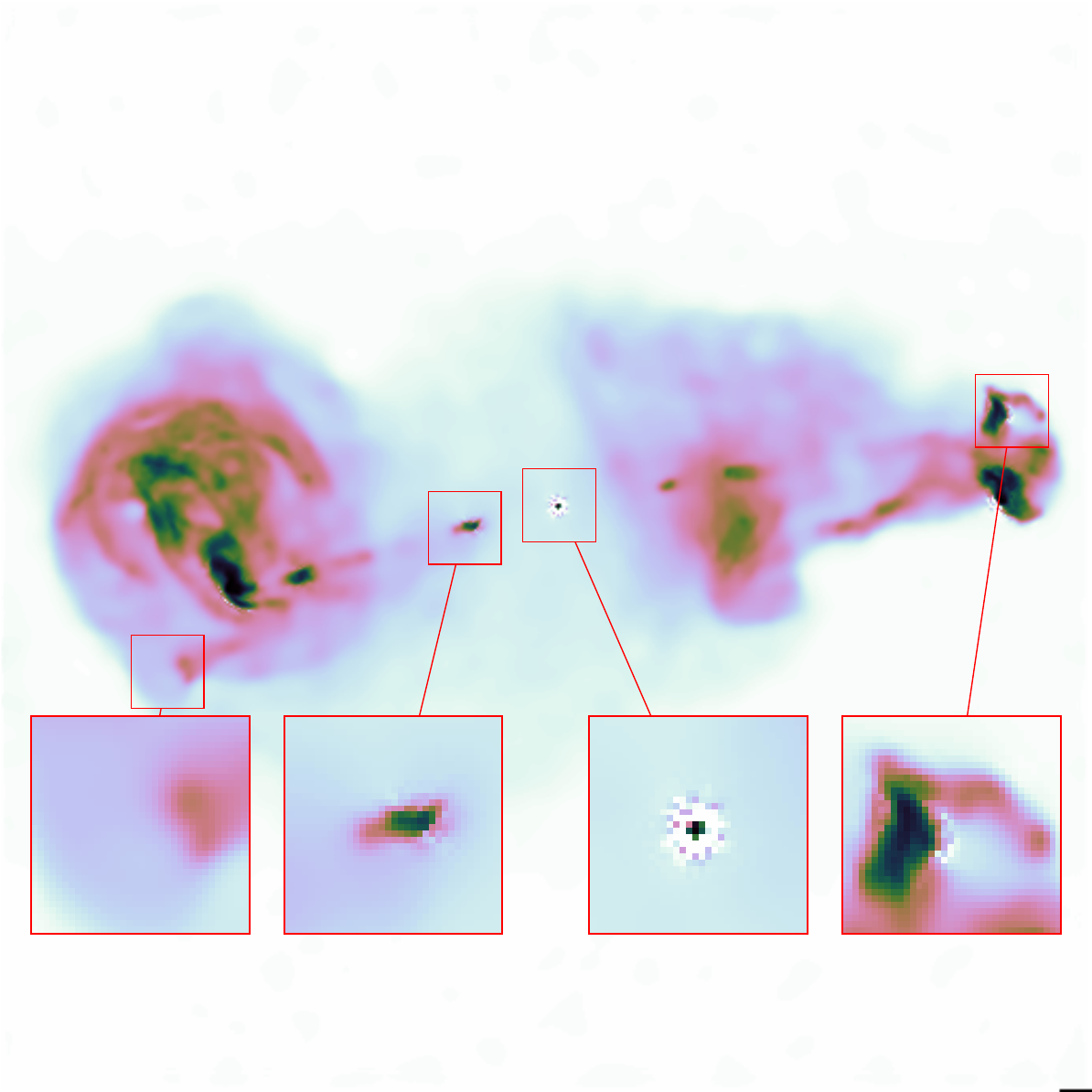}%
         \caption{AIRI-$\ell_2$, $\sigma=\nu_4$ \\ $(25.93\, \text{dB}, 24.62\, \text{dB})$}%
         \label{fig:3c353all:12}%
     \end{subfigure}%
     \hfill%
     \begin{subfigure}[t]{0.249\linewidth}%
         \centering%
         \includegraphics[trim={0 5em 0 10em},width=\textwidth, clip]{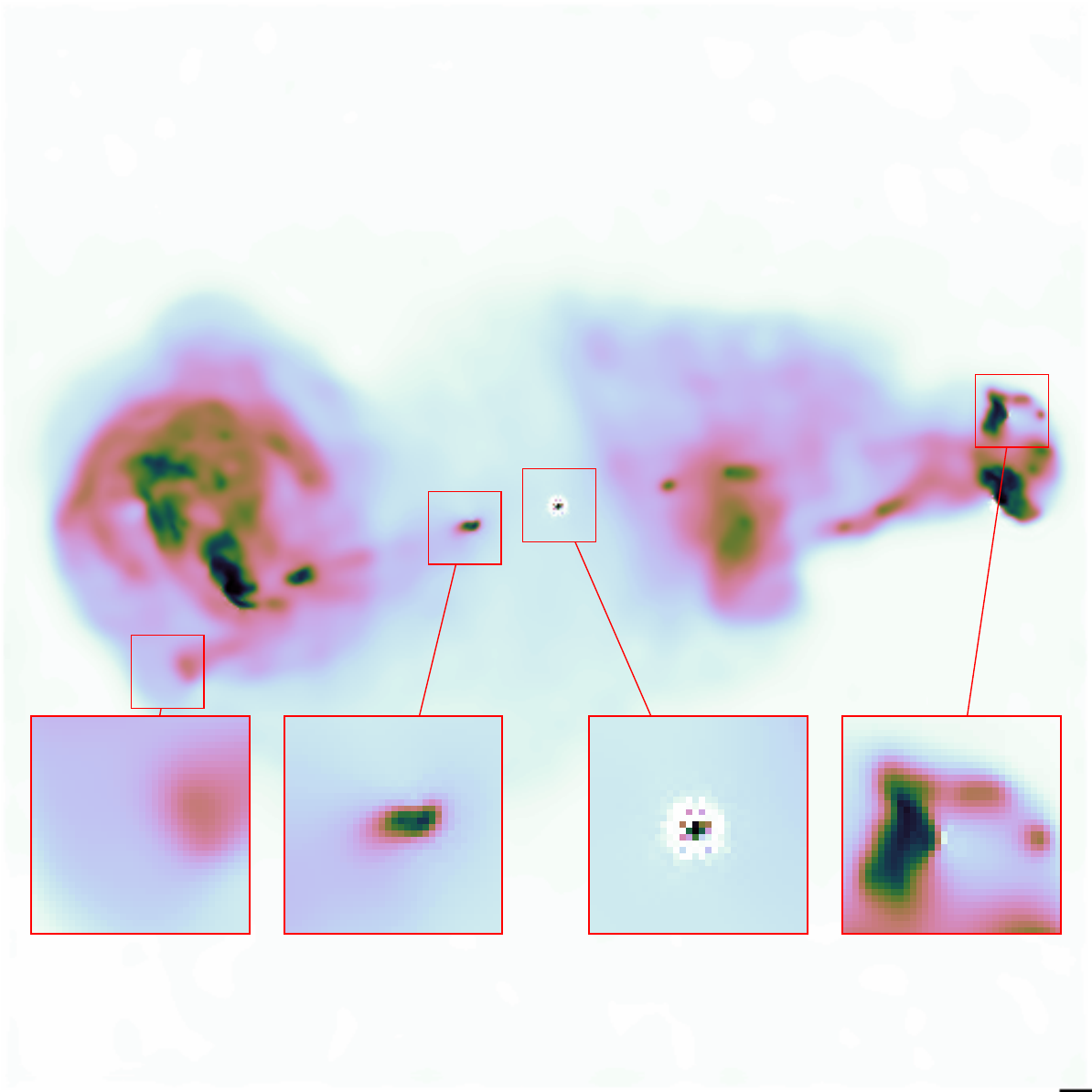}%
         \caption{AIRI-$\ell_2$, $\sigma = 2\sqrt{2}\nu_4$ \\  $(24.57\, \text{dB}, 21.09\, \text{dB})$}%
         \label{fig:3c353all:20}%
     \end{subfigure}%

    \begin{subfigure}[t]{0.249\linewidth}%
         \centering%
         \includegraphics[trim={0 5em 0 10em},width=\textwidth, clip]{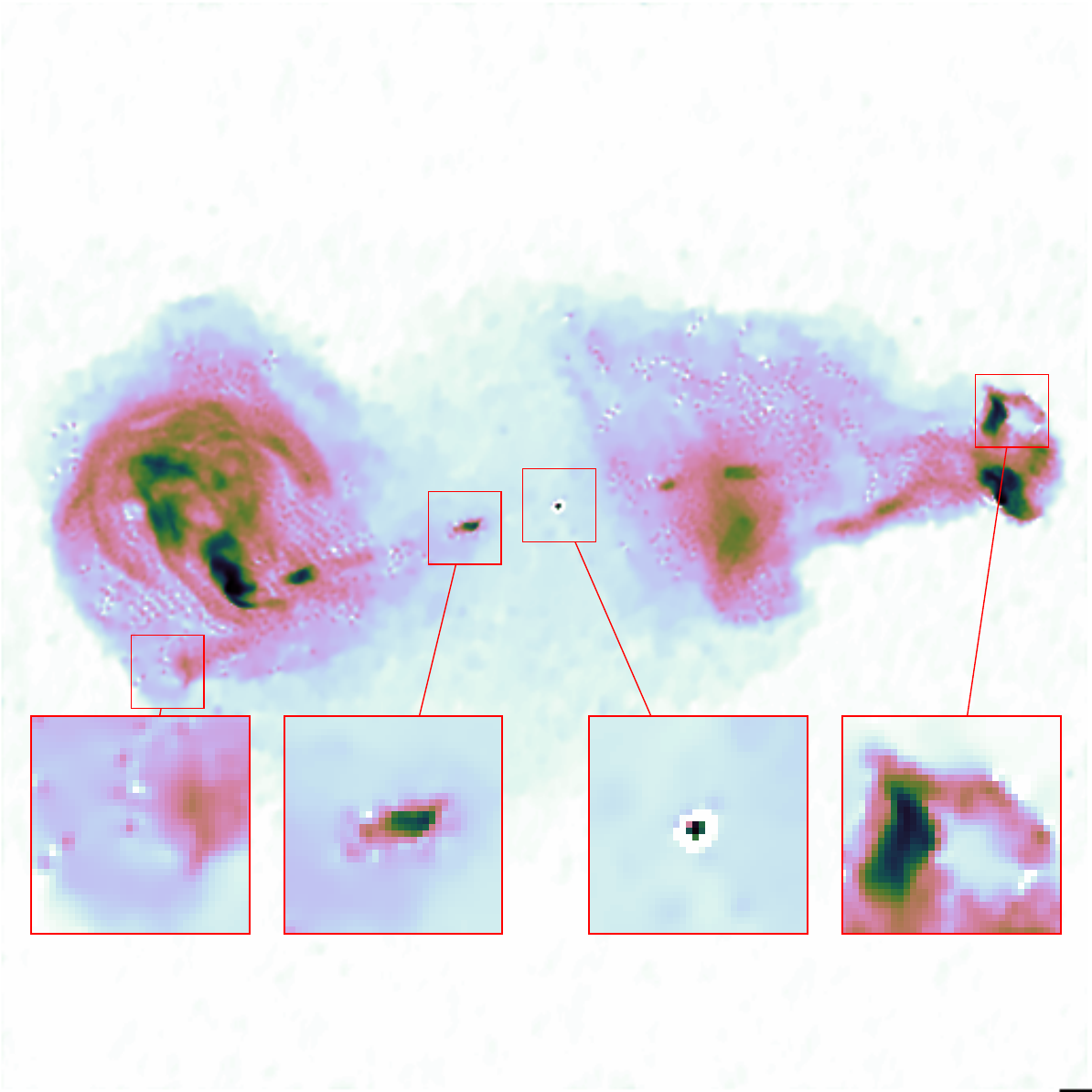}%
         \caption{AIRI-$\ell_1$, $\sigma=\nu_4/4\sqrt{2}$ \\ $(26.64\, \text{dB}, 21.27\, \text{dB})$}%
         \label{fig:3c353all:5}%
     \end{subfigure}%
     \hfill%
     \begin{subfigure}[t]{0.249\linewidth}%
         \centering%
         \includegraphics[trim={0 5em 0 10em},width=\textwidth, clip]{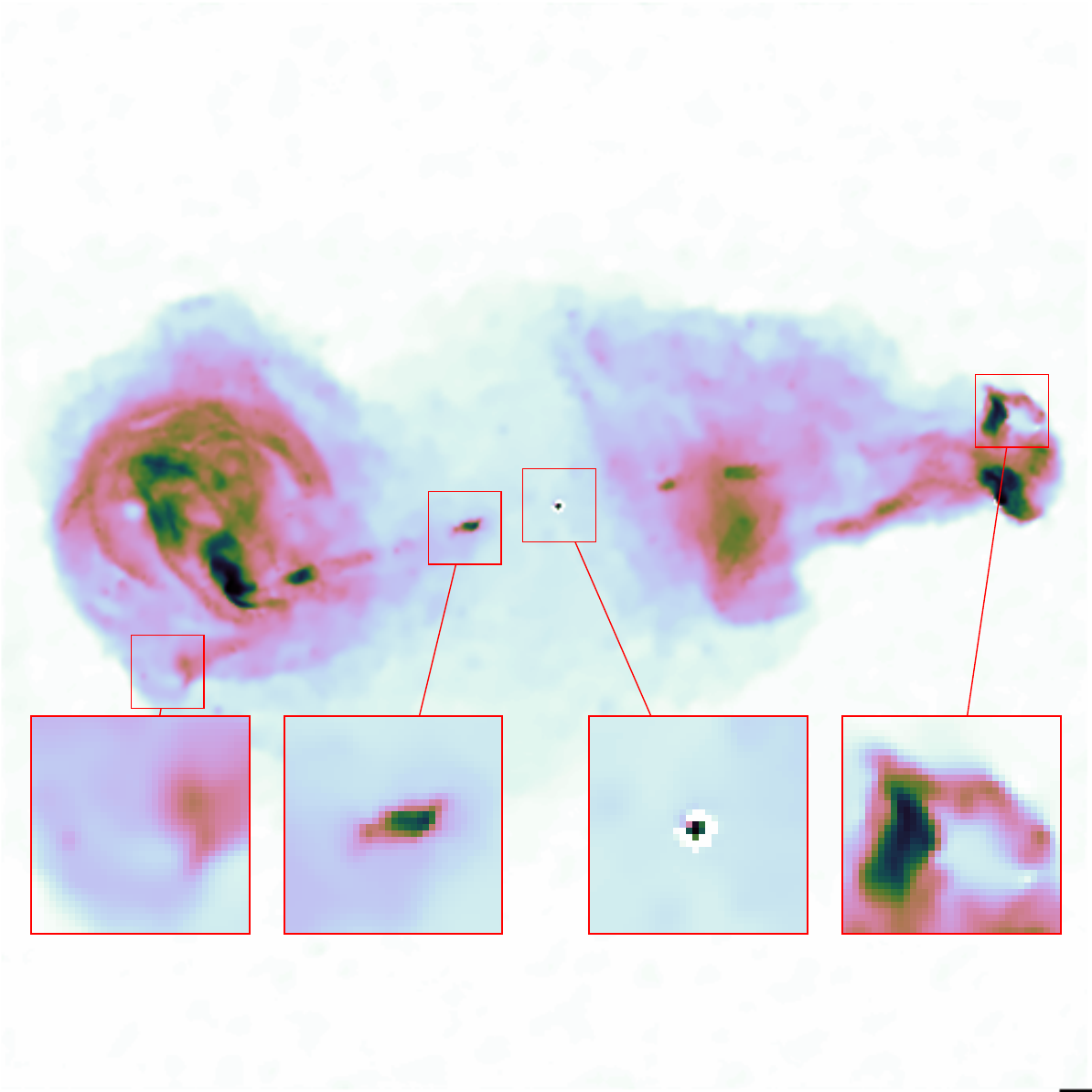}%
         \caption{AIRI-$\ell_1$, $\sigma = \nu_4/2\sqrt{2}$ \\ $(28.31\, \text{dB}, 26.02\, \text{dB})$}%
         \label{fig:3c353all:9}%
     \end{subfigure}%
     \hfill%
     \begin{subfigure}[t]{0.249\linewidth}%
         \centering%
         \includegraphics[trim={0 5em 0 10em},width=\textwidth, clip]{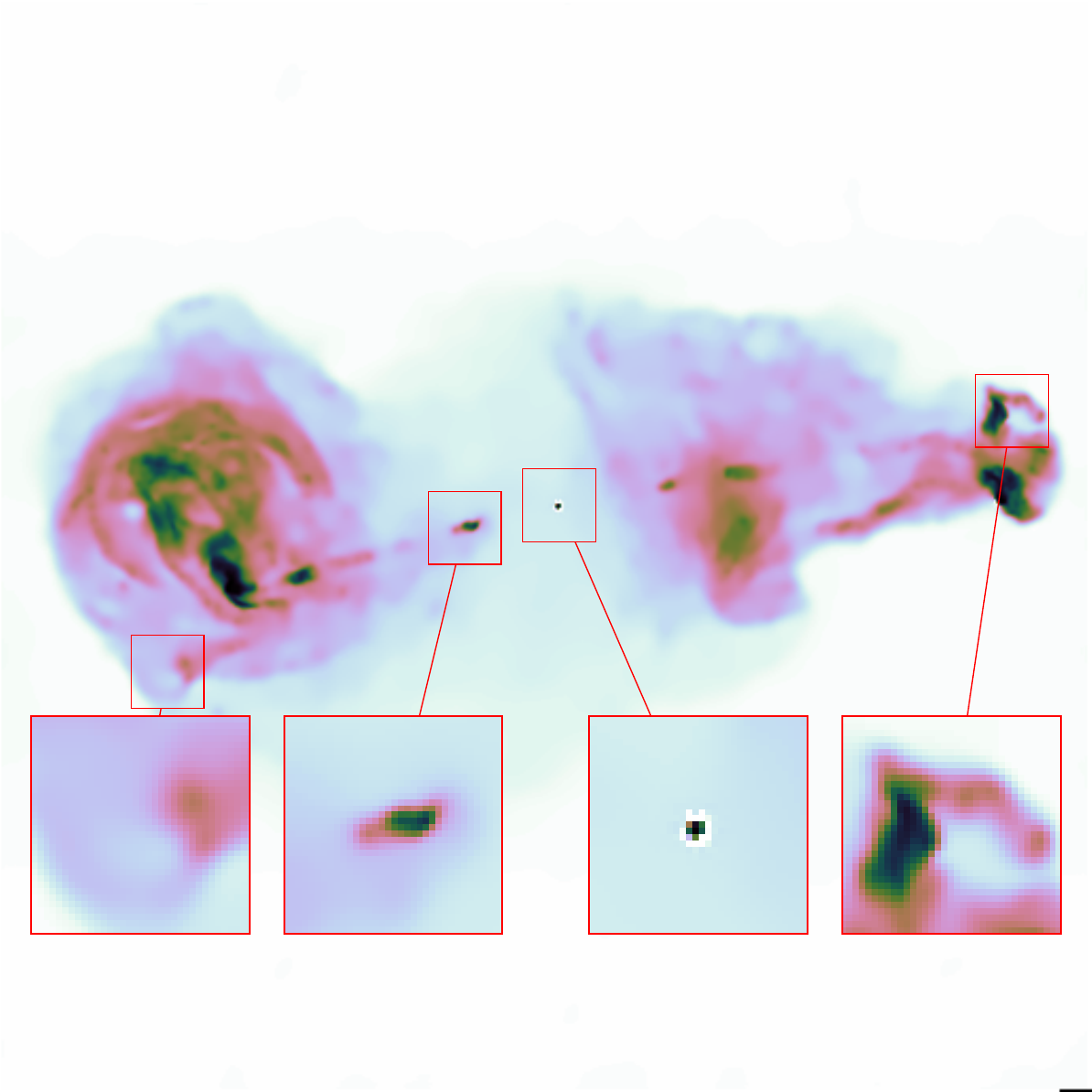}%
         \caption{AIRI-$\ell_1$, $\sigma=\nu_4$ \\ $(28.85\, \text{dB}, 26.83\, \text{dB})$}%
         \label{fig:3c353all:13}%
     \end{subfigure}%
     \hfill%
     \begin{subfigure}[t]{0.249\linewidth}%
         \centering%
         \includegraphics[trim={0 5em 0 10em},width=\textwidth, clip]{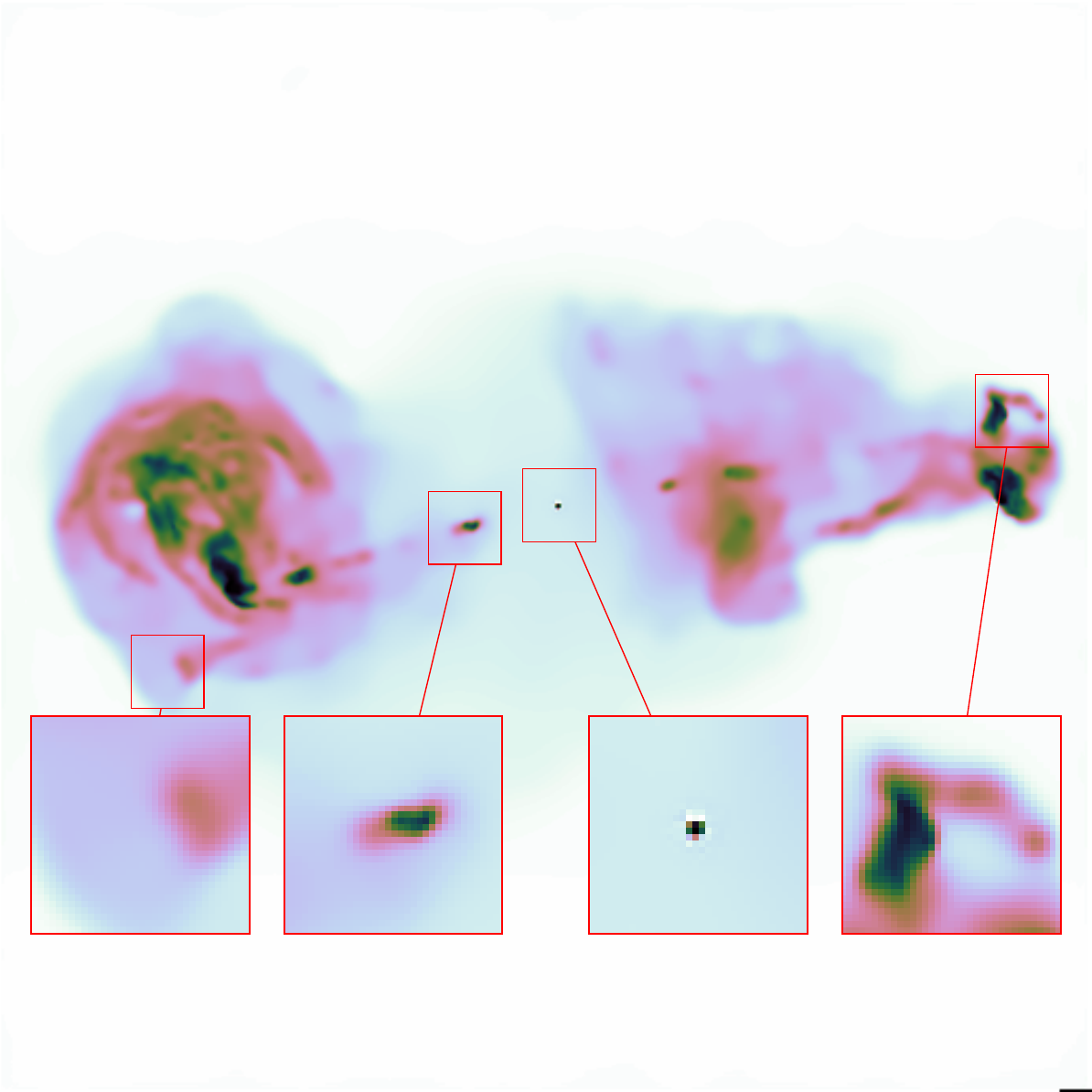}%
         \caption{AIRI-$\ell_1$, $\sigma = 2\sqrt{2}\nu_4$ \\ $(27.13\, \text{dB}, 24.86\, \text{dB})$}%
         \label{fig:3c353all:21}%
     \end{subfigure}%
     
\caption{Experiment 2 results: Impact of the training noise level $\sigma$ (resp. regularization parameter $\lambda$) in imaging with AIRI (resp. uSARA) on simulated RI data using 3c353 image as a groundtruth and the $uv$-pattern from Figure~\ref{fig:samp:seed0} ($\Delta T = 4\,\text{h}$) in comparison with the reference value $\nu_4$ suggested by the AIRI and uSARA heuristics. Top row: the groundtruth image \ref{fig:3c353all:1} and the simulated dirty image \ref{fig:3c353all:2}. Second row: estimated model images obtained with uSARA, with a regularization parameter $\lambda$ increasing with the column index. Third (resp. fourth) row: estimated model images of AIRI-$\ell_2$ (resp.~AIRI-$\ell_1$), with a training noise level $\sigma$ increasing with the column index. Below each image we indicate the reconstruction metrics as (SNR, logSNR). Images are shown in logarithmic scale.}
\label{fig:heuristic}
\end{figure*}

\begin{figure*}%
\captionsetup[subfigure]{justification=centering}%
    \centering%
    \begin{subfigure}[b]{0.249\linewidth}%
         \centering%
         \includegraphics[width=0.9\textwidth]{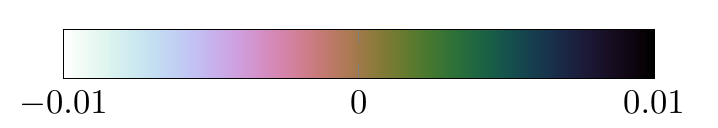}%
    \end{subfigure}%
    \hfill%
    \begin{subfigure}[b]{0.249\linewidth}%
         \centering%
         \phantom{\includegraphics[trim={0 5em 0 10em}, width=0.5\textwidth, clip]{./pictures/colorbar_log_bottom.pdf}}%
    \end{subfigure}%
    \hfill%
    \begin{subfigure}[b]{0.249\linewidth}%
         \centering
          \phantom{\includegraphics[trim={0 5em 0 10em}, width=0.5\textwidth, clip]{./pictures/colorbar_log_bottom.pdf}}
    \end{subfigure}
    \hfill%
    \begin{subfigure}[b]{0.249\linewidth}%
         \centering
          \phantom{\includegraphics[trim={0 5em 0 10em}, width=0.5\textwidth, clip]{./pictures/colorbar_log_bottom.pdf}}
    \end{subfigure}
    
    \begin{subfigure}[b]{0.249\linewidth}%
         \centering%
         \includegraphics[trim={0 5em 0 10em}, width=\textwidth, clip]{./pictures/3c353_residuals/zoom_sara_fb_residual_2.5e-05.pdf}%
         \caption{uSARA, $\gamma\lambda=\nu_4/4\sqrt{2}$}%
         \label{fig:3c353res:0}%
    \end{subfigure}%
    \hfill%
    \begin{subfigure}[b]{0.249\linewidth}%
         \centering%
         \includegraphics[trim={0 5em 0 10em}, width=\textwidth, clip]{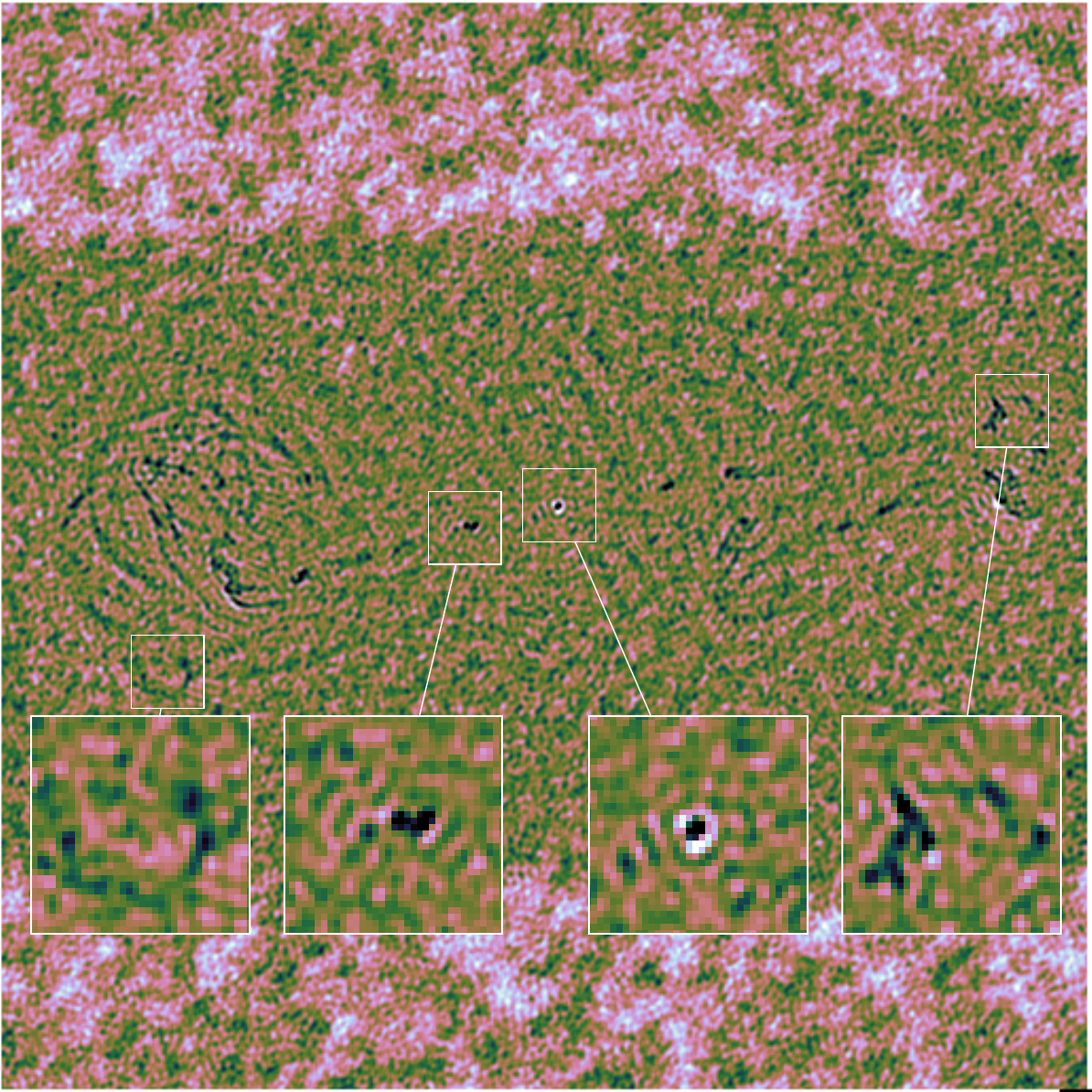}%
         \caption{uSARA, $\gamma\lambda=\nu_4/2\sqrt{2}$}%
         \label{fig:3c353res:1}%
    \end{subfigure}%
    \hfill%
    \begin{subfigure}[b]{0.249\linewidth}%
         \centering%
         \includegraphics[trim={0 5em 0 10em}, width=\textwidth, clip]{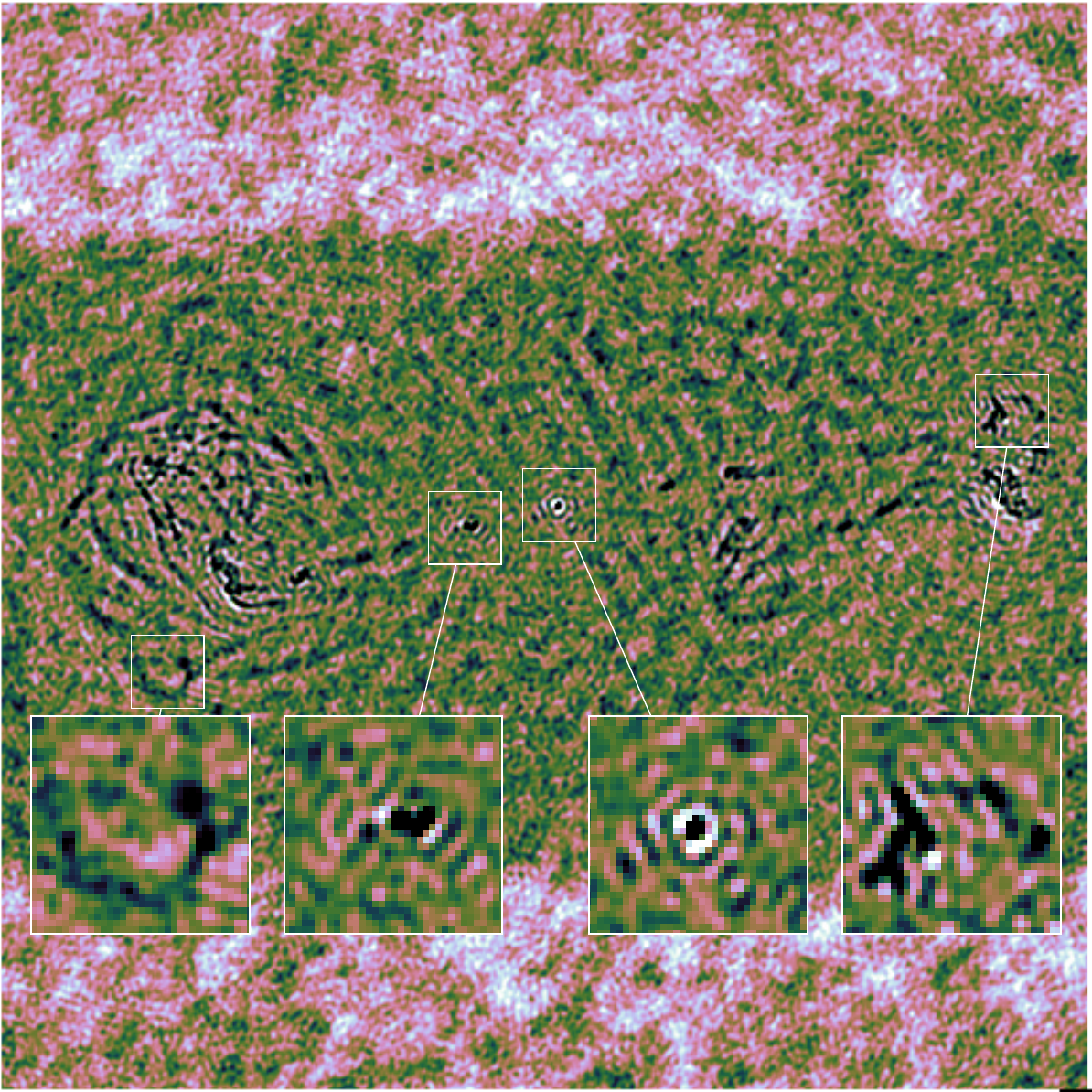}%
         \caption{uSARA, $\gamma\lambda=\nu_4/\sqrt{2}$}%
         \label{fig:3c353res:2}%
    \end{subfigure}%
    \hfill
    \begin{subfigure}[b]{0.249\linewidth}%
         \centering%
         \includegraphics[trim={0 5em 0 10em}, width=\textwidth, clip]{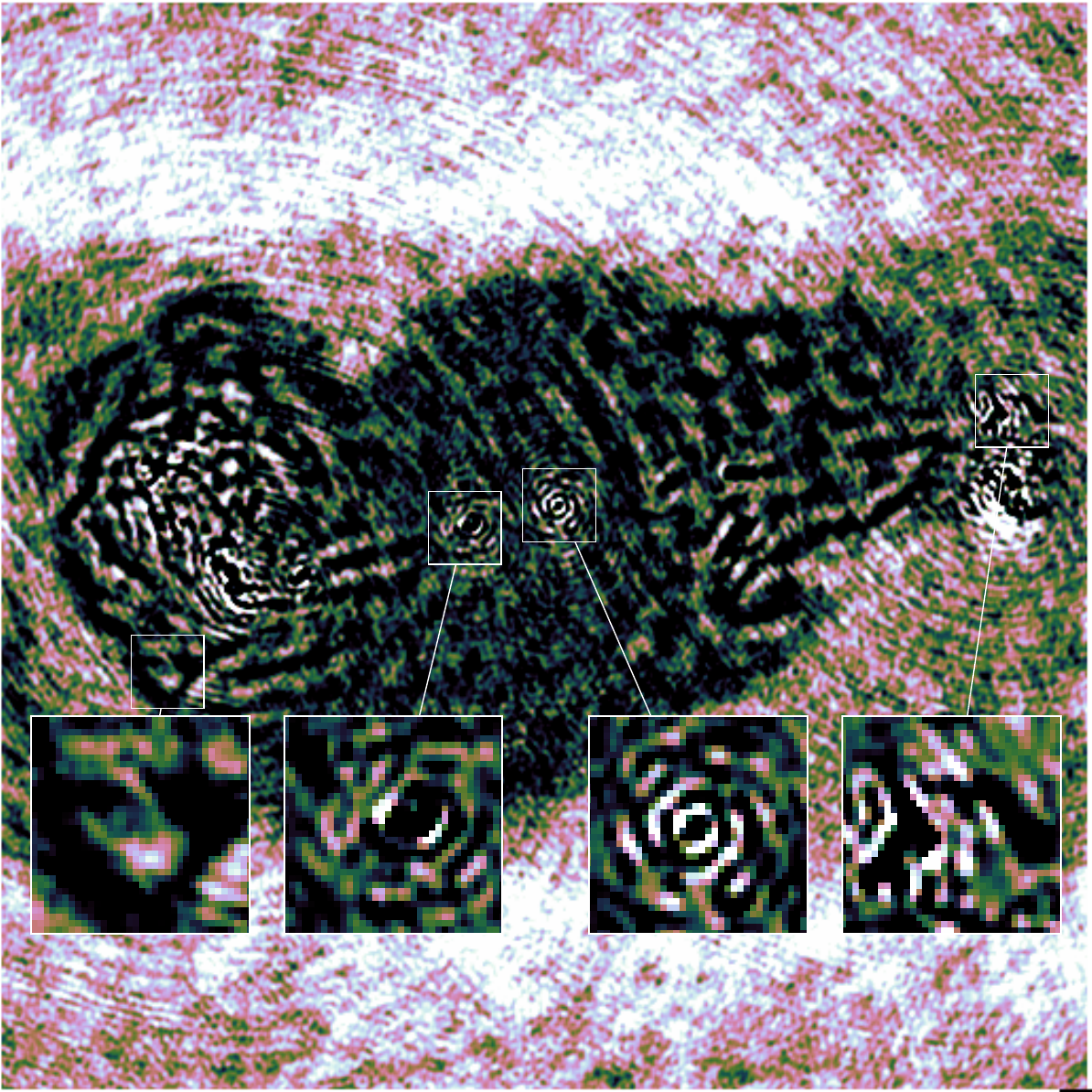}%
         \caption{uSARA, $\gamma\lambda = 2\sqrt{2}\nu_4$}%
         \label{fig:3c353res:3}%
    \end{subfigure}%
     
    \begin{subfigure}[t]{0.249\linewidth}%
         \centering%
         \includegraphics[trim={0 5em 0 10em},width=\textwidth, clip]{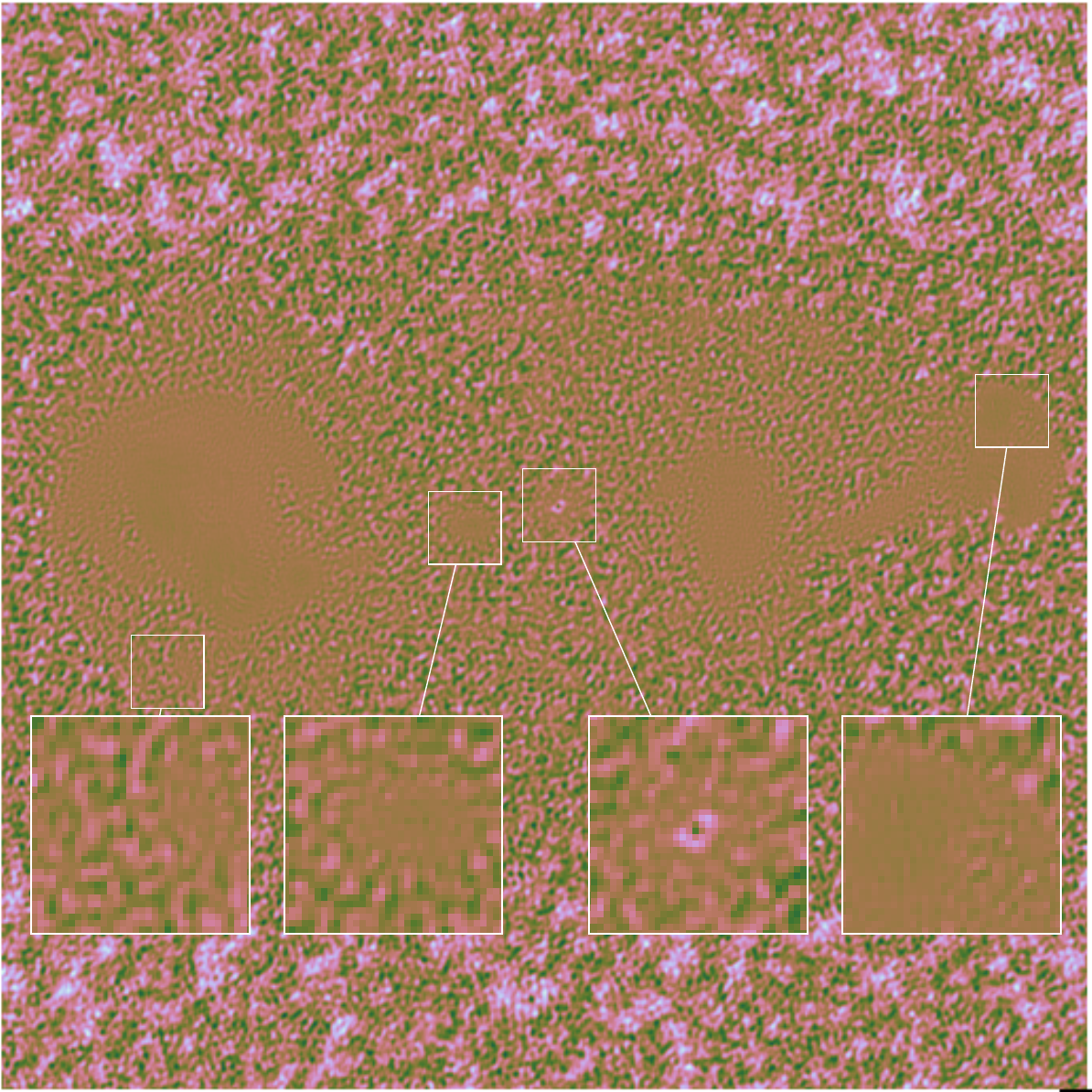}%
         \caption{AIRI-$\ell_2$, $\gamma\lambda=\nu_4/4\sqrt{2}$}%
         \label{fig:3c353res:4}%
    \end{subfigure}%
    \hfill%
    \begin{subfigure}[t]{0.249\linewidth}%
         \centering%
         \includegraphics[trim={0 5em 0 10em},width=\textwidth, clip]{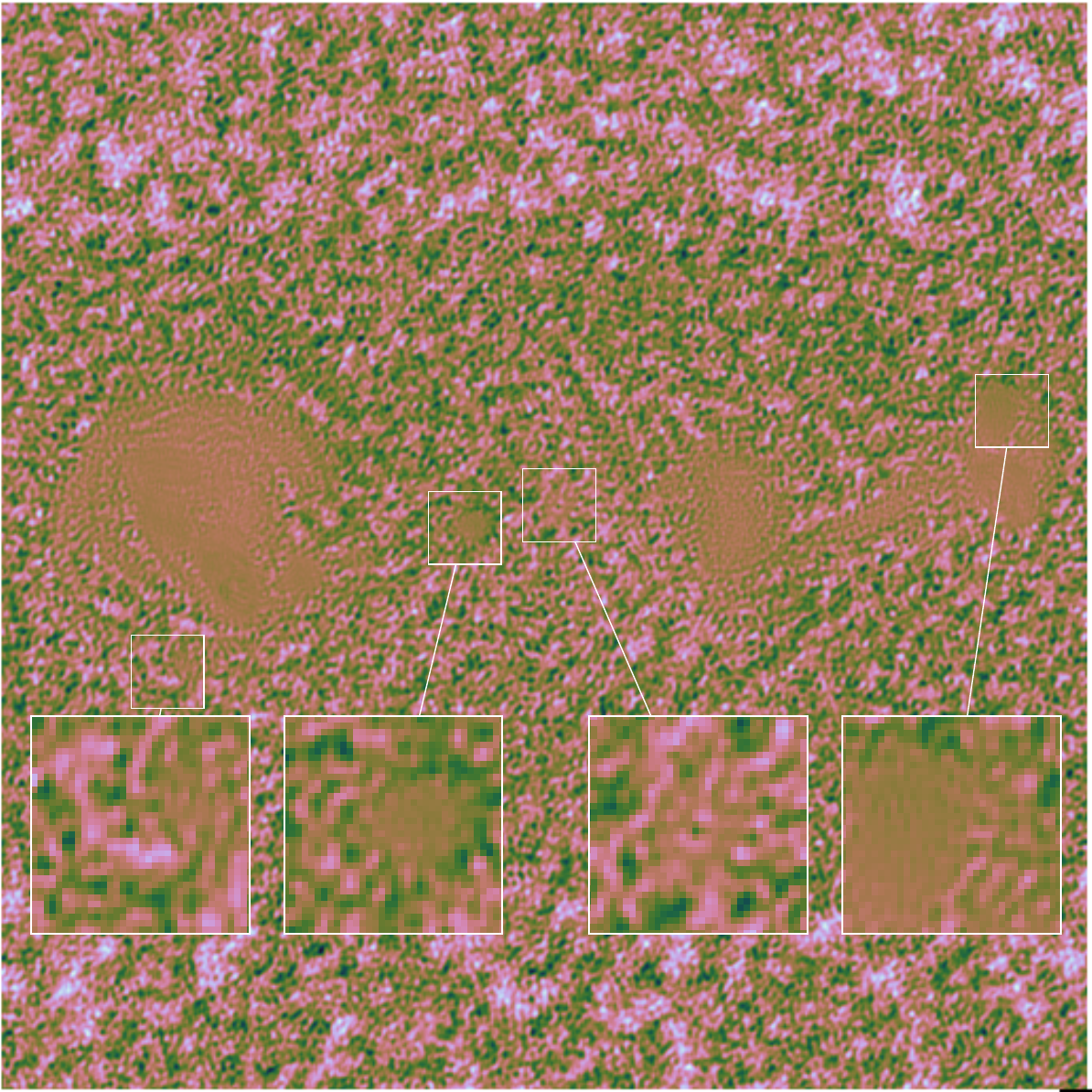}%
         \caption{AIRI-$\ell_2$, $\sigma=\nu_4/2\sqrt{2}$}%
         \label{fig:3c353res:5}%
    \end{subfigure}%
    \hfill%
    \begin{subfigure}[t]{0.249\linewidth}%
         \centering%
         \includegraphics[trim={0 5em 0 10em},width=\textwidth, clip]{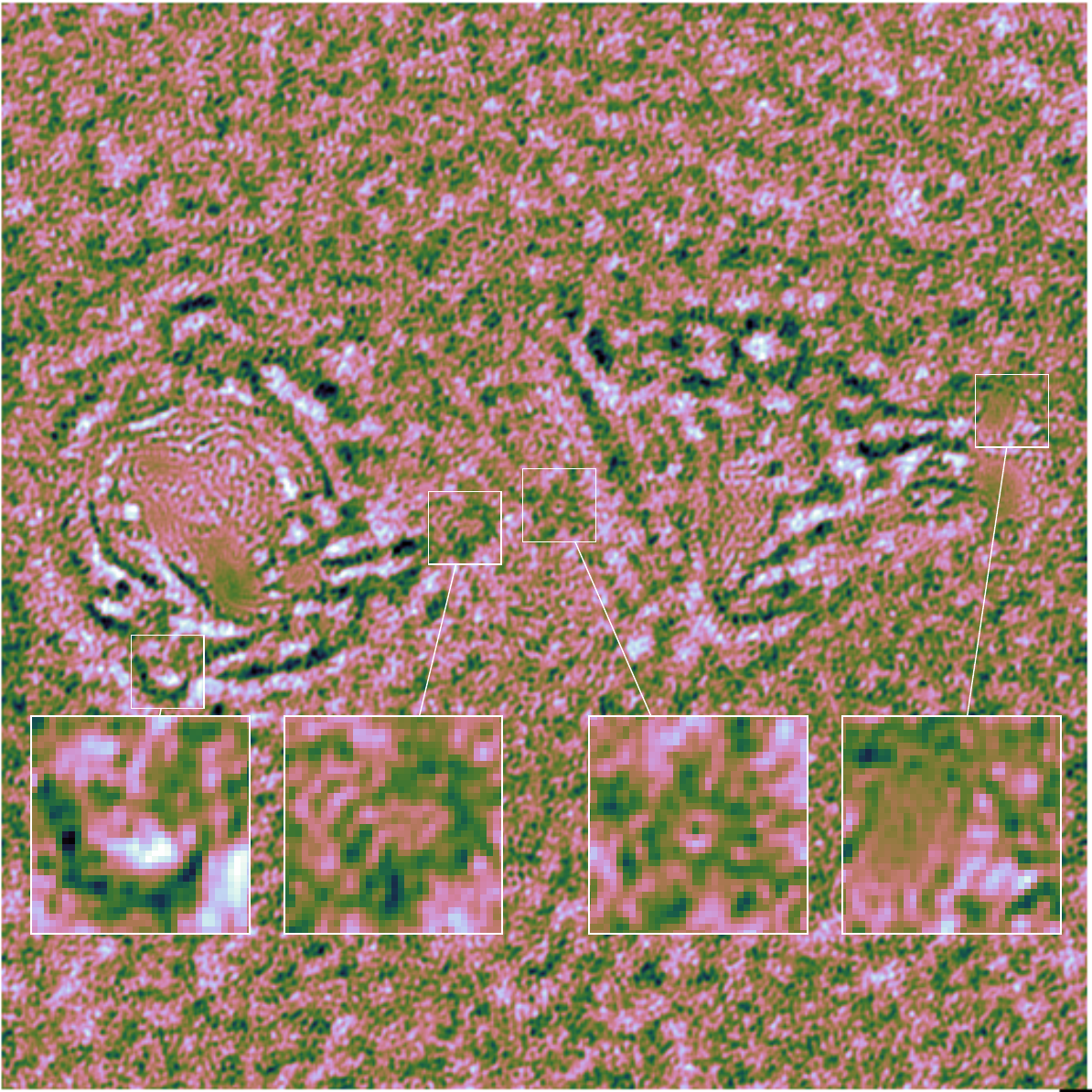}%
         \caption{AIRI-$\ell_2$, $\sigma=\nu_4$}%
         \label{fig:3c353res:6}%
    \end{subfigure}%
    \hfill%
    \begin{subfigure}[t]{0.249\linewidth}%
         \centering%
         \includegraphics[trim={0 5em 0 10em},width=\textwidth, clip]{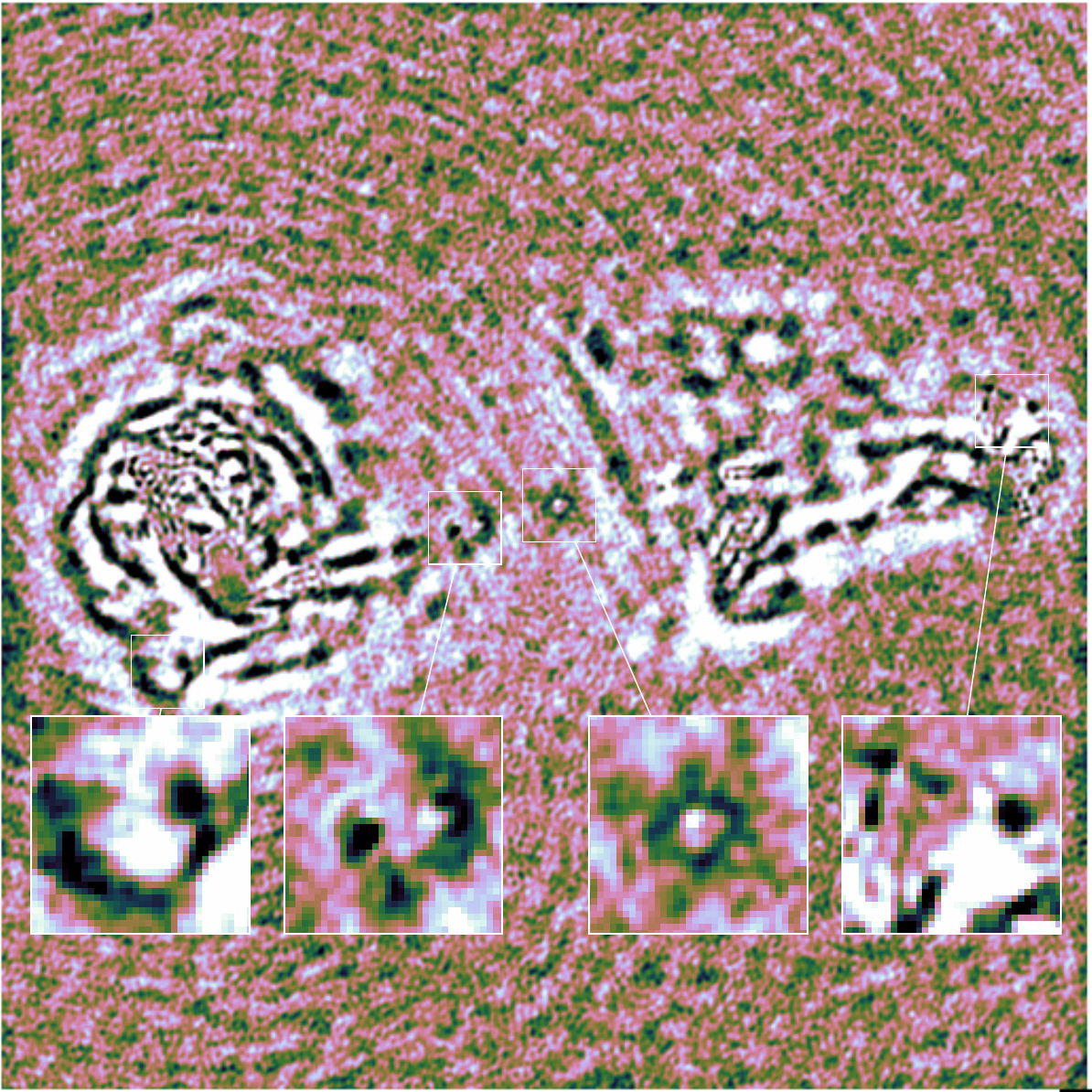}%
         \caption{AIRI-$\ell_2$, $\sigma = 2\sqrt{2}\nu_4$}%
         \label{fig:3c353res:7}%
     \end{subfigure}%
     
    \begin{subfigure}[t]{0.249\linewidth}%
         \centering%
         \includegraphics[trim={0 5em 0 10em},width=\textwidth, clip]{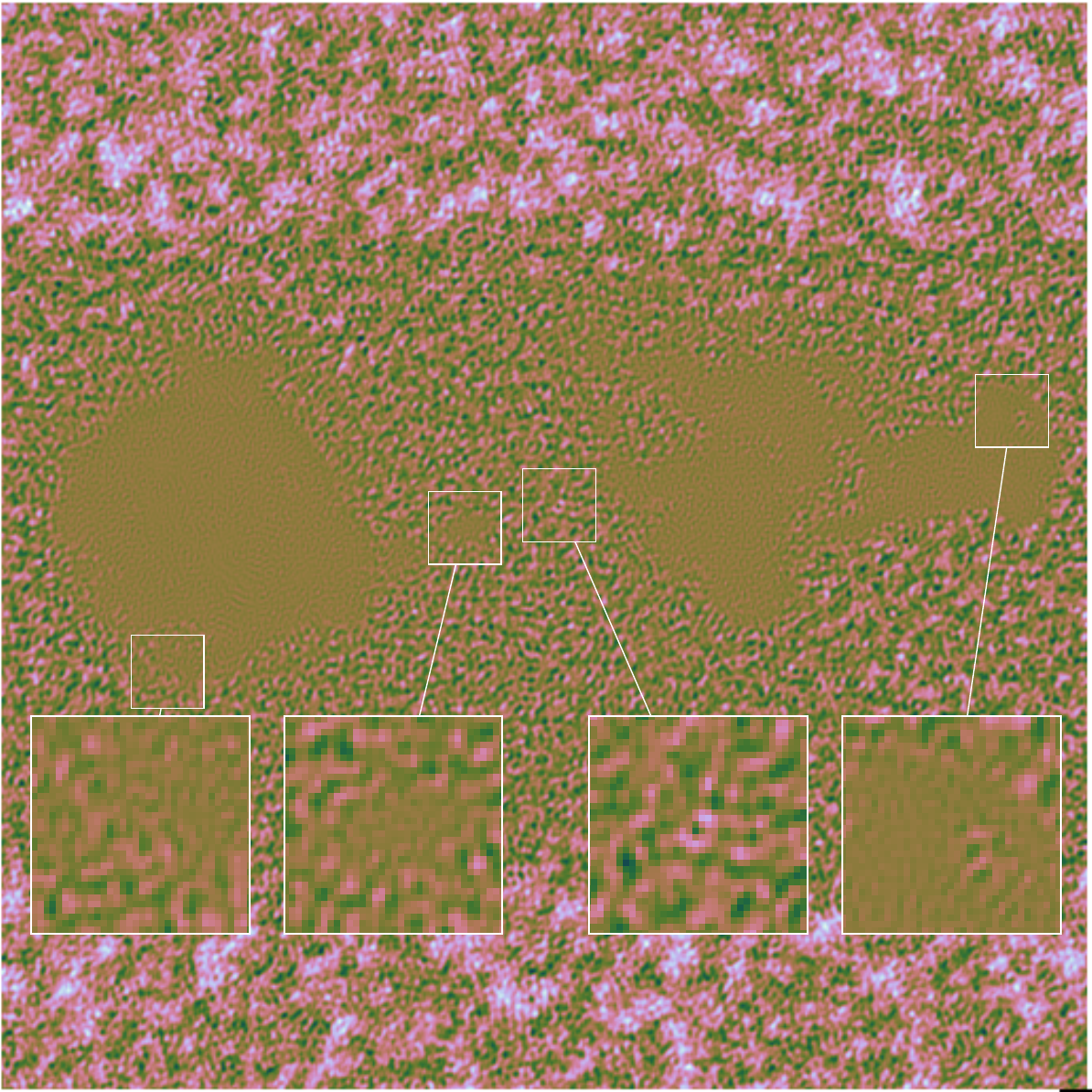}%
         \caption{AIRI-$\ell_1$, $\gamma\lambda=\nu_4/4\sqrt{2}$}%
         \label{fig:3c353res:8}%
    \end{subfigure}%
    \hfill%
    \begin{subfigure}[t]{0.249\linewidth}%
         \centering%
         \includegraphics[trim={0 5em 0 10em},width=\textwidth, clip]{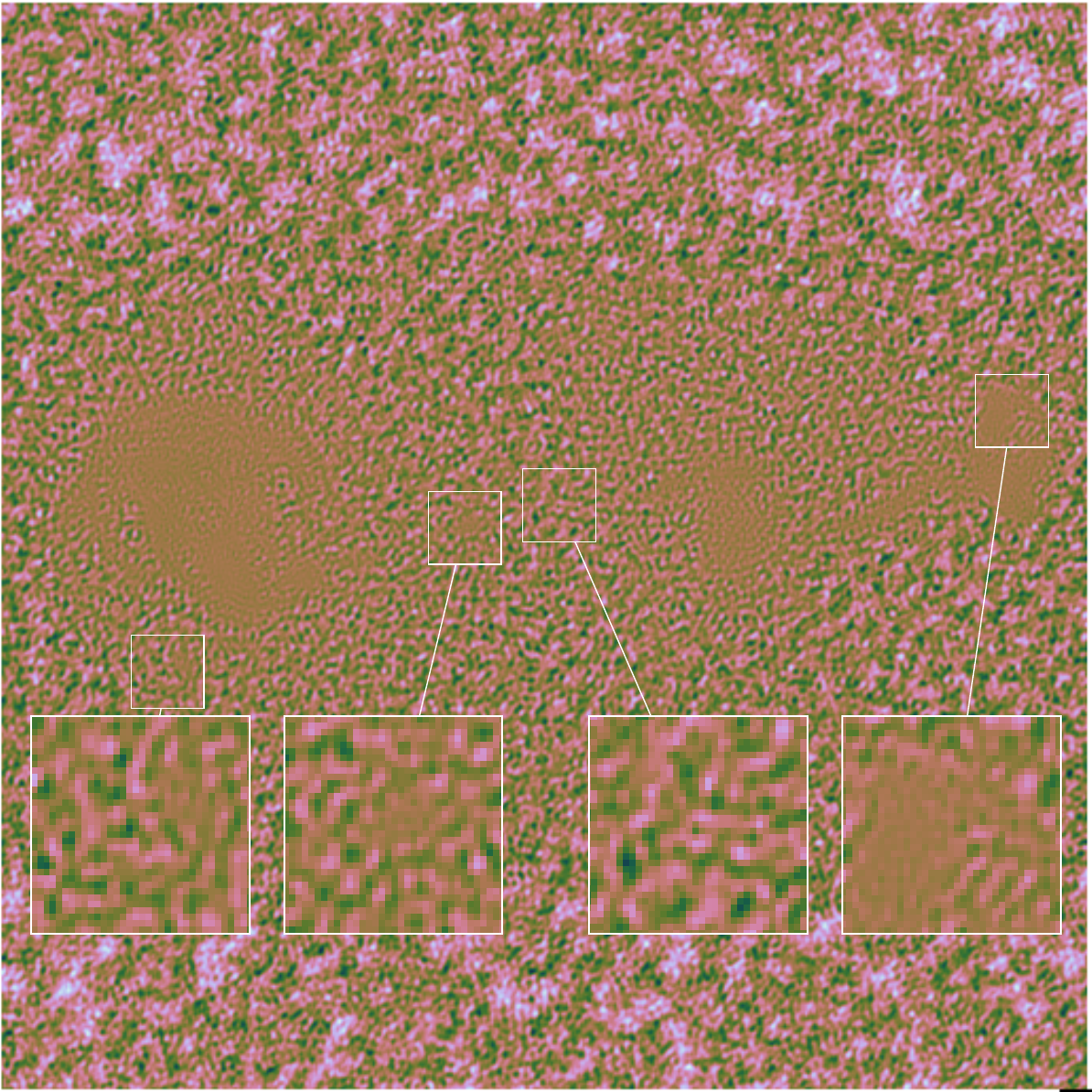}%
         \caption{AIRI-$\ell_1$, $\sigma=\nu_4/2\sqrt{2}$}%
         \label{fig:3c353res:9}%
    \end{subfigure}%
    \hfill%
    \begin{subfigure}[t]{0.249\linewidth}%
         \centering%
         \includegraphics[trim={0 5em 0 10em},width=\textwidth, clip]{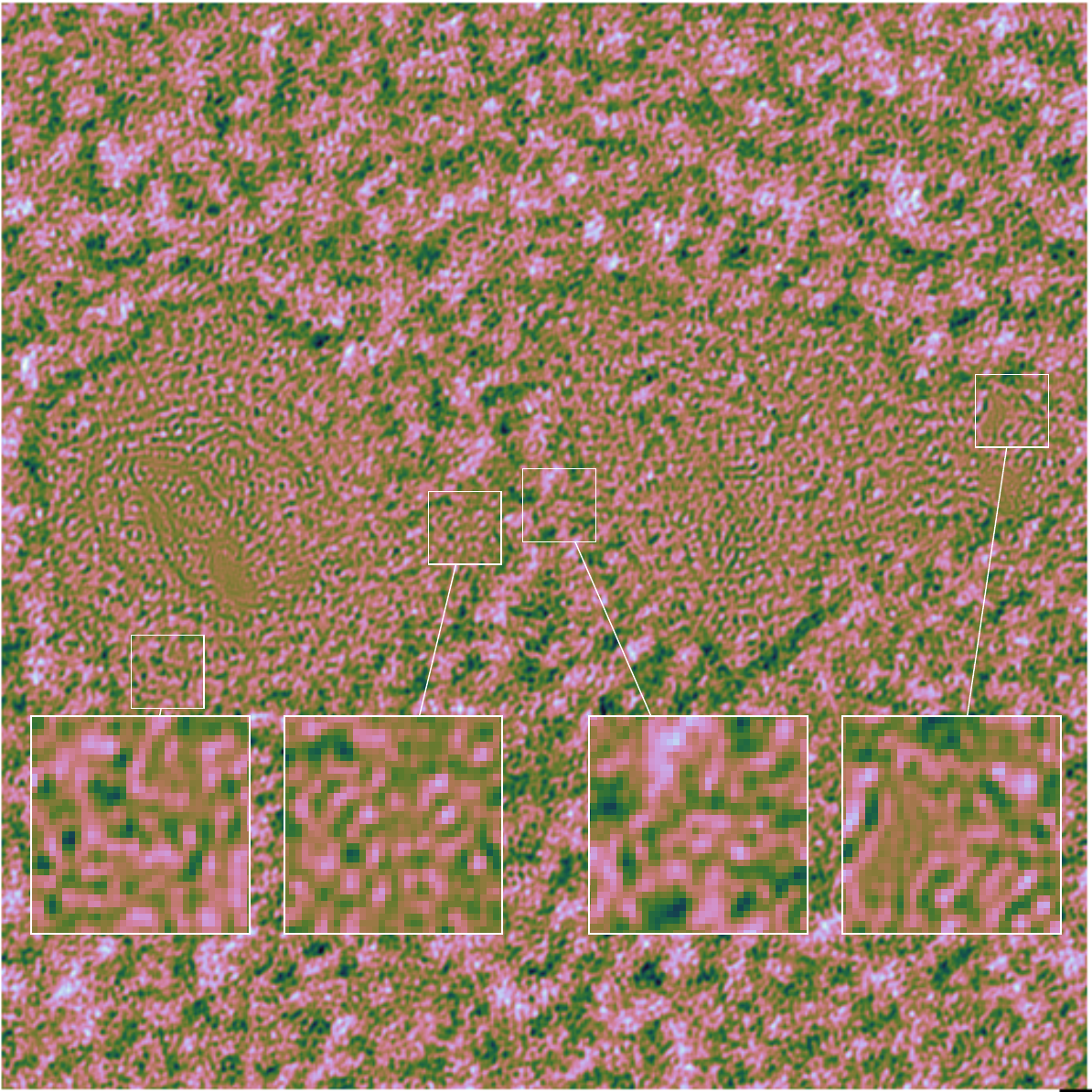}%
         \caption{AIRI-$\ell_1$, $\sigma=\nu_4$}%
         \label{fig:3c353res:10}%
    \end{subfigure}%
    \hfill%
    \begin{subfigure}[t]{0.249\linewidth}%
         \centering%
         \includegraphics[trim={0 5em 0 10em},width=\textwidth, clip]{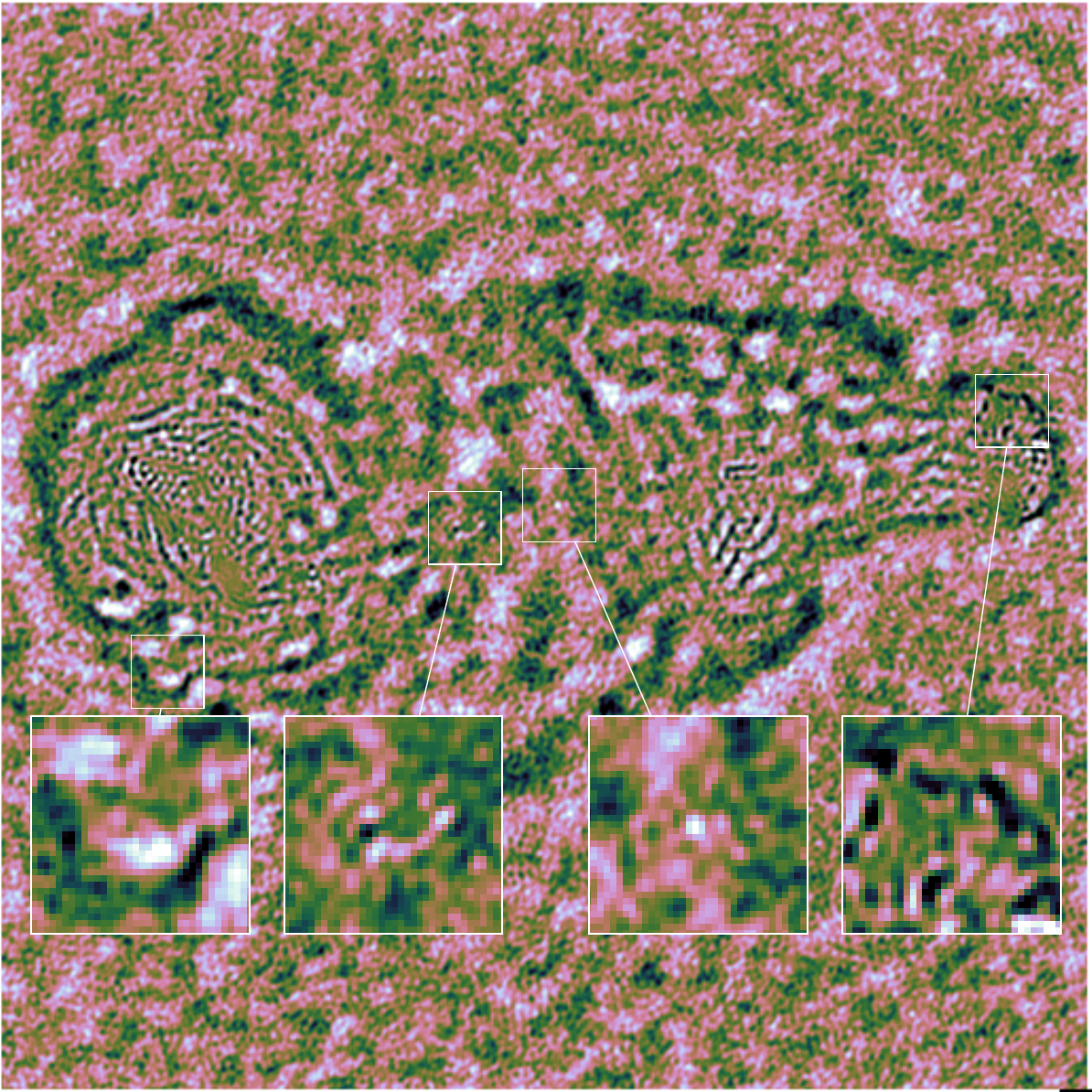}%
         \caption{AIRI-$\ell_1$, $\sigma = 2\sqrt{2}\nu_4$}%
         \label{fig:3c353res:11}%
    \end{subfigure}%
\vspace{-0.5em}
\caption{Experiment 2 results: Residual images associated with the reconstructions of Figure~\ref{fig:heuristic}. Top row: residual images for the estimated model images obtained with uSARA, with a regularization parameter $\lambda$ increasing with the column index. Second (resp. third) row: respective residual images for the estimated model images obtained with AIRI-$\ell_2$ (resp. AIRI-$\ell_1$), with a training noise level $\sigma$ increasing with the column index. All images are shown in linear scale.}
\label{fig:heuristic_residuals}
\end{figure*}

Considering the image of 3c353 as groundtruth and the  $uv$-pattern from Figure~\ref{fig:samp:seed0} ($\Delta T = 4\,\text{h}$) for simulating the measurements, reconstructions for different values of $\sigma$ in AIRI-$\ell_2$ and AIRI-$\ell_1$, and $\lambda$ in uSARA are displayed in Figures \ref{fig:heuristic} and \ref{fig:heuristic_residuals}. More precisely, model images, displayed in logarithmic scale, and their associated SNR and logSNR values are provided in Figure \ref{fig:heuristic}, and residual images are displayed in Figure~\ref{fig:heuristic_residuals} in linear scale. We firstly notice that $\sigma$ and $\lambda$ play a similar role in the reconstruction quality: the higher their value, the smoother the recovered image; the lower their value, the stronger the artefacts. Overall, reconstructions with uSARA  exhibit some wavelet artefacts, particularly noticeable around bright compact sources overlaying faint extended emission. AIRI-$\ell_2$ reconstructions typically exhibit less artefacts, but appear to be generally smoother. AIRI-$\ell_1$ reconstructions contain even less artefacts, with an achieved resolution similar to uSARA. We conclude these observations by highlighting the low-amplitude residual images obtained with low values of $\sigma$ and $\lambda$. However, such results are not a token of a good quality reconstruction. They instead reflect the over-fitting of the data in such settings.

\begin{figure}
\centering
\captionsetup[subfigure]{justification=centering}%
\begin{subfigure}[b]{0.48\linewidth}
     \centering
     \includegraphics[height=.84\textwidth]{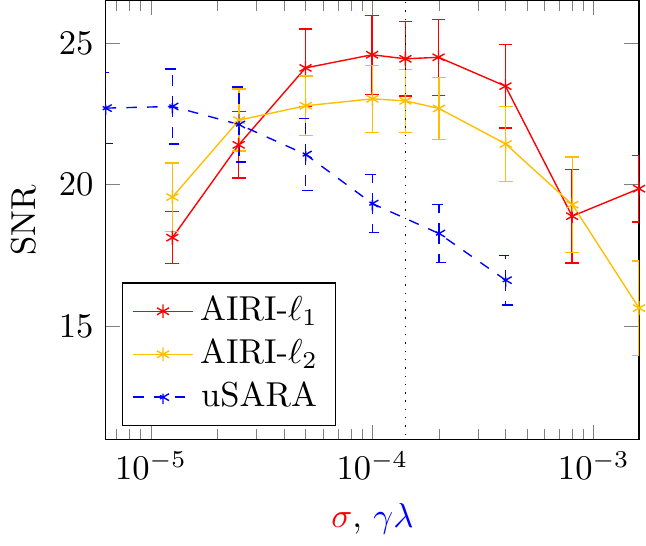}
     \vspace{-2em}
     \caption{Case $\Delta T=4\,\text{h}$}
     \label{fig:snr_1}
 \end{subfigure}
 \hfill
 \begin{subfigure}[b]{0.48\linewidth}
     \centering
     \includegraphics[height=.84\textwidth]{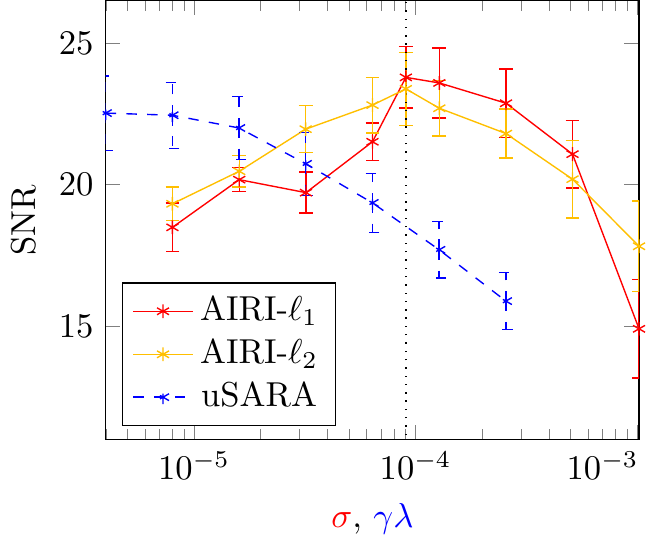}
     \vspace{-2em}
     \caption{Case $\Delta T=8\,\text{h}$}
     \label{fig:snr_2}
\end{subfigure}
 
\begin{subfigure}[b]{0.48\linewidth}
     \centering
     \includegraphics[height=.84\textwidth]{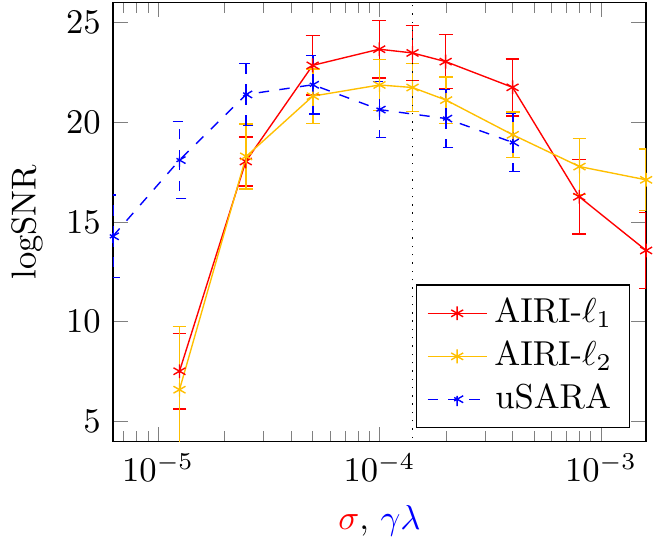}
     \vspace{-2em}
     \caption{Case $\Delta T=4\,\text{h}$}
     \label{fig:logsnr_1}
 \end{subfigure}
 \hfill
 \begin{subfigure}[b]{0.48\linewidth}
     \centering
     \includegraphics[height=.84\textwidth]{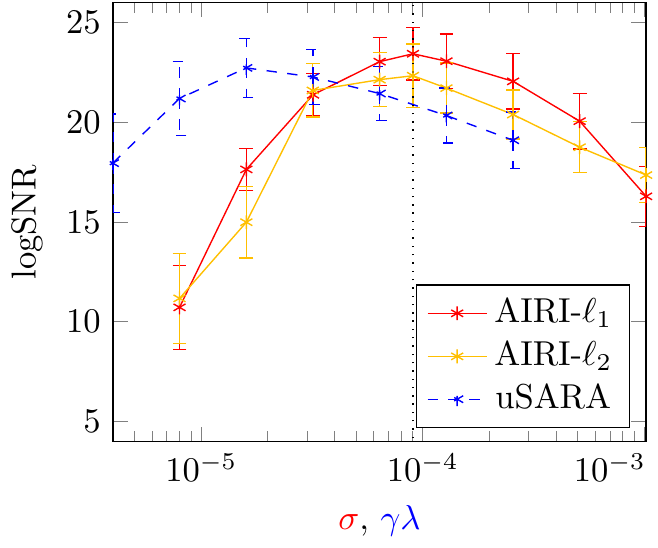}
     \vspace{-2em}
     \caption{Case $\Delta T=8\,\text{h}$}
     \label{fig:logsnr_2}
 \end{subfigure}
\vspace{-1em}
\caption{Experiment 2 results: Reconstruction metrics as a function of the training noise level $\sigma$ in the case of AIRI-$\ell_2$ and AIRI-$\ell_1$, or the thresholding parameter $\gamma\lambda$ in the case of uSARA. Top row: reconstruction SNR in the case of $\Delta T = 4\,\text{h}$ (a) and $\Delta T = 8\,\text{h}$ (b). Bottom row: reconstruction logSNR in the case of $\Delta T = 4\,\text{h}$ (c) and $\Delta T = 8\,\text{h}$ (d). Each point is an average over the 20 simulated observations built from the four different groundtruths from Figure~\ref{fig:gdths} and the five different $uv$-patterns shown in Figures~\ref{fig:samp} and ~\ref{fig:samp:seed2dt4}. Error bars show the 95\% confidence interval. On each graph, the black vertical dotted line indicates the common heuristic value for $\sigma$ and $\gamma\lambda$ (\emph{i.e.}~$\nu_{4}$ for $\Delta T = 4\,\text{h}$ and $\nu_{8}$ for $\Delta T = 8\,\text{h}$). Values for $\sigma$ and $\gamma\lambda$ are sampled a factor $\sqrt{2}$ around the heuristic and by steps of $2$ thereafter.}
\vspace{-1.5em}
\label{fig:exp1_metrics}
\end{figure}

Figure~\ref{fig:exp1_metrics} shows the reconstruction SNR and logSNR graphs obtained when varying $\sigma$ and  $\gamma\lambda$ for reconstructions with AIRI and uSARA respectively, for both the total observation durations $\Delta T = 4\,\text{h}$ and $\Delta T = 8\,\text{h}$. Each point on each of the four graphs gives the average value and $95\%$ confidence interval over 20 reconstructions arising when considering the four test images in Figure~\ref{fig:gdths} and five different pointing directions from Figures~\ref{fig:samp} and~\ref{fig:samp:seed2dt4}.

Firstly, for both AIRI-$\ell_2$ and AIRI-$\ell_1$, we observe that the SNR and logSNR metrics clearly peak at the proposed heuristic \eqref{eq:heuristic} (shown as the dashed vertical black line) for both $\Delta T = 4\,\text{h}$ and $\Delta T = 8\,\text{h}$, suggesting that equating the training dynamic range to the target dynamic range is an accurate procedure. This result is also confirmed by the visual analysis in Figure~\ref{fig:heuristic}. For uSARA, the SNR and logSNR results and the visual analysis in Figure~\ref{fig:heuristic} suggest that the relation \eqref{eq:heuristic_uSARA} is a useful reference, but rather provides an upper-bound on the optimal value for the soft-thresholding parameter, roughly 6 to 11 times smaller when looked at from an SNR standpoint, or 3 to 6 times smaller from a logSNR perspective. In general, we find that the logSNR provides better coherence with the visual analysis, suggesting a optimal soft-thresholding value 3 to 6 times below the reference. Interestingly, the $1/3$ correction factor brings a closer-to-optimal value, but does not solve the discrepancy between the SNR and logSNR metrics. In contrast to AIRI-$\ell_2$ and AIRI-$\ell_1$, uSARA thus exhibits a regularization parameter of which the value cannot be set simultaneously automatically and optimally. This represents a significant advantage of AIRI over uSARA.

Secondly, considering each method at its best, SNR and logSNR values for both $\Delta T = 4\, \text{h}$ and $\Delta T = 8\,\text{h}$ confirm a superior performance of AIRI-$\ell_1$ over both AIRI-$\ell_2$ and uSARA, the latter two achieving similar reconstruction qualities, with virtually same logSNR, and a slight SNR advantage for AIRI-$\ell_2$.

Thirdly, these results are in line with the visual analysis when considering each method at its best (see Figures \ref{fig:3c353all:7}, \ref{fig:3c353all:12}, \ref{fig:3c353all:13}, and Figures \ref{fig:3c353res:1}, \ref{fig:3c353res:6}, \ref{fig:3c353res:10}). AIRI-$\ell_2$ and uSARA exhibit similar reconstruction qualities with the former, smoother but with less artefacts than the latter. AIRI-$\ell_1$ exhibits less artefacts than AIRI-$\ell_2$, while preserving the resolution offered by uSARA. AIRI-$\ell_1$ provides the best residuals, uSARA residuals clearly contain sources discarded by the model and relating to the artefacts, while AIRI-$\ell_2$'s higher residuals correlate with the sub-optimal resolution.

Fourthly, we recall that AIRI-$\ell_2$, AIRI-$\ell_1$, and uSARA are underpinned by the same FB structure, and only differ by their choice of denoiser. From this perspective, the results of the current experiment are in line with those of the previous experiment, with a better regularization denoiser naturally leading to better overall reconstruction.

Finally, we observe that the confidence intervals around the mean SNR and logSNR values in Figure~\ref{fig:exp1_metrics} are rather small for both AIRI-$\ell_2$ and AIRI-$\ell_1$, and similar to those of uSARA. This is a first illustration that in the PnP approach there is no issue of generalizability to measurement conditions not considered during training as, by construction, the denoisers are blind to the measurement setting (with regards to the noise level, see discussion around \eqref{eq:heuristic}), in contrast to end-to-end networks.

\subsection{Experiment 3: validating AIRI in  precision \& cost}
\label{ssect:exp3}

\subsubsection{Simulation setup}
In this section, we validate AIRI-$\ell_2$ and AIRI-$\ell_1$ more extensively, both in reconstruction quality and computational cost, across a wider range of observation durations, and in comparison with, uSARA, SARA, CLEAN, as well as UNets trained in an end-to-end fashion. We consider the same four groundtruth images and experimental setup as for the previous experiment (see Section~\ref{ssect:simul_setup}), with the same set of five pointing directions (see $uv$-coverages in Figures~\ref{fig:samp} and \ref{fig:samp:seed2dt4}). In order to further assess robustness to varying measurement conditions, the set of observation durations is enlarged to $\Delta T \in\{1\,\text{h},2\,\text{h},4\,\text{h},8\,\text{h}\}$ (see Figure~\ref{fig:samplings_seed2}). We note that a different UNet is trained for each observation duration to alleviate the challenge of generalizability with respect to the sampling pattern. The exponentiation parameter for the UNet training database, resulting from the procedure in Section~\ref{ssect:dr_enhancement}, is $a=10^3$.

We recall that our groundtruth images have peak values at $1$, yielding the same heuristic value for $\sigma$ and $\gamma\lambda$. These values are still within a $25\%$ variation range across pointing directions for all four $\Delta T$ values, and a single heuristic value is considered for all five $uv$-patterns associated with the same $\Delta T$. The resulting four average $\nu_{\Delta T}$ are: $\nu_1=2.9\times 10^{-4}$, $\nu_2=2.2\times 10^{-4}$, $\nu_4=1.4\times 10^{-4}$, and $\nu_8=9.3\times 10^{-5}$. Given the result of the previous experiment the values of $\sigma$ for AIRI-$\ell_2$ and AIRI-$\ell_1$ are taken exactly at the heuristic, while the value of $\gamma\lambda$ for uSARA is fine-tuned manually to optimize the logSNR of each reconstruction, again resulting in soft-thresholding values ranging between $3$ and $6$ times below the heuristic, depending on $\Delta T$. Moreover, setting $\kappa = 10^{-9}$ (resp. $\kappa = 10^{-5}$) for the AIRI-$\ell_2$ (resp. AIRI-$\ell_1$) denoiser in \eqref{eq:training_loss}, with $\varepsilon = 5\times 10^{-2}$, ensured the stability of Algorithm~\ref{algo:pnp}. Following the procedure detailed in Section~\ref{ssect:dr_enhancement}, the exponentiation parameter for on-the-fly dynamic range enhancement of the database  is $a=10^3$. 

At the level of Algorithms~\ref{algo:sara_fb}, \ref{algo:dfb}, and \ref{algo:pnp}, the stepsize $\gamma$, convergence criteria $\left(\xi_{i}\right)_{1\leq i \leq 3}$ and maximum number of iterations are set to the same values as in the previous experiment. 

Finally, AIRI-$\ell_2$, AIRI-$\ell_1$, uSARA, and SARA are run in Matlab, while CLEAN is implemented in the highly optimized C++ WSClean package, and the UNets are implemented in Python using PyTorch. uSARA, SARA, and CLEAN implementations utilize 10 CPU cores (Intel Xeon E5-2695 2.1 GHz). The parallel PDFB algorithmic structure underpinning SARA (see Section~\ref{ssect:sara}) optimizes the distribution of the operations involved at each iteration across those 10 CPU cores: 1 for the data-fidelity term, and 1 for each of the 9 bases of the average sparsity dictionary $\bm{\Psi}$ \citep{onose2016scalable}. For uSARA, the average sparsity dictionary is parallelized via a faceting procedure using 9 facets \citep{Prusa2012,thouvenin2021parallel}. AIRI-$\ell_2$ and AIRI-$\ell_1$ utilize 10 CPU cores and 1 GPU (NVIDIA Tesla V100). The latter is for the application of the denoiser, as commonly done with DNNs in deep learning. The DNN can be run on CPU, but this slows down the reconstructions significantly, as will be evident from the experimental results. Implementations of uSARA, SARA, and CLEAN leveraging GPUs are not investigated here. The UNets are run utilizing 1 GPU (NVIDIA Tesla V100).

\subsubsection{Reconstruction quality}

\begin{figure}
\centering
\begin{subfigure}[b]{0.49\linewidth}
     \centering
     \includegraphics[width=\textwidth, clip]{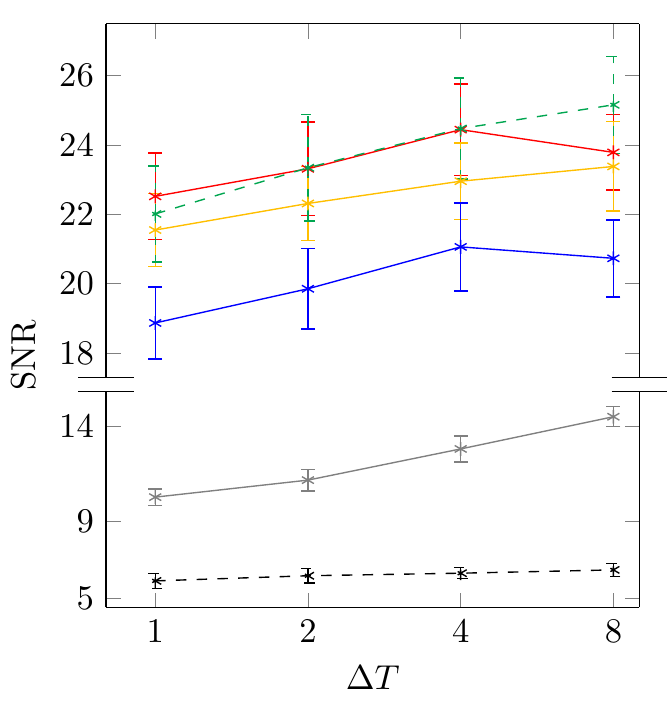}
 \end{subfigure}
 \hfill
 \begin{subfigure}[b]{0.5\linewidth}
     \centering
     \includegraphics[width=\textwidth, clip]{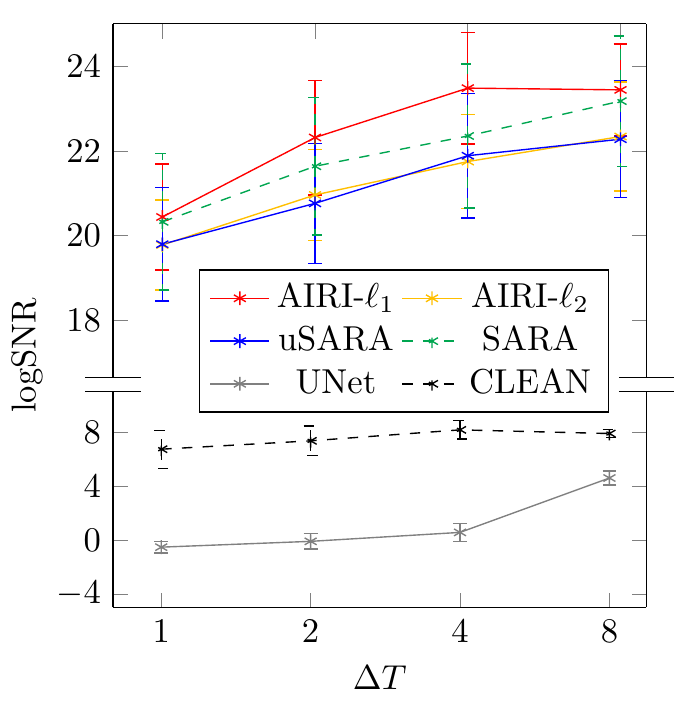}
 \end{subfigure}
 \vspace{-2em}
\caption{Experiment 3 results: Reconstruction SNR (left) and logSNR (right) as a function of the observation duration for AIRI-$\ell_2$, AIRI-$\ell_1$, uSARA, SARA, CLEAN and the end-to-end UNets. Each point is an average over the 20 simulated observations built from the four different groundtruths from Figure~\ref{fig:gdths} and the five different  $uv$-patterns shown in Figures~\ref{fig:samp} and \ref{fig:samp:seed2dt4}. For AIRI-$\ell_2$ and AIRI-$\ell_1$, $\sigma$ is chosen at the heuristic. For uSARA, $\gamma\lambda$ is fine-tuned manually to optimize the logSNR, resulting in values between $3$ and $6$ times below the heuristic (in line with Figure \ref{fig:exp1_metrics}). Error bars show the $95\%$ confidence interval.}
\vspace{-1em}
\label{fig:exp2:metrics}
\end{figure}

Figure~\ref{fig:exp2:metrics} shows the average reconstruction SNR and logSNR as a function of the observation duration $\Delta T$. Each point, on each of the two graphs, represents an average over 20 reconstructions arising when considering the four test images in Figure~\ref{fig:gdths} and five pointing directions, while error bars give the $95\%$ confidence intervals. As expected, the reconstruction SNR increases with $\Delta T$, as the data size increases with $\Delta T$. The results extend the conclusion of the previous experiment of a superior performance of AIRI-$\ell_1$ over both AIRI-$\ell_2$ and uSARA  across $\Delta T \in\{1\,\text{h},2\,\text{h},4\,\text{h},8\,\text{h}\}$. In SNR, AIRI-$\ell_1$ and AIRI-$\ell_2$ are respectively around 3dB and 2dB above uSARA. In logSNR, AIRI-$\ell_1$ is between 1dB and 2dB above both AIRI-$\ell_2$ and uSARA. The comparison with SARA reveals that AIRI-$\ell_1$ is on par with SARA in SNR and slightly superior in logSNR across the range of measurement settings considered. As previously, we observe that the confidence interval around the mean SNR and logSNR values for both AIRI-$\ell_2$ and AIRI-$\ell_1$ is similar to that of uSARA and SARA, reflecting the robustness of the proposed method. The reconstruction quality of both CLEAN and the UNets trained end-to-end is much inferior to those of AIRI and SARA approaches, for both SNR and logSNR and all measurement conditions considered. Interestingly, notice that the UNets significantly improve over CLEAN in linear scale, but perform significantly worse in logarithmic scale, which confirms the difficulty of the UNets to recover very faint emissions. This phenomenon was also noticed during training.

We recall that SARA and uSARA leverage the same prior model, encapsulated in the averaged sparsity proximal operator. However, SARA relies on a constrained data-fidelity term (see objective \eqref{eq:SARA}), while uSARA uses an unconstrained formulation enforcing less abruptly data fidelity, in the sense that it imposes a soft penalty rather than a hard constraint (see objective \eqref{eq:uSARA}). A by-product of our analysis is to confirm the superiority of the former over the latter. 

\begin{landscape}
\begin{figure}%
\captionsetup[subfigure]{justification=centering}%
    \centering
        \begin{subfigure}[t]{0.165\linewidth}%
         \centering%
         \includegraphics[width=0.9\textwidth]{./pictures/colorbar_log_bottom.pdf}%
     \end{subfigure}%
    \hfill%
     \begin{subfigure}[t]{0.165\linewidth}%
         \centering%
         \phantom{\includegraphics[trim={0 5em 0 10em}, width=0.5\textwidth, clip]{./pictures/colorbar_log_bottom.pdf}}%
     \end{subfigure}%
     \hfill%
     \begin{subfigure}[t]{0.165\linewidth}%
         \centering
          \phantom{\includegraphics[trim={0 5em 0 10em}, width=0.5\textwidth, clip]{./pictures/colorbar_log_bottom.pdf}}
     \end{subfigure}
     \hfill%
     \begin{subfigure}[t]{0.165\linewidth}%
         \centering
          \phantom{\includegraphics[trim={0 5em 0 10em}, width=0.5\textwidth, clip]{./pictures/colorbar_log_bottom.pdf}}
     \end{subfigure}
     \hfill%
     \begin{subfigure}[t]{0.165\linewidth}%
         \centering
          \phantom{\includegraphics[trim={0 5em 0 10em}, width=0.5\textwidth, clip]{./pictures/colorbar_log_bottom.pdf}}
     \end{subfigure}
     
     \begin{subfigure}[t]{0.165\linewidth}%
         \centering%
         \includegraphics[trim={0 8em 0 7em}, width=\textwidth, clip]{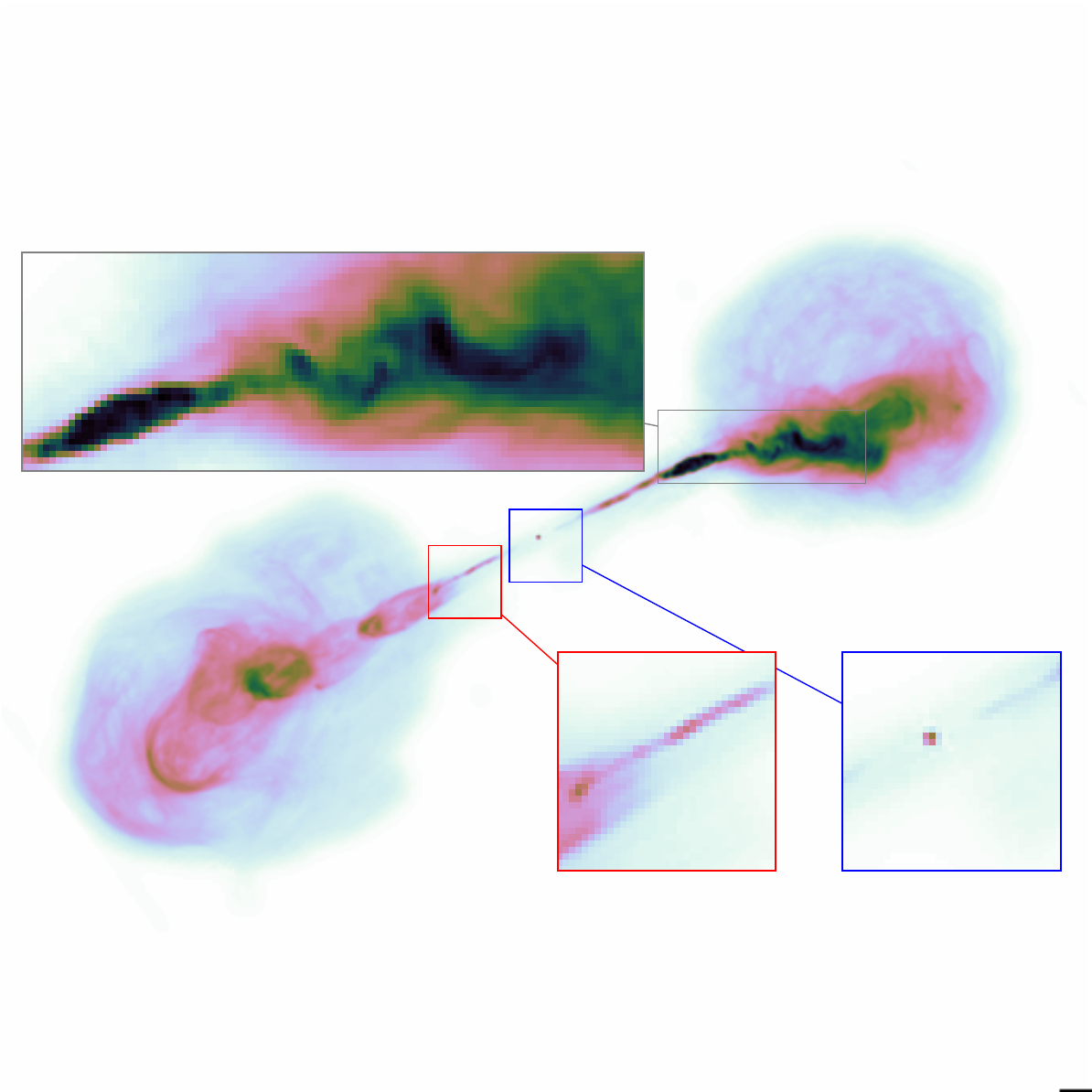}%
         \caption{Groundtruth  \\ \phantom{(SNR, logSNR)}}%
         \label{fig:hA:1}
     \end{subfigure}%
     \hfill
      \begin{subfigure}[t]{0.165\linewidth}%
         \centering%
         \includegraphics[trim={0 8em 0 7em},width=\textwidth, clip]{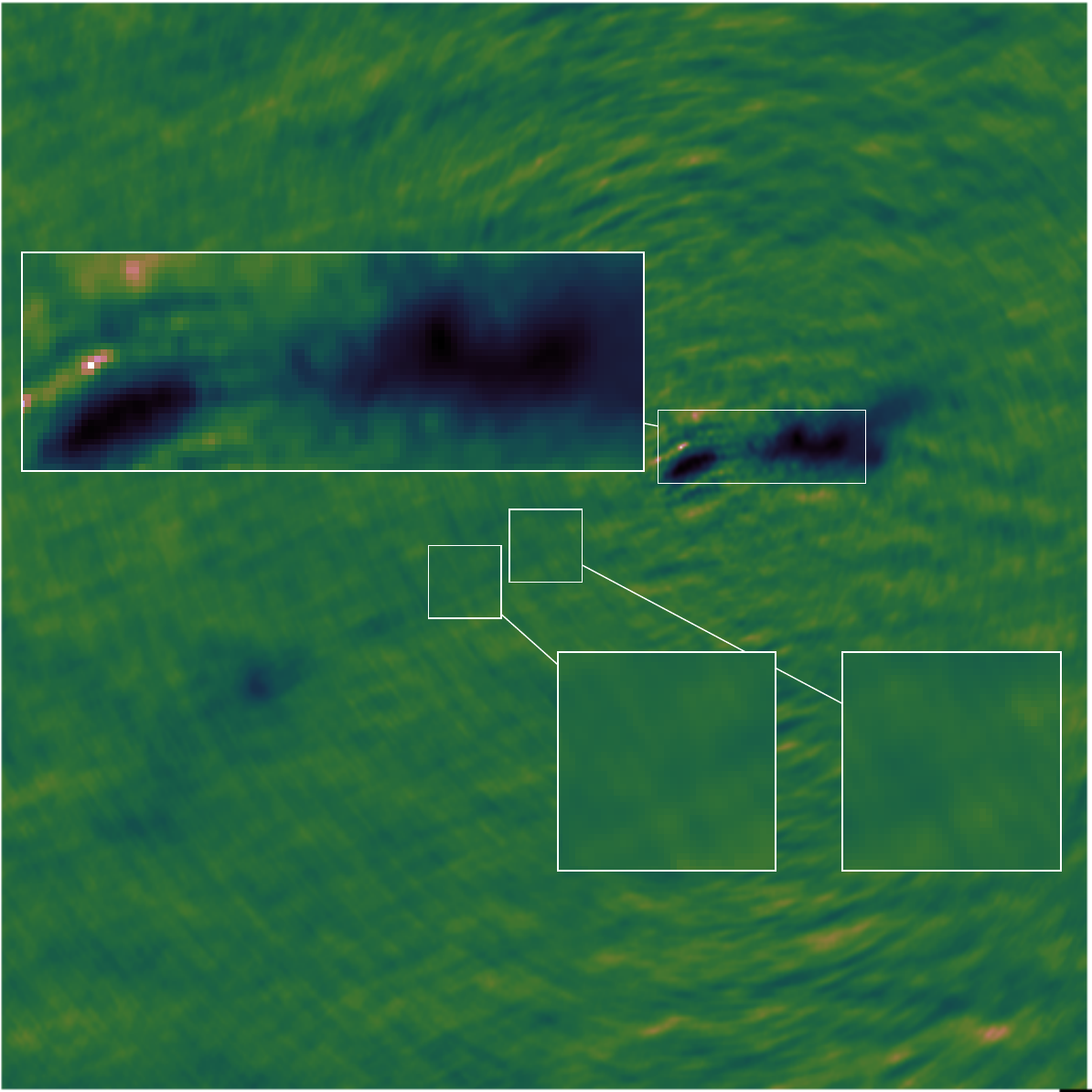}%
         \caption{Dirty image, $\Delta T=1\,\text{h}$ \\ \phantom{(SNR, logSNR)}}%
         \label{fig:hA:2}%
     \end{subfigure}%
     \hfill
     \begin{subfigure}[t]{0.165\linewidth}%
         \centering%
         \includegraphics[trim={0 8em 0 7em}, width=\textwidth, clip]{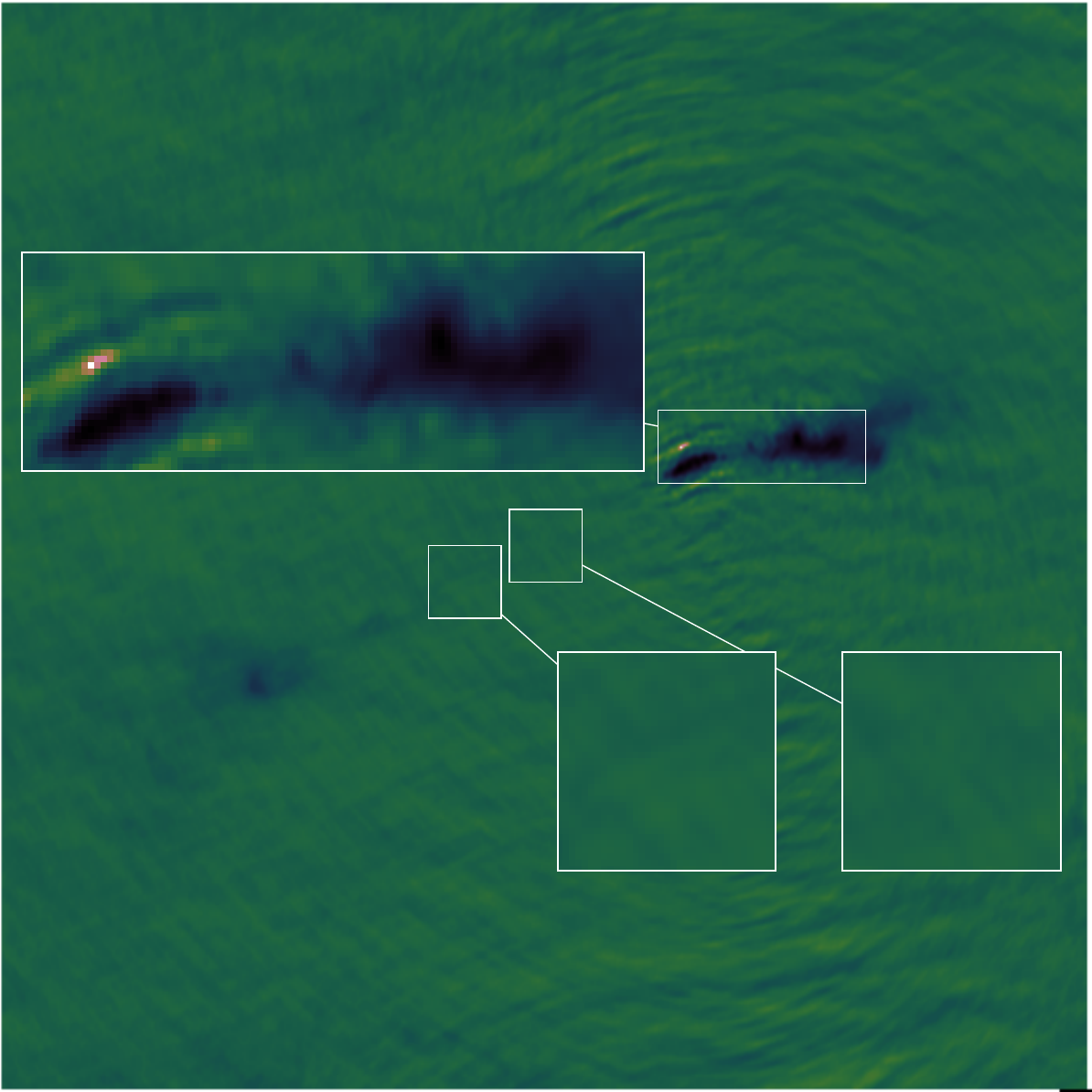}%
         \caption{Dirty image, $\Delta T=2\,\text{h}$ \\ \phantom{(SNR, logSNR)}}%
         \label{fig:hA:3}%
     \end{subfigure}%
     \hfill
     \begin{subfigure}[t]{0.165\linewidth}%
         \centering%
         \includegraphics[trim={0 8em 0 7em}, width=\textwidth, clip]{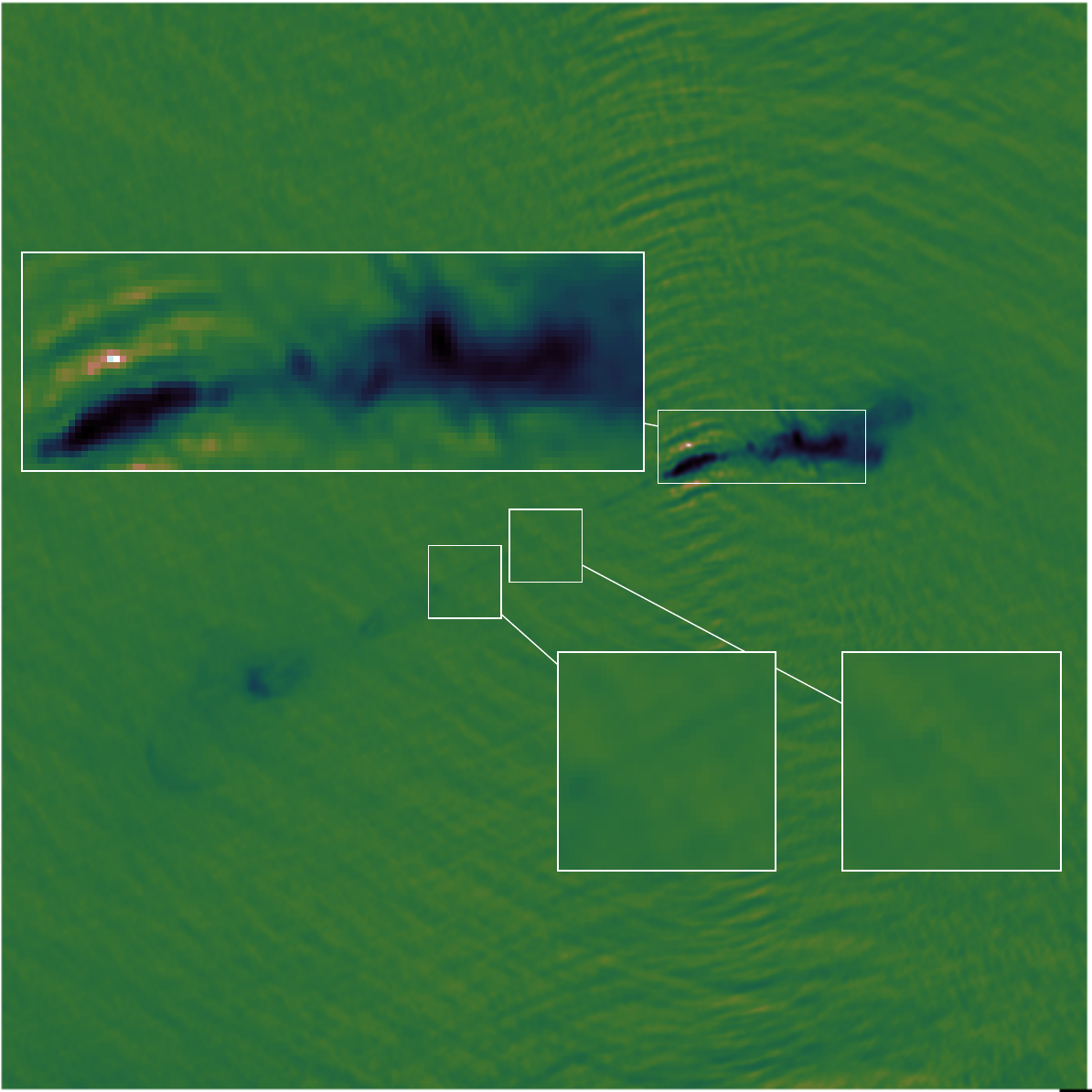}%
         \caption{Dirty image, $\Delta T=4\,\text{h}$ \\ \phantom{(SNR, logSNR)}}%
         \label{fig:hA:4}
     \end{subfigure}%
     \hfill
     \begin{subfigure}[t]{0.165\linewidth}%
         \centering%
         \phantom{\includegraphics[trim={0 8em 0 7em}, width=\textwidth,clip]{pictures/hercA/zoom_dirty_4.pdf}}%
     \end{subfigure}%
     \hfill
     \begin{subfigure}[t]{0.165\linewidth}%
         \centering%
         \phantom{\includegraphics[trim={0 8em 0 7em}, width=\textwidth,clip]{pictures/hercA/zoom_dirty_4.pdf}}%
     \end{subfigure}%

     \begin{subfigure}[t]{0.165\linewidth}%
         \centering%
         \includegraphics[trim={0 8em 0 7em}, width=\textwidth,, clip]{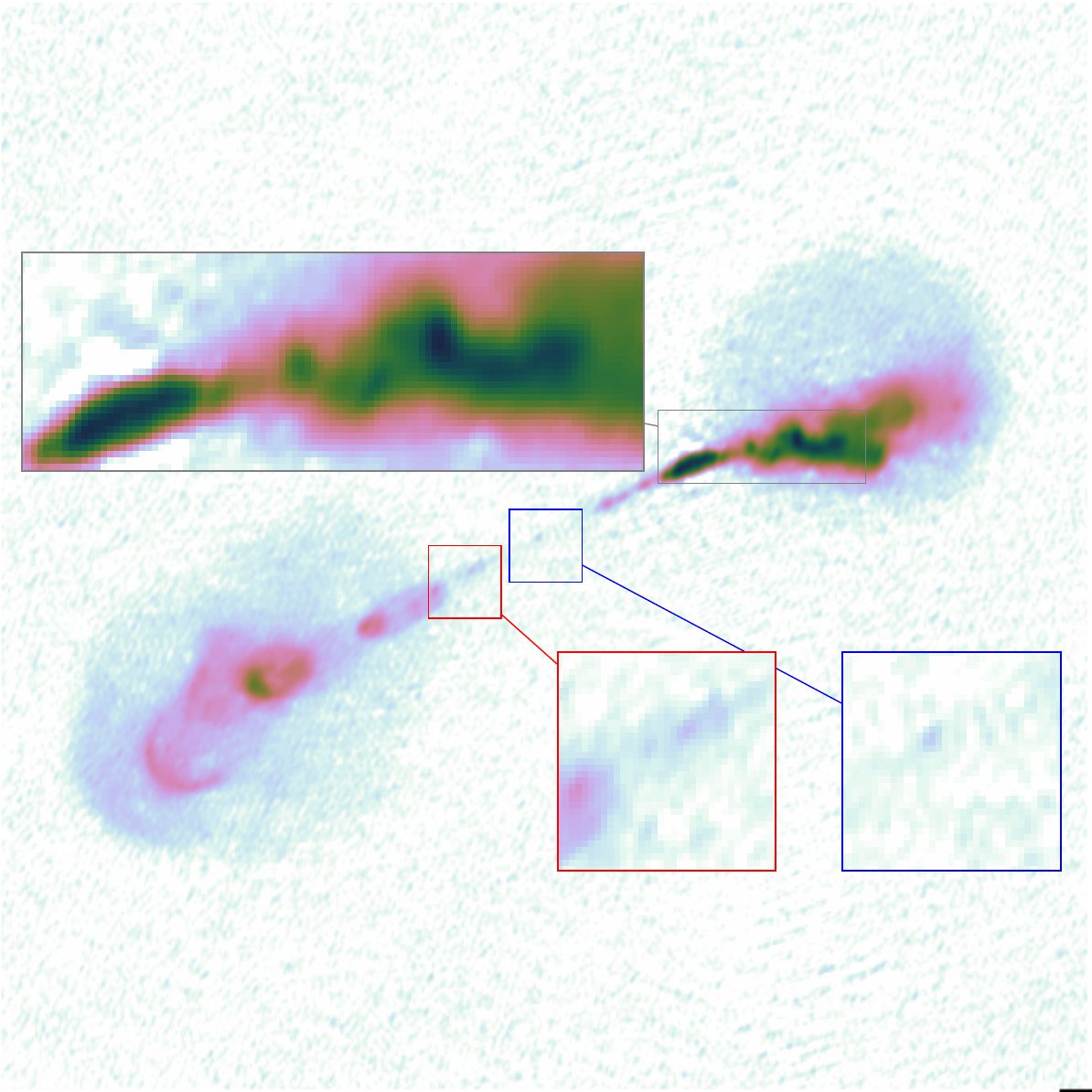}%
         \caption{CLEAN, $\Delta T = 1\,\text{h}$ \\ $(4.90\, \text{dB}, 9.26\, \text{dB}, 4\,\text{min})$}%
         \label{fig:hA:5}%
     \end{subfigure}%
    \hfill
          \begin{subfigure}[t]{0.165\linewidth}%
         \centering%
         \includegraphics[trim={0 8em 0 7em}, width=\textwidth,, clip]{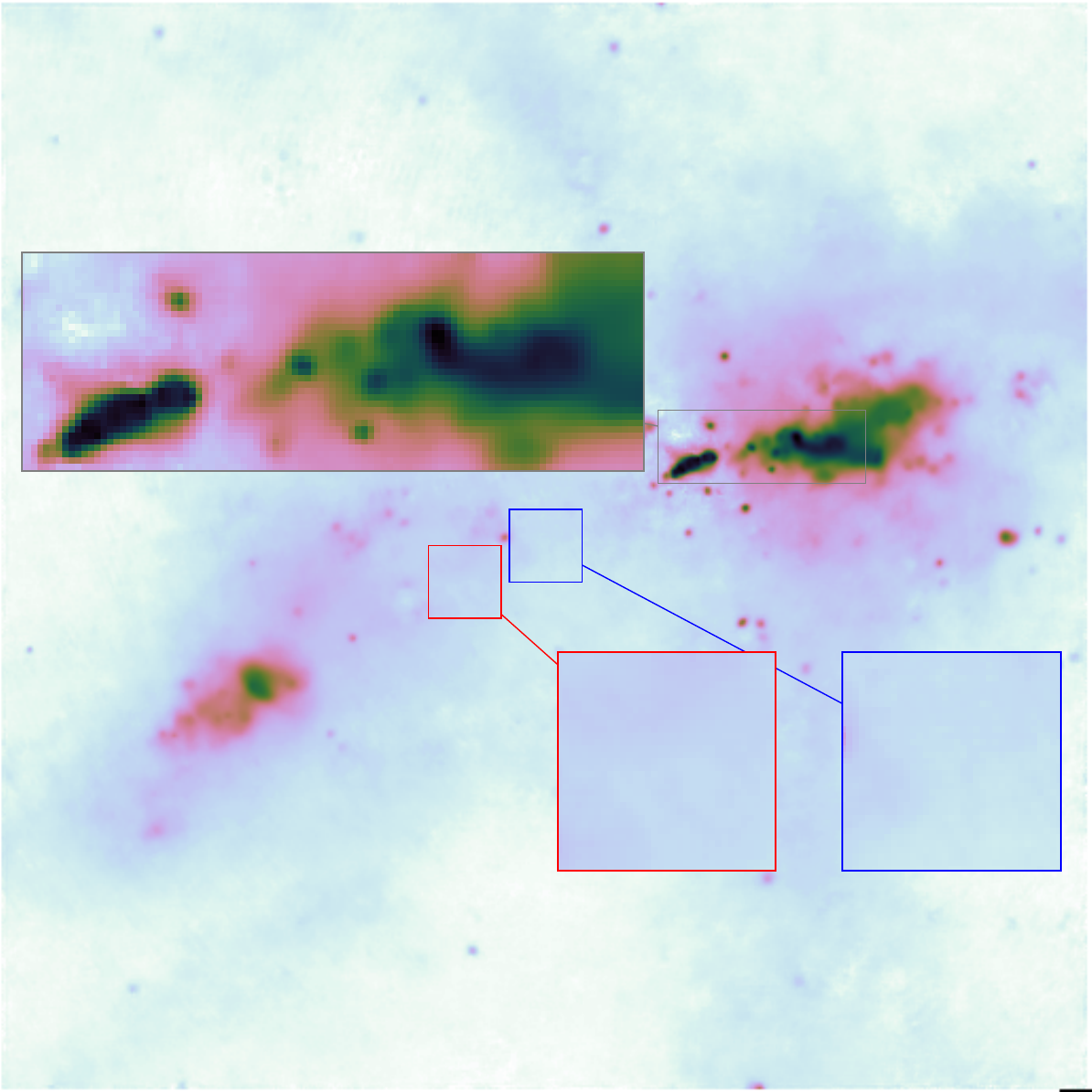}%
         \caption{UNet, $\Delta T = 1\,\text{h}$ \\ $(11.05\, \text{dB}, 1.24\, \text{dB}, 7\,\text{ms})$}%
         \label{fig:hA:5bis}%
     \end{subfigure}%
    \hfill
          \begin{subfigure}[t]{0.165\linewidth}%
         \centering%
         \includegraphics[trim={0 8em 0 7em}, width=\textwidth,, clip]{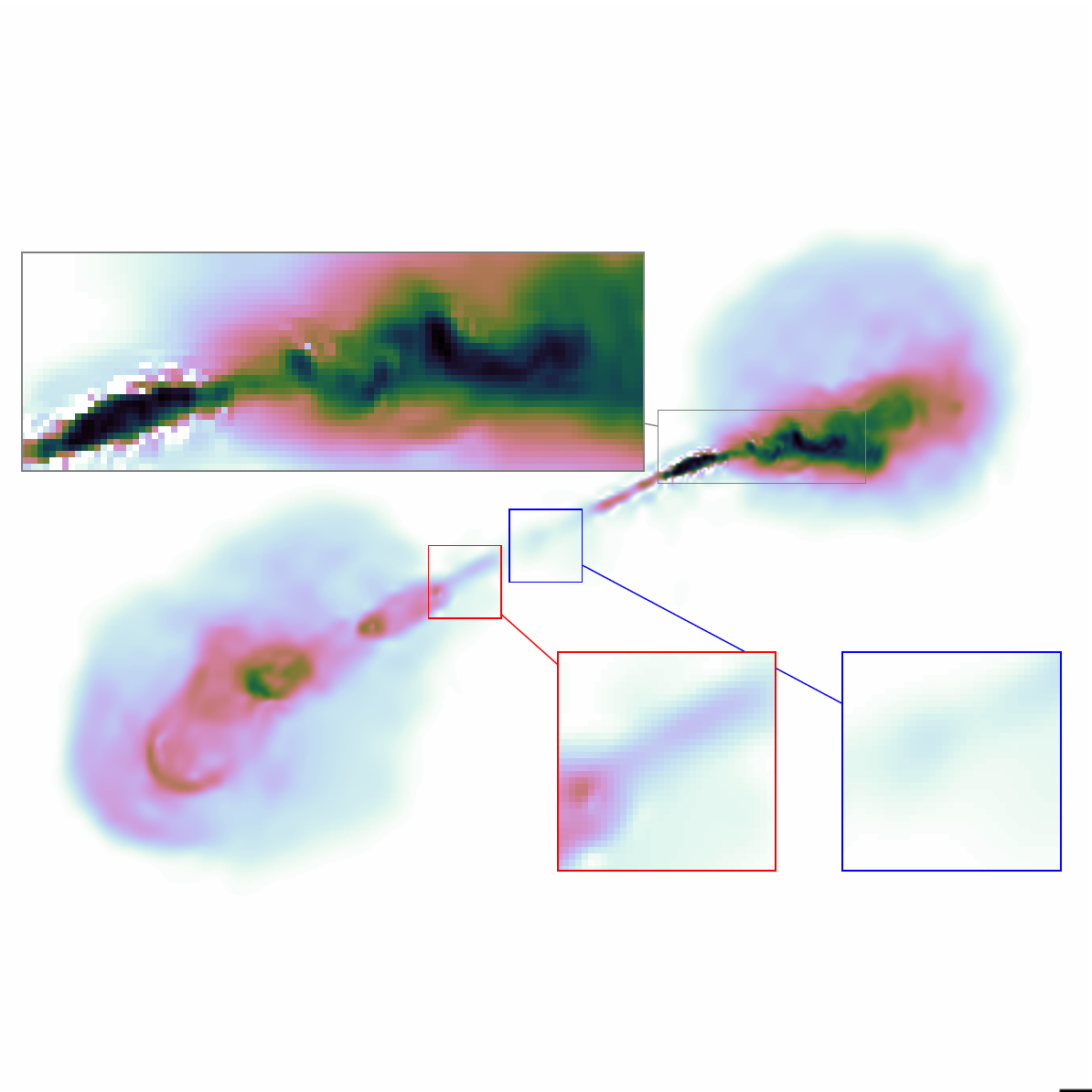}%
         \caption{SARA, $\Delta T = 1\,\text{h}$ \\ $(23.78\, \text{dB}, 22.99\, \text{dB}, 38 \text{min})$}%
         \label{fig:hA:6}%
     \end{subfigure}%
    \hfill
          \begin{subfigure}[t]{0.165\linewidth}%
         \centering%
         \includegraphics[trim={0 8em 0 7em}, width=\textwidth,, clip]{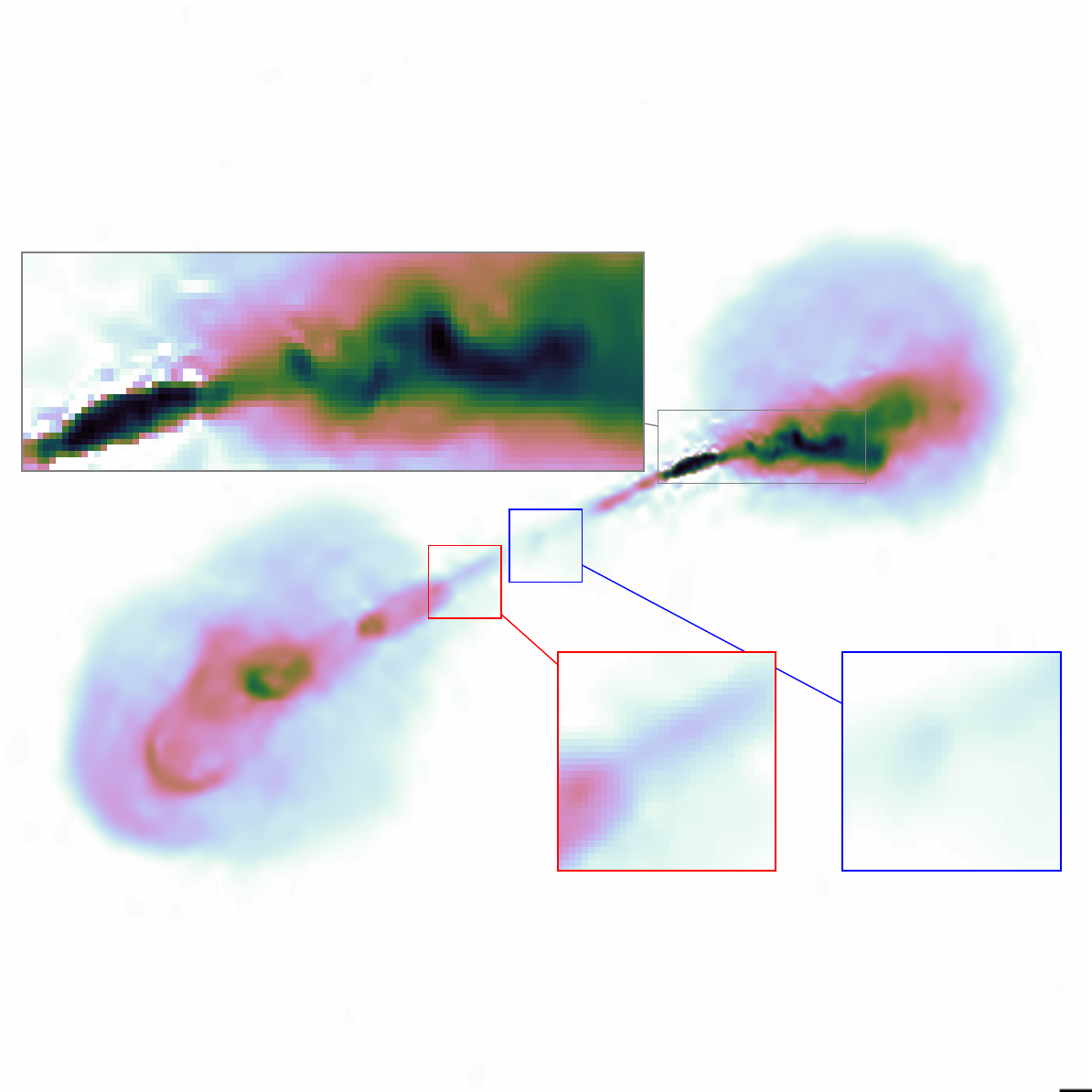}%
         \caption{uSARA, $\Delta T = 1\,\text{h}$ \\ $(19.73\, \text{dB}, 21.47\, \text{dB}, 3.5 \text{h})$}%
         \label{fig:hA:7}%
     \end{subfigure}%
    \hfill
          \begin{subfigure}[t]{0.165\linewidth}%
         \centering%
         \includegraphics[trim={0 8em 0 7em}, width=\textwidth,, clip]{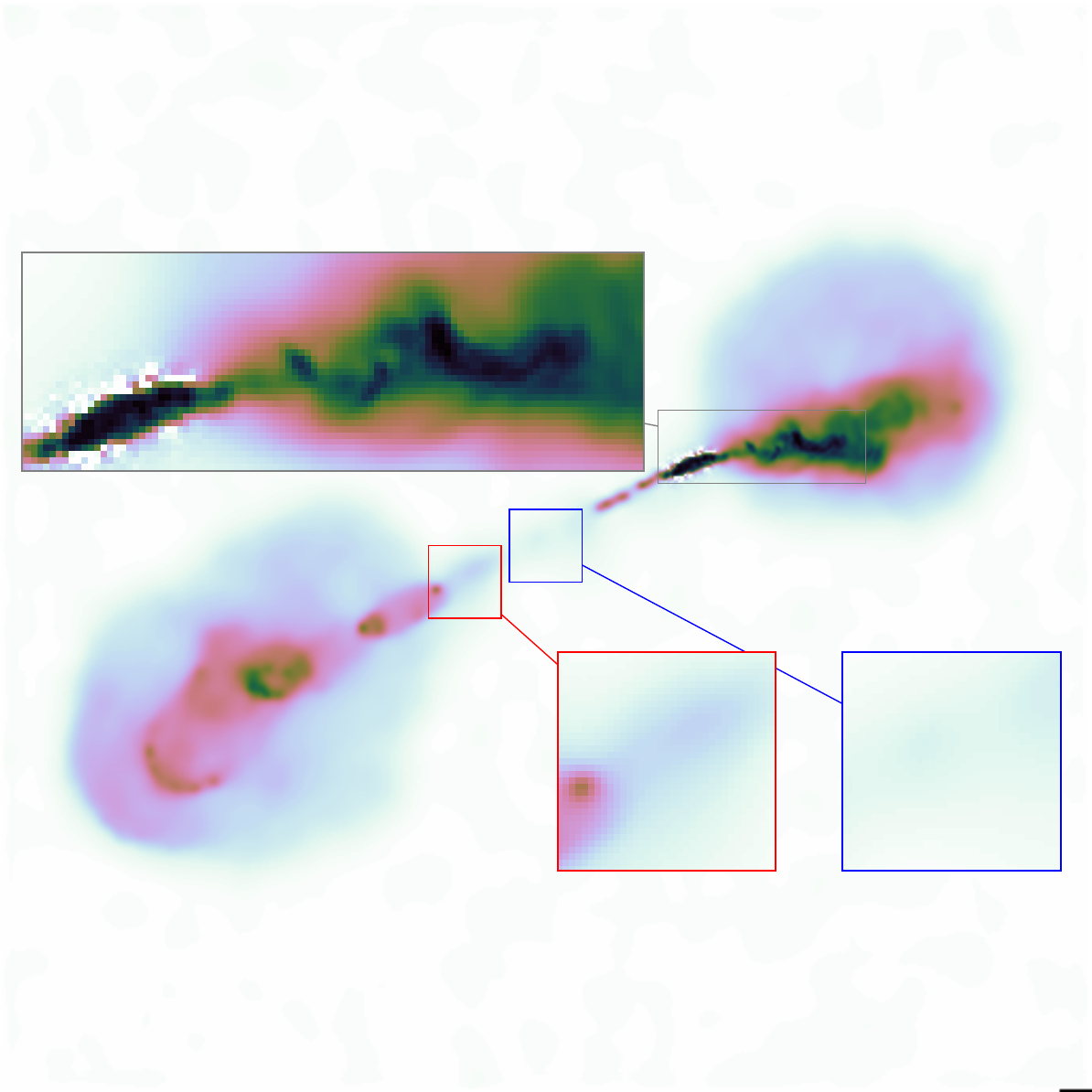}%
         \caption{AIRI-$\ell_2$, $\Delta T = 1\,\text{h}$ \\ $(22.40\, \text{dB}, 21.22\, \text{dB}, 12 \text{min})$}%
         \label{fig:hA:8}%
     \end{subfigure}%
    \hfill
          \begin{subfigure}[t]{0.165\linewidth}%
         \centering%
         \includegraphics[trim={0 8em 0 7em}, width=\textwidth,, clip]{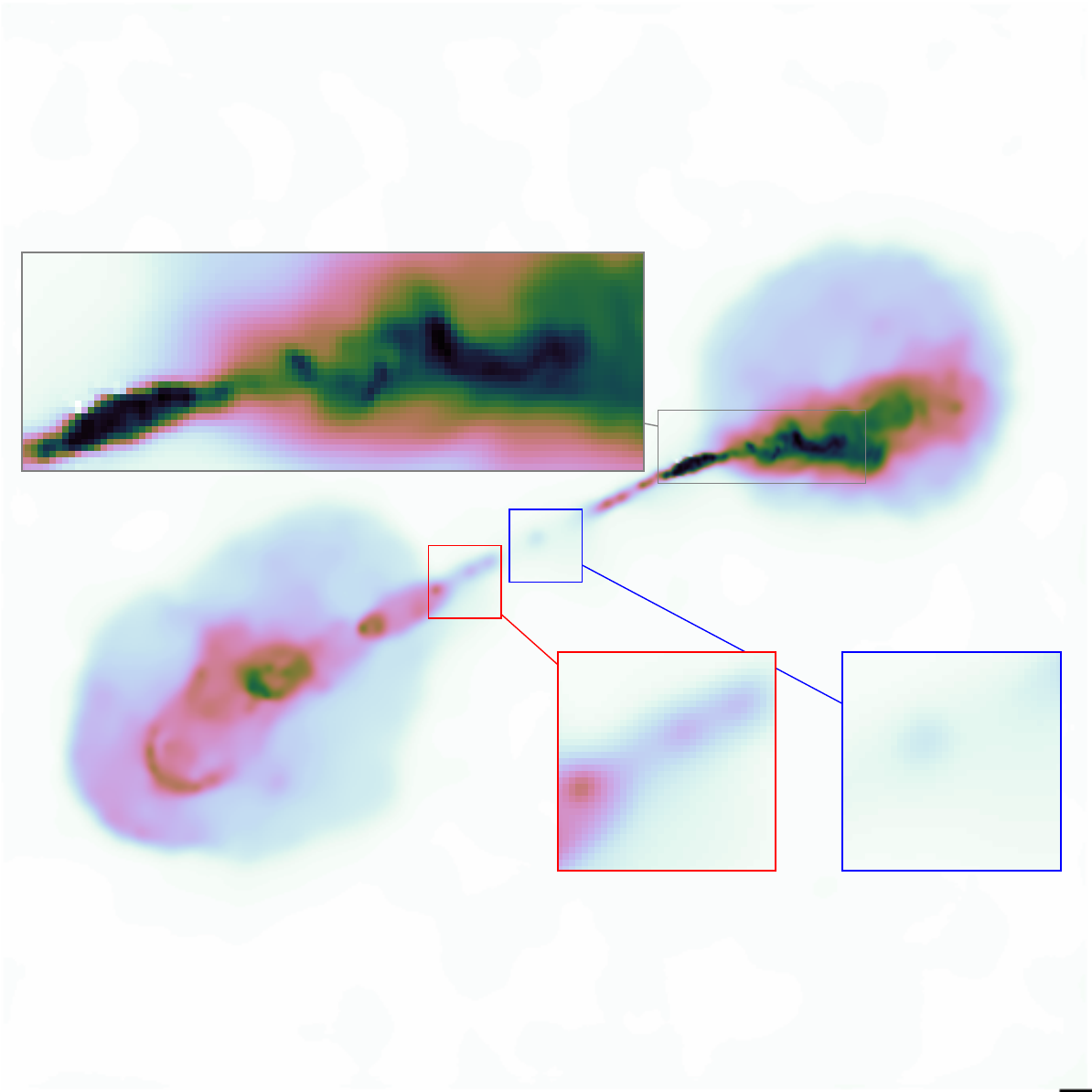}%
         \caption{AIRI-$\ell_1$, $\Delta T = 1\,\text{h}$ \\ $(24.13\, \text{dB}, 21.80\, \text{dB}, 12\,\text{min})$}%
         \label{fig:hA:9}%
     \end{subfigure}%
     
    \begin{subfigure}[t]{0.165\linewidth}%
         \centering%
         \includegraphics[trim={0 8em 0 7em}, width=\textwidth,, clip]{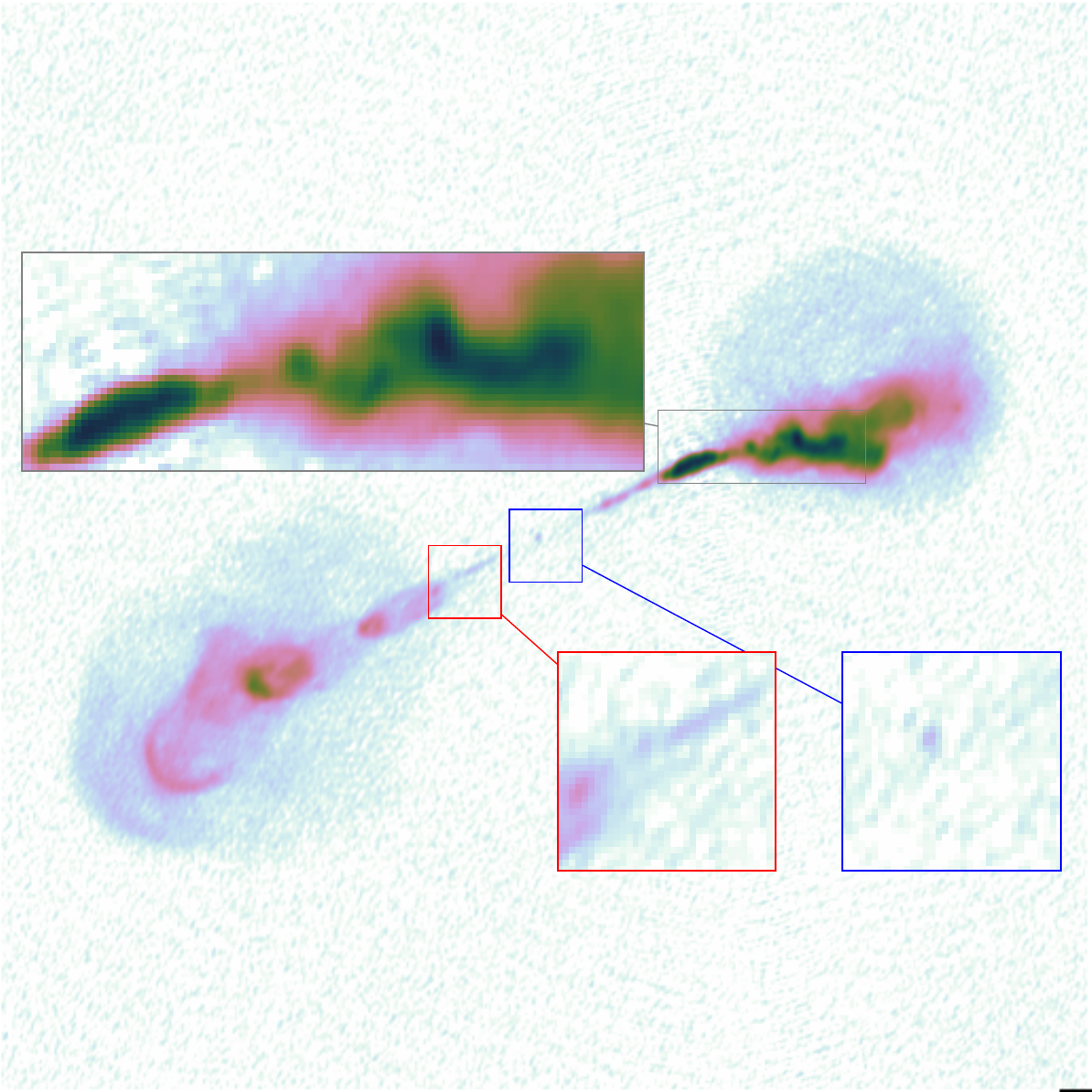}%
         \caption{CLEAN, $\Delta T = 2\,\text{h}$ \\$(5.05\, \text{dB}, 8.96\, \text{dB}, 3\,\text{min})$}%
         \label{fig:hA:10}%
     \end{subfigure}%
     \hfill
      \begin{subfigure}[t]{0.165\linewidth}%
         \centering%
         \includegraphics[trim={0 8em 0 7em},width=\textwidth,, clip]{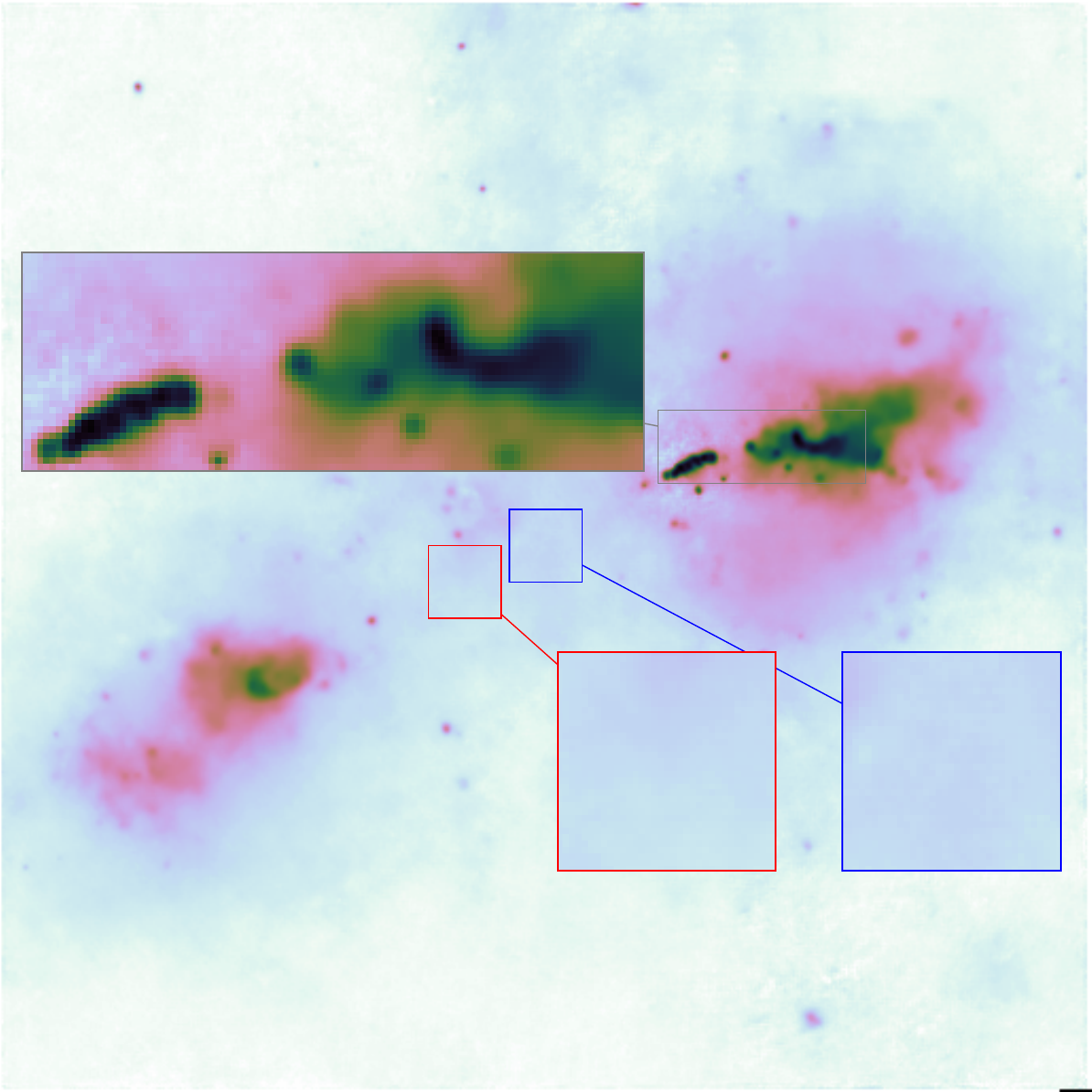}%
         \caption{UNet, $\Delta T = 2\,\text{h}$ \\ $(12.22\, \text{dB}, 1.19\, \text{dB}, 7\,\text{ms})$}%
         \label{fig:hA:11bis}%
     \end{subfigure}%
     \hfill
      \begin{subfigure}[t]{0.165\linewidth}%
         \centering%
         \includegraphics[trim={0 8em 0 7em},width=\textwidth,, clip]{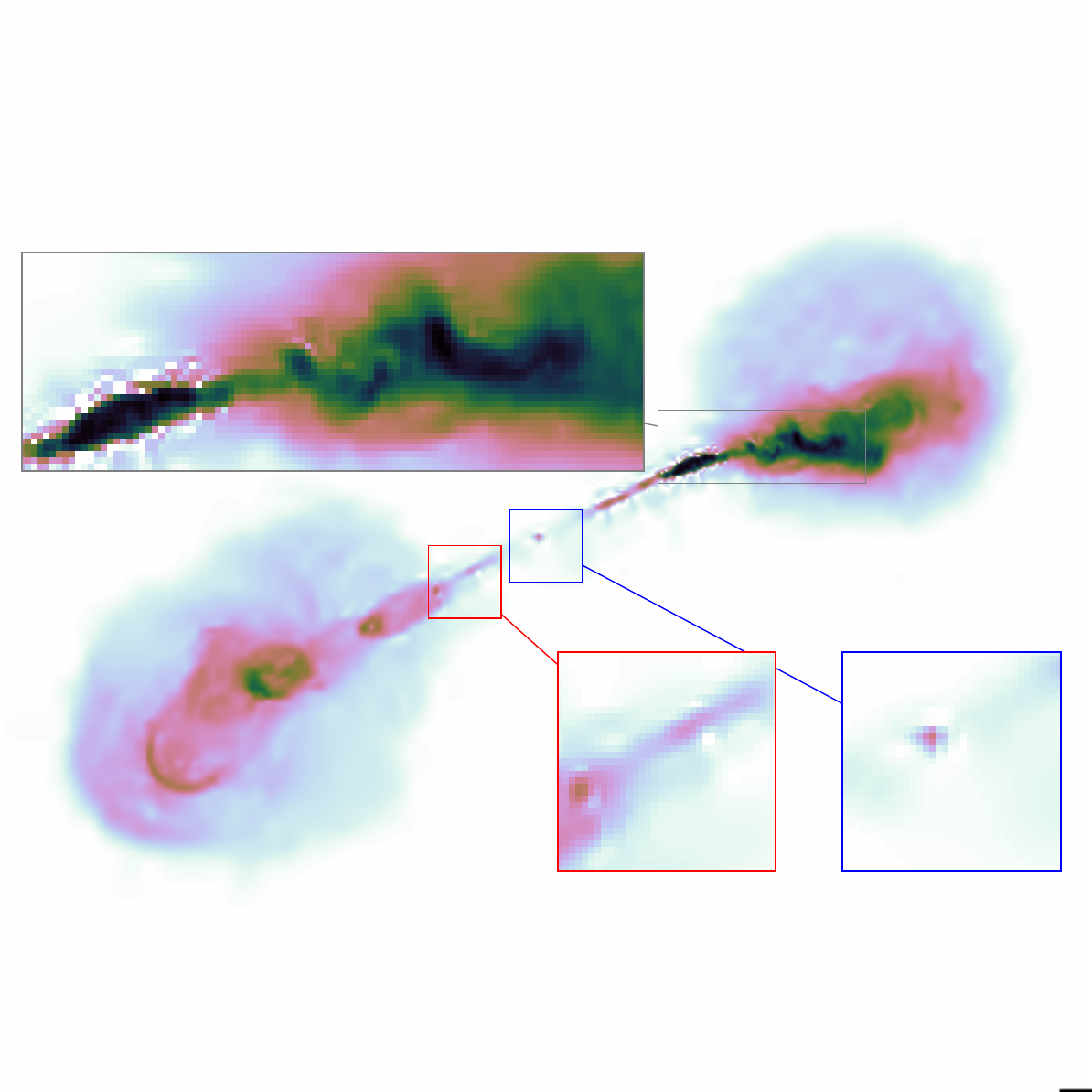}%
         \caption{SARA, $\Delta T = 2\,\text{h}$ \\ $(25.44\, \text{dB}, 23.63\, \text{dB}, 49\,\text{min})$}%
         \label{fig:hA:12}%
     \end{subfigure}%
     \hfill
      \begin{subfigure}[t]{0.165\linewidth}%
         \centering%
         \includegraphics[trim={0 8em 0 7em},width=\textwidth,, clip]{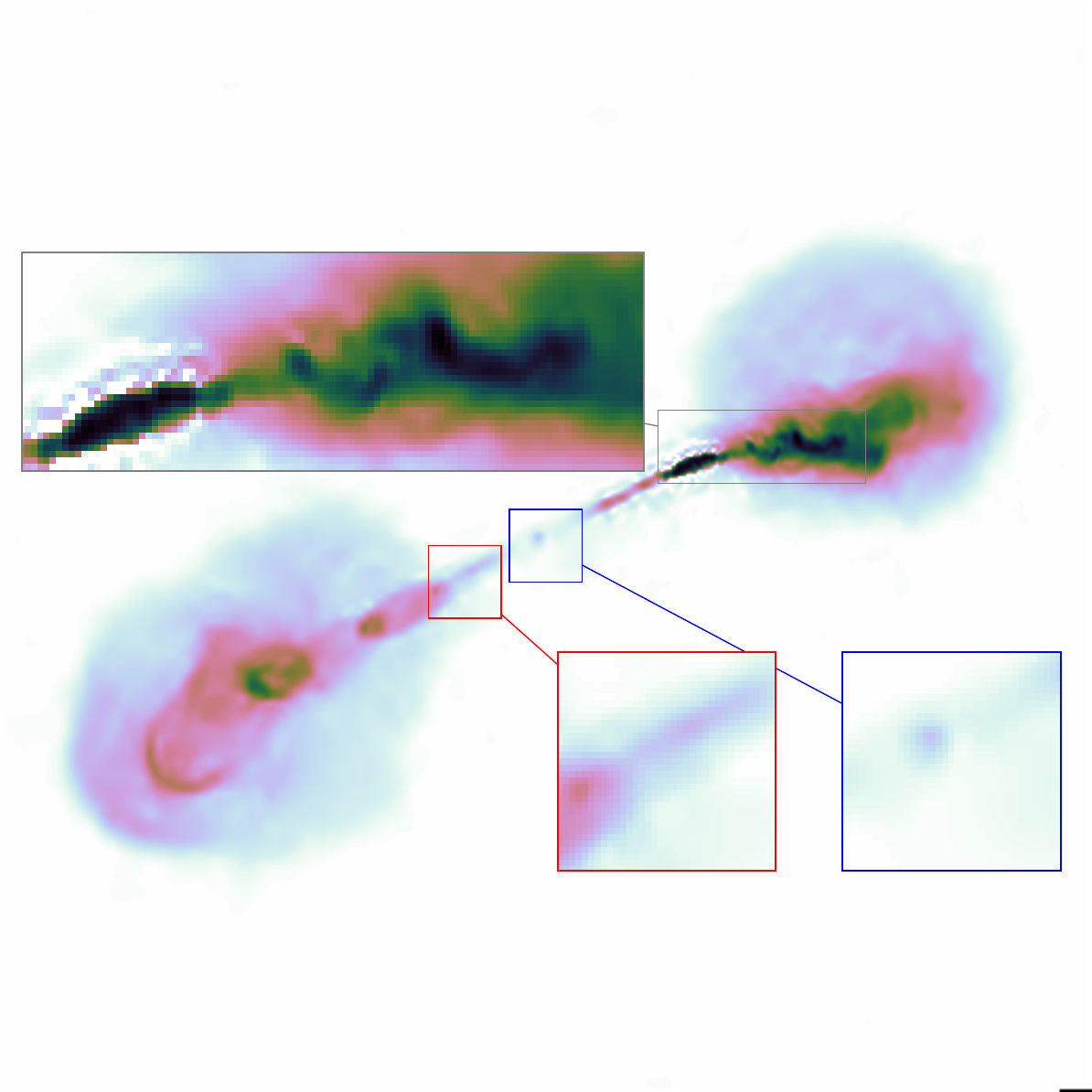}%
         \caption{uSARA, $\Delta T = 2\,\text{h}$ \\$(20.53\, \text{dB}, 22.61\, \text{dB}, 2.9\,\text{h})$}%
         \label{fig:hA:11}%
     \end{subfigure}%
     \hfill
      \begin{subfigure}[t]{0.165\linewidth}%
         \centering%
         \includegraphics[trim={0 8em 0 7em},width=\textwidth,, clip]{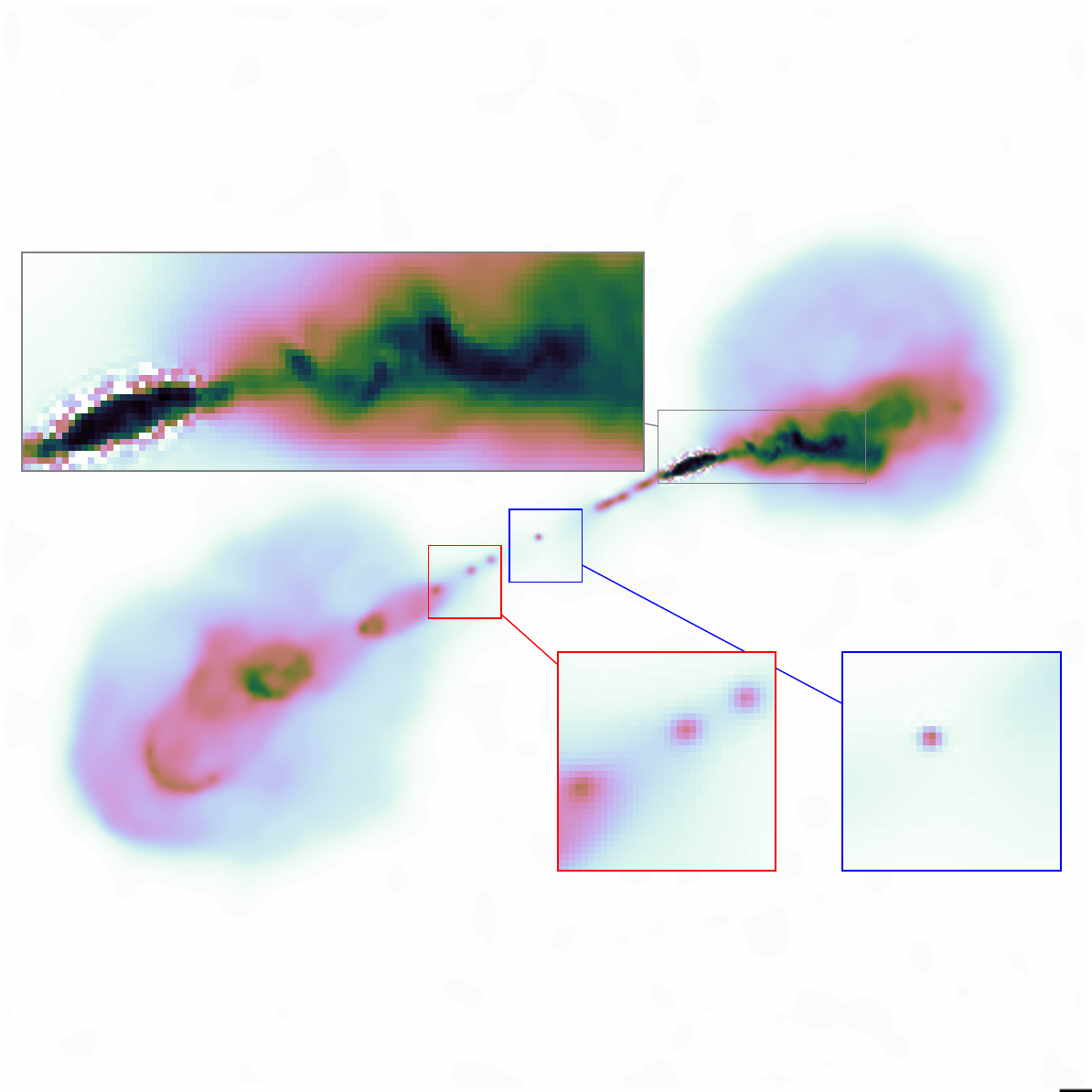}%
         \caption{AIRI-$\ell_2$, $\Delta T = 2\,\text{h}$ \\ $(23.54\, \text{dB}, 23.15\, \text{dB}, 20\,\text{min})$}%
         \label{fig:hA:13}%
     \end{subfigure}%
     \hfill
      \begin{subfigure}[t]{0.165\linewidth}%
         \centering%
         \includegraphics[trim={0 8em 0 7em},width=\textwidth,, clip]{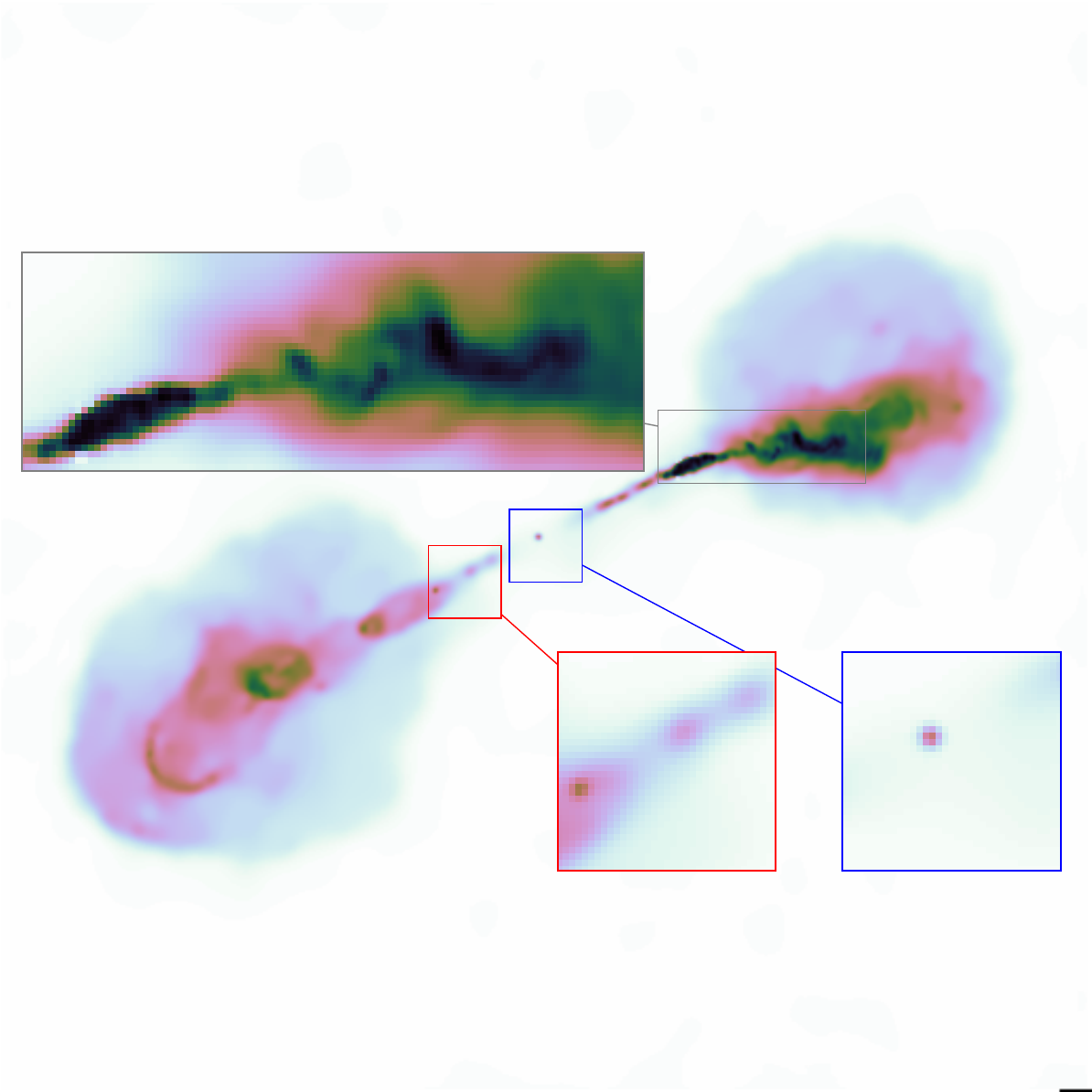}%
         \caption{AIRI-$\ell_1$, $\Delta T = 2\,\text{h}$ \\$(25.30\, \text{dB}, 24.76\, \text{dB}, 20\,\text{min})$}%
         \label{fig:hA:14}%
     \end{subfigure}%

      \begin{subfigure}[t]{0.165\linewidth}%
         \centering%
         \includegraphics[trim={0 8em 0 7em}, width=\textwidth,, clip]{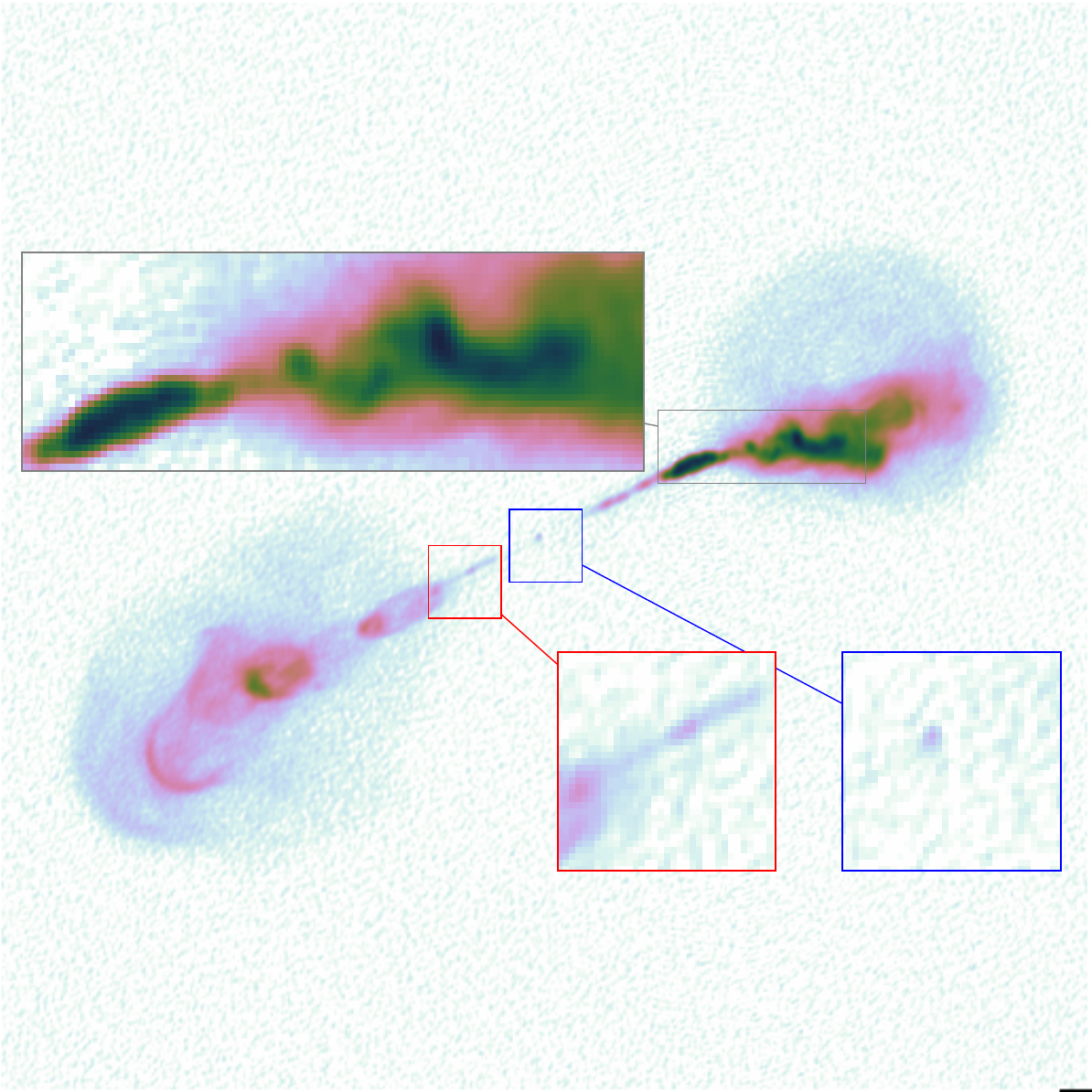}%
         \caption{CLEAN, $\Delta T = 4\,\text{h}$ \\$(5.19\, \text{dB}, 8.95\, \text{dB}, 4\,\text{min})$}%
         \label{fig:hA:15}%
     \end{subfigure}%
     \hfill
      \begin{subfigure}[t]{0.165\linewidth}%
         \centering%
         \includegraphics[trim={0 8em 0 7em}, width=\textwidth,, clip]{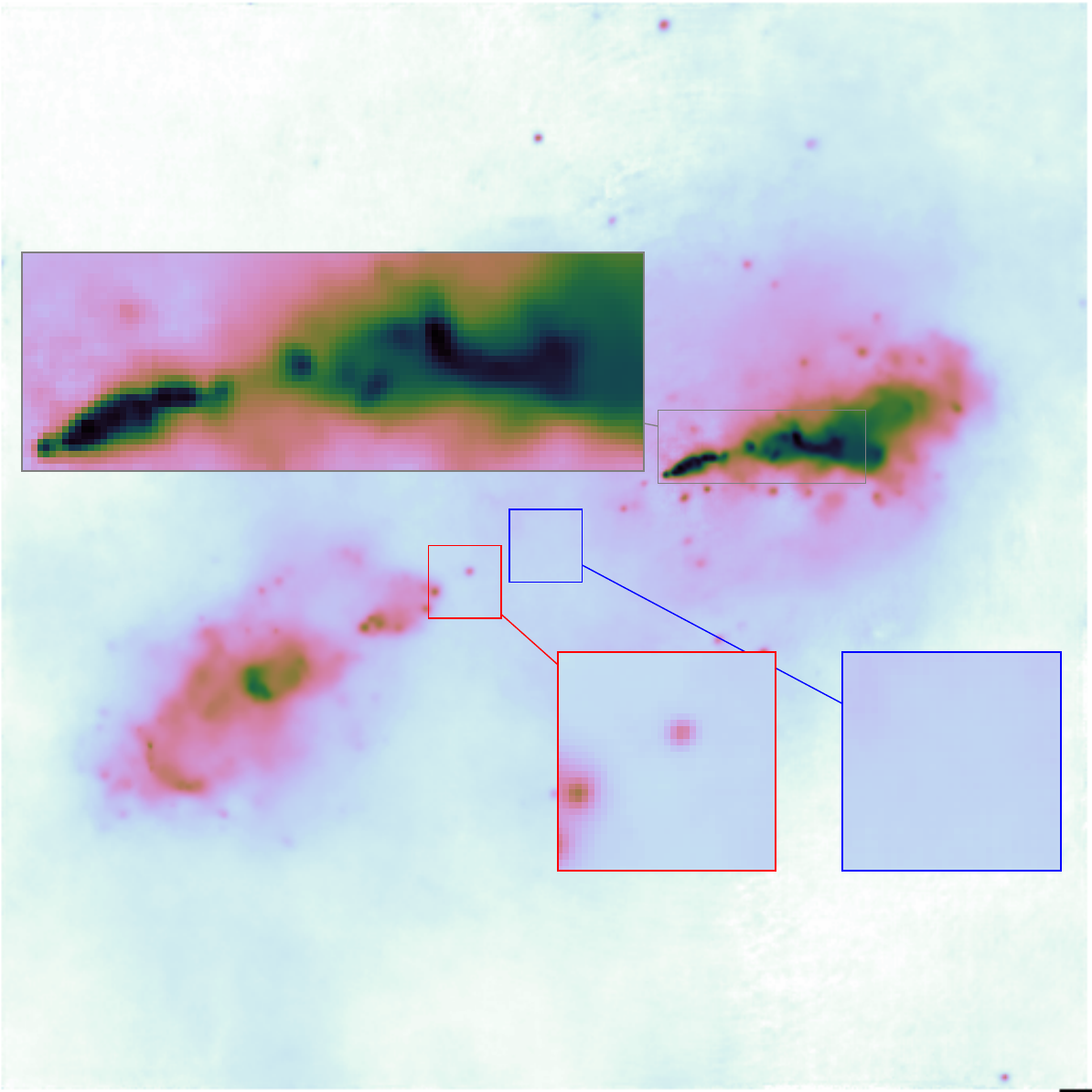}%
         \caption{UNet, $\Delta T = 4\,\text{h}$ \\ $(13.81\, \text{dB}, 1.0\, \text{dB}, 7\,\text{ms})$}%
         \label{fig:hA:15bis}%
     \end{subfigure}%
     \hfill
      \begin{subfigure}[t]{0.165\linewidth}%
         \centering%
         \includegraphics[trim={0 8em 0 7em}, width=\textwidth,, clip]{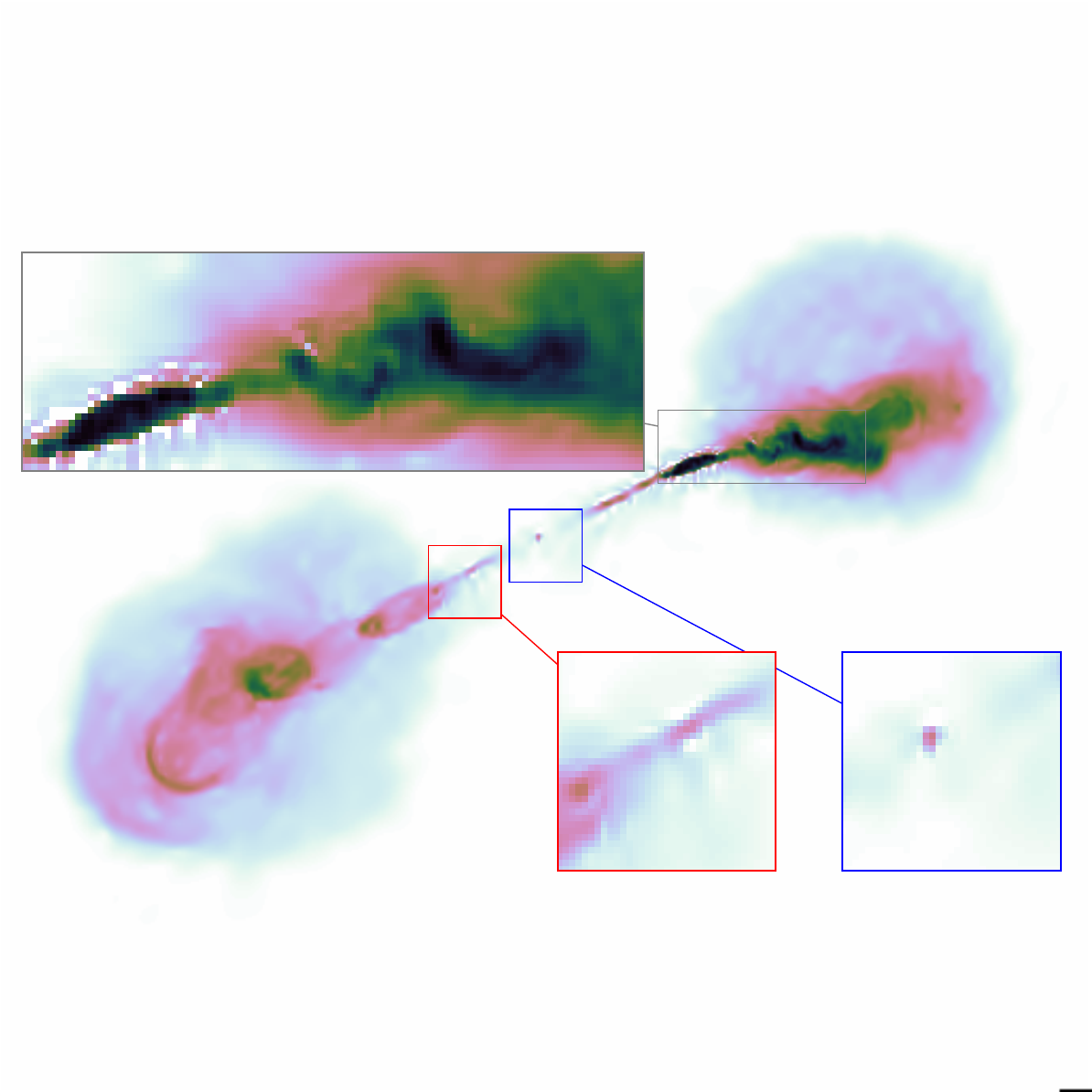}%
         \caption{SARA, $\Delta T = 4\,\text{h}$ \\ $(26.34\, \text{dB}, 24.69\, \text{dB}, 1.3\,\text{h})$}%
         \label{fig:hA:17}%
     \end{subfigure}%
     \hfill
      \begin{subfigure}[t]{0.165\linewidth}%
         \centering%
         \includegraphics[trim={0 8em 0 7em}, width=\textwidth,, clip]{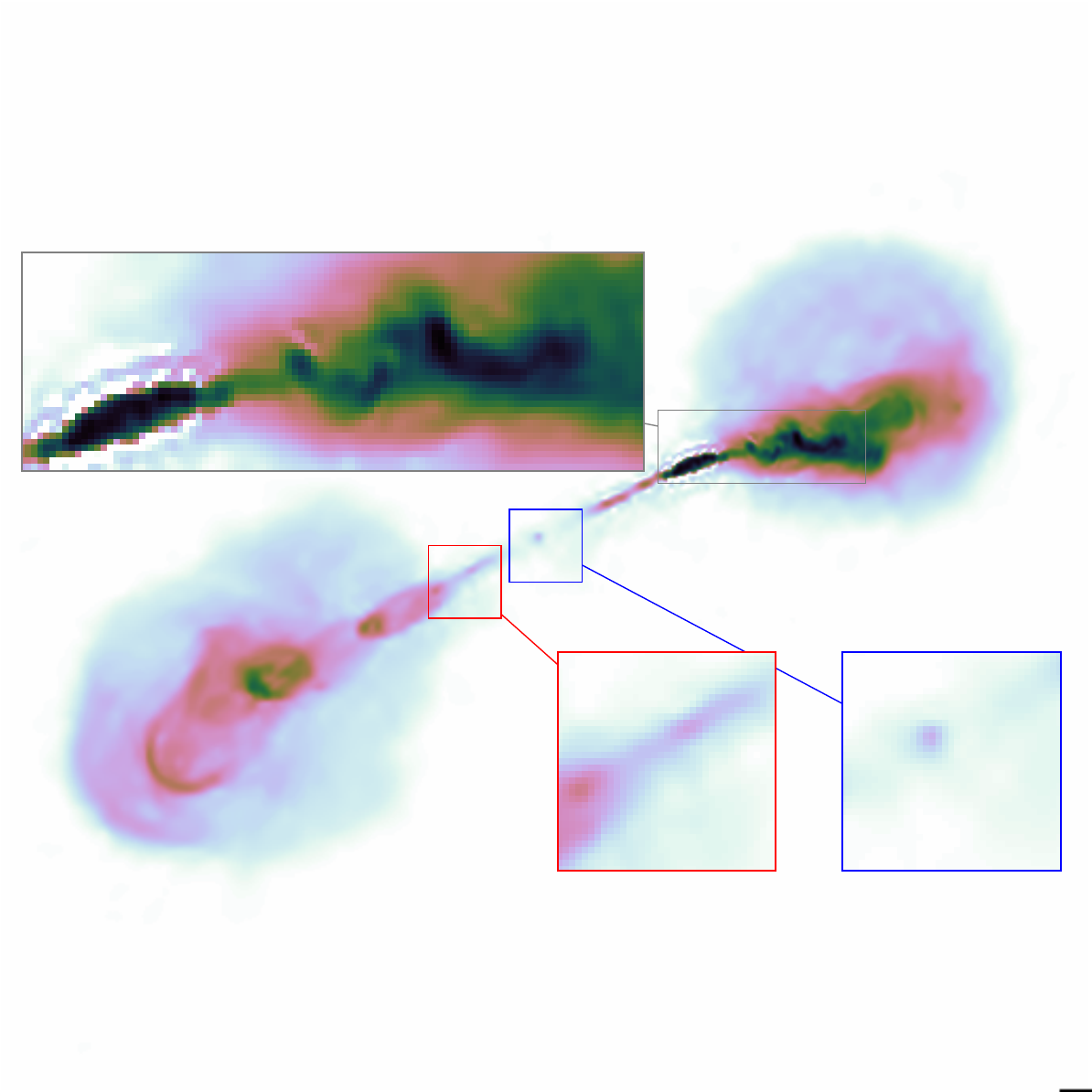}%
         \caption{uSARA, $\Delta T = 4\,\text{h}$ \\ $(22.42\, \text{dB}, 23.97\, \text{dB}, 3.4\,\text{h})$}%
         \label{fig:hA:16}%
     \end{subfigure}%
     \hfill
        \begin{subfigure}[t]{0.165\linewidth}%
         \centering%
         \includegraphics[trim={0 8em 0 7em}, width=\textwidth,, clip]{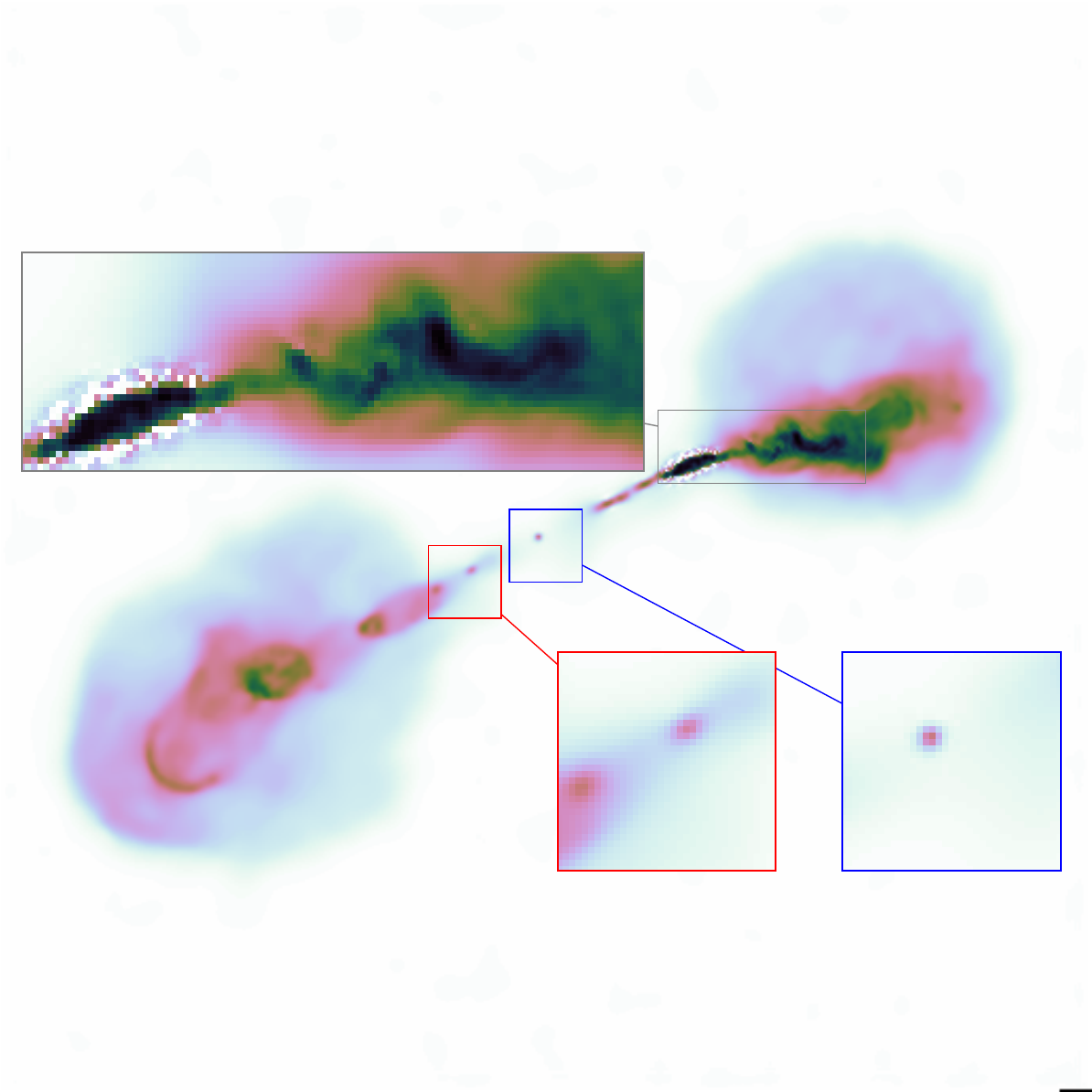}%
         \caption{AIRI-$\ell_2$, $\Delta T = 4\,\text{h}$ \\ $(24.41\, \text{dB}, 23.77\, \text{dB}, 39\,\text{min})$}%
         \label{fig:hA:18}%
     \end{subfigure}%
     \hfill
      \begin{subfigure}[t]{0.165\linewidth}%
         \centering%
         \includegraphics[trim={0 8em 0 7em}, width=\textwidth,, clip]{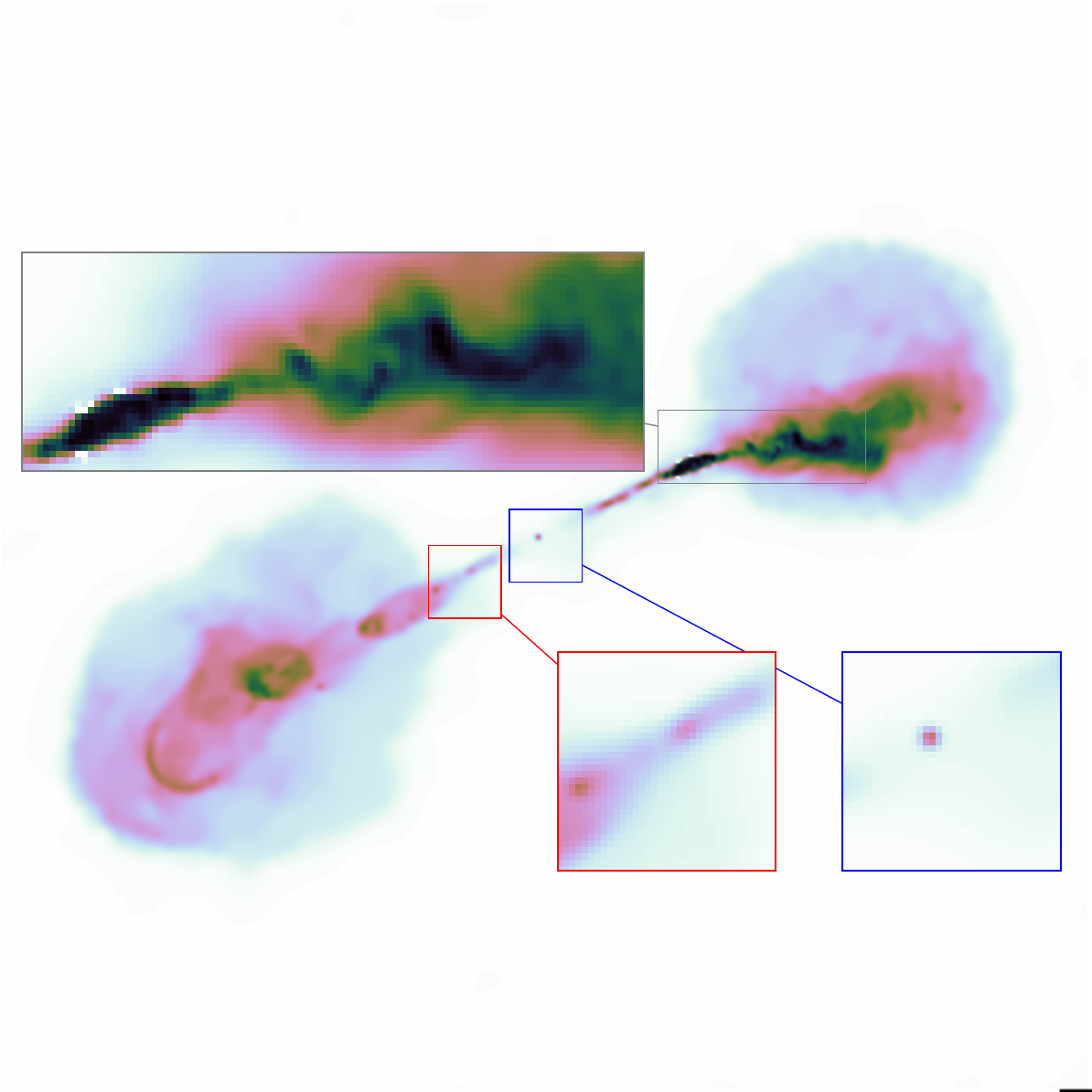}%
         \caption{AIRI-$\ell_1$, $\Delta T = 4\,\text{h}$ \\$(26.41\, \text{dB}, 25.81\, \text{dB}, 39\,\text{min})$}%
         \label{fig:hA:19}%
     \end{subfigure}%
      
\caption{Experiment 3 results: Influence of the observation duration $\Delta T$ on the reconstruction quality for the different algorithms with the pointing position from Figure~\ref{fig:samplings_seed2}; the associated $uv$-patterns are displayed in Figures~\ref{fig:samp:seed2dt1}, \ref{fig:samp:seed2dt2}, and \ref{fig:samp:seed2dt4}. Second, third and last row show reconstruction results for $\Delta T = 1\,\text{h}$, $\Delta T = 2\,\text{h}$ and $\Delta T = 4\,\text{h}$, respectively. For these three rows, each column from left to right shows estimated model images obtained with CLEAN, UNets, SARA, uSARA, AIRI-$\ell_2$, and AIRI-$\ell_1$, respectively. Below each reconstruction we indicate the reconstruction metrics and time (SNR, logSNR, reconstruction time). All images are shown in logarithmic scale.}
\label{fig:comparisonshercA}
\end{figure}
\end{landscape}

Figure~\ref{fig:comparisonshercA} shows the reconstructed Hercules~A images for the considered methods and the $uv$-coverages shown in Figures~\ref{fig:samp:seed2dt1}, \ref{fig:samp:seed2dt2}, and \ref{fig:samp:seed2dt4}. The visual analysis confirms the quantitative results from Figure~\ref{fig:exp2:metrics}. We acknowledge that the AIRI-$\ell_2$ and AIRI-$\ell_1$ algorithms can create artefacts by over-emphasising small structures (see the red zoom box in Figure~\ref{fig:hA:13}). SARA exhibits very similar quality to AIRI-$\ell_1$, but not always with the same resolution-to-artefacts tradeoff. The images suggest that the poor reconstruction of CLEAN stems from a strong offset in intensity values (see in particular the difference between the estimated higher intensity values in the upper rectangular zoom from Figures~\ref{fig:hA:5},\ref{fig:hA:10}, and \ref{fig:hA:15}, and that of the groundtruth). Also note the important residual noise hiding low intensity features of the image, as well as remaining artefacts for $\Delta T=1\,\text{h}$ and $\Delta T=2\,\text{h}$. This figure also illustrates clearly the strengths and shortcomings of the UNets trained in an end-to-end manner. The relatively good SNR and low logSNR, together with the visual results, confirm that the UNets manage to recover quite accurately the high intensity emission but struggle to recover faint emission. Visually, the results of the UNets present some hallucinated point-like sources \citep{muckley2021results}, which we believe is the consequence of our synthetic training database containing a large number of such sources.

\subsubsection{Computational cost and reconstruction times}

\begin{figure}
  \begin{subfigure}[b]{0.49\linewidth}
         \centering%
         \includegraphics[width=\textwidth]{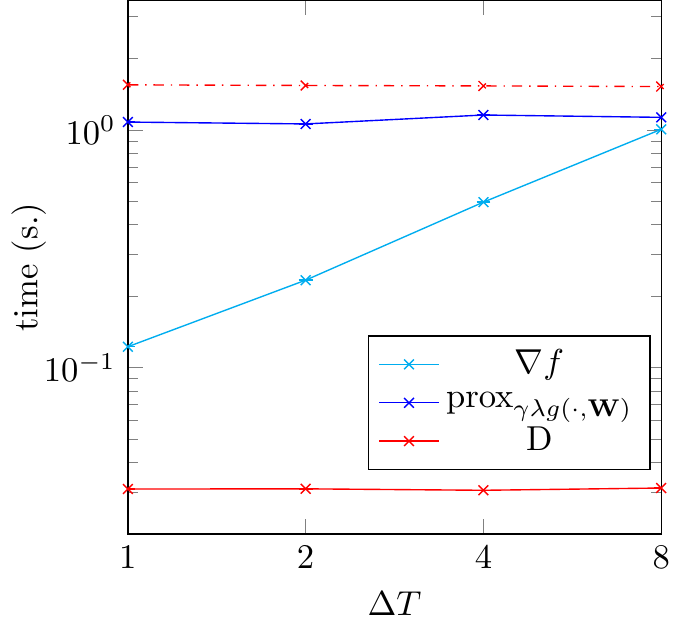}%
         \caption{Average time per iteration.}
         \label{fig:exp3:dt1}
     \end{subfigure}
     \hfill%
     \begin{subfigure}[b]{0.49\linewidth}%
         \centering%
         \includegraphics[width=0.94\textwidth]{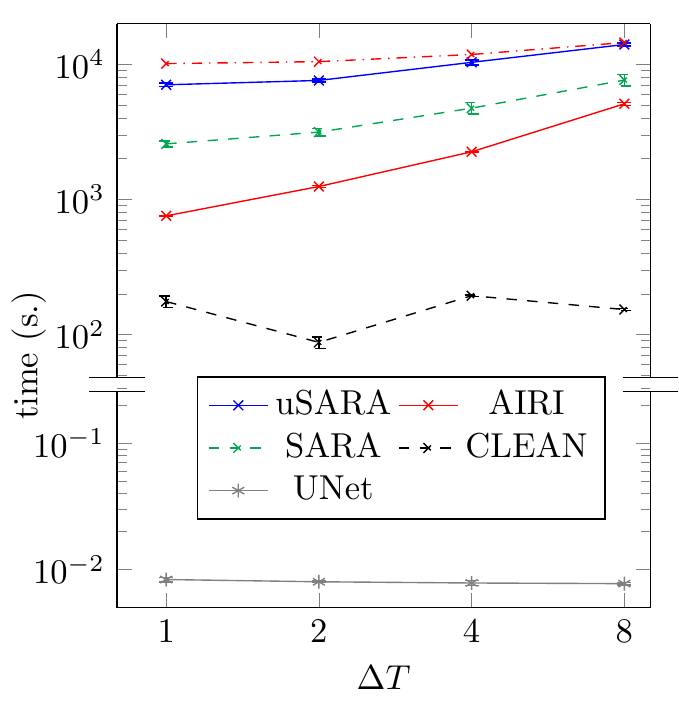}%
         \caption{Total timing for each method.}
         \label{fig:exp3:dt2}
     \end{subfigure}
\caption{Experiment 3 results: reconstruction times for AIRI-$\ell_2$, AIRI-$\ell_1$, uSARA, SARA, CLEAN, and the end-to-end UNets (on GPU)}. AIRI-$\ell_2$ and AIRI-$\ell_1$ exhibit identical behaviours, with only AIRI-$\ell_2$ results are reported and referred to as AIRI on the graphs. Each point is an average over the 20 simulated observations. Error bars show the (very small and virtually invisible) 95\% confidence interval. For AIRI-$\ell_2$ and AIRI-$\ell_1$, all reconstruction times are reported for both GPU (solid red lines) and CPU implementation (dashed red lines) of $\operatorname{D}$. Left: average time per iteration in seconds (s.) for the forward and backward operators involved in Algorithm~\ref{algo:sara_fb} ($\nabla f$ and $\operatorname{prox}_{\gamma \lambda g}$) and Algorithm~\ref{algo:pnp} ($\nabla f$ and denoiser $\operatorname{D}$), as a function of the observation duration. Right: average total reconstruction times for AIRI-$\ell_2$, AIRI-$\ell_1$, uSARA, SARA, CLEAN, and the end-to-end UNets, as a function of the observation duration.
\vspace{-1em}
\label{fig:exp3}
\end{figure}

The computational costs for the various approaches considered are reported in Figure~\ref{fig:exp3}. Each point on the graphs is an average over the 20 simulated observations built from the four different groundtruths from Figure~\ref{fig:gdths} and the five different $uv$-patterns shown in Figures~\ref{fig:samp} and \ref{fig:samp:seed2dt4}. Error bars show the 95\% confidence intervals.

Figure~\ref{fig:exp3:dt1} displays the average time per iteration, independently  for each of the forward and backward steps of both uSARA and AIRI, as a function of the observation duration. We recall once more that uSARA relies on the same algorithmic structure as AIRI-$\ell_2$ and AIRI-$\ell_1$. In both cases, the underpinning FB algorithm alternates between a (forward) gradient-descent step enforcing data fidelity and a (backward) regularization step enforcing a prior model via the application of either the average sparsity proximal operator (uSARA, Algorithm~\ref{algo:sara_fb}) or a learned denoiser (AIRI, Algorithm~\ref{algo:pnp}). We note that, in such an algorithmic structure, the computational cost of the regularization step scales linearly with the image size, while that of the gradient-descent step scales linearly with the data size. 

Firstly, we observe from the figure that the computation time of the gradient step naturally scales linearly with the observation duration, which is indeed proportional to data size. Secondly, we acknowledge that the computation of the proximal operator within uSARA is slow, which is due to its sub-iterative nature. For the image and data sizes considered in our simulations, and given our CPU implementation, it always dominates the computation time of the gradient, on average per iteration of Algorithm~\ref{algo:sara_fb}. This conclusion holds despite the fact that we have carefully optimized the external parameters of Algorithm~\ref{algo:sara_fb} and Algorithm~\ref{algo:dfb} to optimize the reconstruction time ($K$ in Algorithm~\ref{algo:sara_fb},  and the tolerance criterion $\xi_2$ and dual variable $\bm{v}_0$ in Algorithm~\ref{algo:dfb}). In that regard, we note that an efficient initialization for the dual variable $\bm{v}_0$ consists in setting it as the dual variable from the previous use of Algorithm~\ref{algo:dfb}. Using such an initialization, the number of sub-iterations for Algorithm~\ref{algo:dfb} progressively decreases with the iteration count in Algorithm~\ref{algo:sara_fb}. Thirdly, using a standard GPU implementation of the AIRI-$\ell_2$ and AIRI-$\ell_1$ denoisers, their application is two orders of magnitude faster than the computation of the proximal operator. A CPU implementation of the denoisers makes them even slower than the proximal operator in uSARA. For the image and data sizes at stake, the AIRI denoisers on GPU are also significantly faster than the computation of the gradient, on average per iteration of Algorithm~\ref{algo:sara_fb}.

Figure~\ref{fig:exp3:dt2} displays the total reconstruction times for uSARA and AIRI as a function of the observation duration, and in comparison with SARA and CLEAN. The AIRI-$\ell_2$ and AIRI-$\ell_1$ reconstruction times with denoiser on GPU scale almost linearly with the observation duration, due to the domination of the gradient step. For uSARA, and for AIRI-$\ell_2$ and AIRI-$\ell_1$ with denoiser on CPU, the dependency on observation duration is less severe given the domination of the regularization step at each iteration. Overall, given the speed up in the regularization step, the total reconstruction time of AIRI with denoiser on GPU is significantly ($5$ to $10$ times) lower than the one of uSARA. AIRI-$\ell_2$ and AIRI-$\ell_1$ with denoiser on CPU are much slower, in fact slightly slower than uSARA itself. SARA offers reconstruction time between those of AIRI and uSARA. CLEAN remains significantly faster than AIRI ($4$ to $15$ times), but the fastest method among all is by far the UNet trained end-to-end. We underline that this is to be expected since, not only UNets were run on GPU, but the reconstruction with the UNet only requires a single inference step, as opposed to the other methods that are iterative.

To summarize, the application of an AIRI denoiser (on GPU) is orders of magnitude faster than that of the  average sparsity proximal operator of uSARA (on CPU). Substituting the former by the latter does not only provide superior imaging quality, but significantly reduces the computational cost of the regularization step. For image and data sizes where the uSARA regularization step dominates over the gradient-descent data-fidelity step, a cost reduction in the former, substituting the proximal operator for a DNN denoiser leads to significant speed up of the algorithm. 

\section{Conclusion \& Future Work}
\label{sect:conclusion}

\subsection{Conclusion for both AIRI \& uSARA}

In this work, we have proposed a new algorithmic framework for scalable precision RI imaging, dubbed AIRI. The approach consists in encapsulating a prior image model in a DNN denoiser, and plugging it in lieu of the proximal regularization operator of an optimization algorithm of the SARA family for image reconstruction. AIRI inherits the robustness and interpretability of optimization approaches and the learning power and speed of networks, while avoiding generalizability limitations of end-to-end networks, for which training involves the details of the measurement model.

More specifically, we have firstly designed a realistic low dynamic range training database from optical images. We have also established a method to train firmly nonexpansive denoisers at a noise level inferred from the target dynamic ranges of reconstruction, with either an $\ell_2$ or $\ell_1$ loss, including a procedure for on-the-fly database dynamic range enhancement. Finally, the AIRI-$\ell_2$ and AIRI-$\ell_1$ algorithms resulting from plugging the denoisers into the FB algorithm were validated in simulation for reconstruction of images involving complex structure with diffuse and faint emission across the field of view (at a dynamic range of $10^4$). The benchmark algorithms included CLEAN, the state-of-the-art SARA algorithm, as well as its new unconstrained version uSARA, which only differs from AIRI-$\ell_2$ and AIRI-$\ell_1$ by the use of the average sparsity proximal regularization operator instead of a learned denoiser in the FB algorithm.

Our results show that the first incarnations of AIRI are already competitive with, if not superior to, the advanced optimization algorithms of the SARA family in imaging quality, well beyond the capabilities of CLEAN and of the UNets trained in an end-to-end fashion. In particular, while the AIRI denoisers are trained on the same database as the UNets, the PnP solutions do not exhibit hallucinated artefacts, which illustrates the robustness of AIRI compared to the end-to-end UNets. With regards to computation time, the AIRI approach was shown to provide a significant acceleration potential over uSARA and SARA, thanks to the inference speed of DNN denoisers over the averaged sparsity proximal operator, but remain significantly slower than CLEAN, itself orders of magnitude slower than the UNets. The results also confirm the validity of our heuristic to set the training noise level $\sigma$, consisting in enforcing equality of the training and target dynamic range of the reconstruction, and the validity of the database exponentiation approach (with parameter $a$), consisting in enforcing equality of the final nominal dynamic range of the database and target dynamic range of reconstruction. This provides a significant advantage over uSARA, for which the regularization parameter $\lambda$ requires manual fine-tuning, even though the fact that the corresponding heuristic (and its $1/3$-corrected version) provides an appropriate reference upper-bound for the optimal value is an interesting result. 

\subsection{Current limitations \& future work}

In what follows, we discuss current limitations of the AIRI framework, and future work. This also includes considerations for further developments of uSARA
and end-to-end DNNs.

Firstly, the modified DnCNN denoiser architecture used, the $\ell_2$ and $\ell_1$ losses considered, as well as the set of optical images used to build a training database constitute the most basic choices leading to our first AIRI-$\ell_2$ and AIRI-$\ell_1$ incarnations of the framework. We anticipate that future work, building a richer database from RI observation, and considering more advanced losses, such as adversarial losses \citep{wang2018esrgan}, and more advanced network architectures, such as UNets \citep{hurault2022proximal}, can provide a quantum jump in imaging precision with AIRI. We also acknowledge that our approach to enforce the firm nonexpansiveness of the denoiser is only approximate, and relies on fine-tuning the regularization parameter $\kappa$ in the training loss. Simpler and more precise approaches to ensure firm nonexpansiveness should be investigated in the future (\emph{e.g.}~leveraging 1-Lipschitz layers). We note however that,  in the context of our experiments, the proposed approach has, in practice, provided a robust way to ensure convergence of the resulting PnP algorithms. 

Secondly, fully optimized and parallelized implementations of AIRI algorithms, but also SARA, or uSARA, should be investigated, leveraging GPUs not only for DNN denoisers, but also for the implementation of the linear (measurement or regularization) operators involved. This should lead to further acceleration of these approaches, and possibly participate to closing the gap with CLEAN in terms of reconstruction time. We also note that accelerated versions of uSARA are achievable, in particular leveraging preconditioning strategies \citep{repetti2021variable}.

Thirdly, with regards to the details of the target imaging modality, we have investigated monochromatic intensity imaging on small fields of view only, with mild, if any, super-resolution factor with respect to the nominal resolution of the observation, as set by the largest baseline. Extensions of the framework to further super-resolution functionality and to wideband polarization imaging should be contemplated. Also, a by-product of the present analysis is to confirm the superiority of SARA over uSARA, due to the constrained approach to data fidelity in SARA, versus the unconstrained approach underpinning uSARA. In this context, the superior results of AIRI over uSARA are entirely related to the capability to encapsulate a better prior model than the average sparsity model in learned DNN denoisers. On the one hand, a first possible enhancement of the AIRI approach would consist in developing a PnP version of SARA (as opposed to uSARA), simultaneously taking advantage of networks to learn prior models and of the constrained data-fidelity formulation. On the other hand, to date, the unconstrained formulation underpinning uSARA, AIRI-$\ell_2$, and AIRI-$\ell_1$ is a critical building block of the only optimization-based algorithm proposed for RI imaging capable of handling jointly DDE calibration and imaging \citep{repetti2017non,dabbech2021cygnus}. This is a strong justification for the study of unconstrained versus constrained formulations of the imaging module. 

Fourthly, the proposed UNet end-to-end implementation showed interesting results, albeit significantly outperformed by the proposed AIRI algorithms. A more thorough study should be undertaken, leveraging more advanced architectures, such as unfolded networks that would enable to incorporate knowledge about the measurement operator within the architecture.

Finally, the present work is simulation-based only. \citet{dabbech2022first}, \citet{wilber2022first}, and \citet{wilber2022second} validate AIRI and uSARA on real wide-field high-resolution high-dynamic range data from MeerKAT and ASKAP, relying on a highly parallelised implementation of the measurement operator tailored to the wide-field setting.

\section*{Acknowledgements}
The research of M.~T.~was funded by Heriot-Watt under the James Watt Scholarship scheme. The work of M.~T., A.~D., and Y.~W.~was supported by EPSRC under grants EP/T028270/1 and ST/W000970/1. The computing resources came from the Cirrus UK National Tier-2 HPC Service at EPCC (http://www.cirrus.ac.uk) funded by the University of Edinburgh and EPSRC (EP/P020267/1), partly through time allocation under the SUSA project, and partly through GPU resources directly provided by EPCC (Adrian Jackson). Credits for the 32 images used in our training database go to NOIRLab/NSF/AURA/H.Schweiker/WIYN/T.A.Rector (University of Alaska Anchorage). The radio images where taken from: for Hercules~A: R.~Perley and W.~Cotton (NRAO/AUI/NSF), for 3c353: NRAO/VLA, for Centaurus~A: NRAO/AUI/NSF/Univ.Hertfordshire/M.Hardcastle. 

\section*{Data Availability}

AIRI code will be made available as part of a later release of and joint parallel toolbox for AIRI and uSARA.


\bibliographystyle{mnras}
\bibliography{bibliography} 




\appendix
%


\section{On the chosen RI imaging model}
\label{appendix:ri_model}
We discuss two points regarding the generality of the measurement model in \eqref{eq:invpb}. Firstly, at the dynamic range of interest for current and future observations,  direction-dependent effects (DDEs) of atmospheric and instrumental origin complicate the RI measurement equation. More precisely, at each time instant, the visibility associated with a pair of antennas is formed from the modulation of the radio sky with the product of a DDE pattern specific to each antenna. The DDEs are typically unknown and need to be calibrated, leading to a blind deconvolution problem, which requires the design of joint calibration and imaging approaches, typically alternating between imaging and calibration modules \citep{repetti2017non,dabbech2021cygnus}. One exception to this, is the so-called $w$-term induced by the projection of the baseline associated with each visibility on the line of sight, which becomes non-negligible on wide fields of view, and thus acts as a known DDE. In practice, DDEs can be integrated into extended convolution kernels in $\bm{\mathrm{G}}$ \citep{dabbech2017w}. In this work, we assume a measurement operator $\bm{\Phi}$ not affected by any known or unknown DDE, considering a pure imaging problem on small fields of view.

Secondly, the measurement model \eqref{eq:invpb} assumes noise with constant variance across visibilities. In full generality, the noise is not white, and a diagonal whitening matrix $\bm{\Theta} \in \mathbb{R}^{m \times m}$ with entries equal to the inverse noise standard deviation per visibility can be applied to the original measurement vector, ensuring that the resulting visibility vector is indeed affected by white noise. In this case problem \eqref{eq:invpb} still holds, with a measurement operator $ \bm{\Phi} = \bm{\Theta\mathrm{GFZ}}$, and $\bm{y}$ and $\bm{e}$ the visibility and noise vectors after application of the whitening matrix. This operation is known as natural weighting. This scheme is used by the algorithms of the SARA family to ensure the negative log-likelihood interpretation of its data-fidelity terms \citep{carrillo2012sparsity}. More complex weighting schemes, such as uniform and Briggs weighting, are used by CLEAN \citep{briggs95}, which involve additional multiplicative terms in $\bm{\Theta}$. Both approaches add to natural weighting by downweighting visibilities in regions of the $uv$-plane with high sampling density, with the aim to reduce the sidelobes of the dirty beam and improve the resolution of the CLEAN reconstruction, though at the expense of sensitivity. Interestingly, \citet{onose2017accelerated} showed, for algorithms of the SARA family specifically, that, while keeping to natural weighting, using sampling density information in acceleration strategies such as preconditioning are very efficient at bringing jointly optimal resolution and dynamic range. In other words, the resolution-versus-sensitivity trade-off disappears asymptotically at convergence. In this work, without loss of generality, our simulations consider i.i.d.~Gaussian random measurement noise so that $\bm{\Theta}=\mathbf{I}$ for natural weighting, where $\mathbf{I}$ denotes the identity matrix. All algorithms discussed (SARA, uSARA, AIRI) use natural weighting, while CLEAN is implemented with uniform weighting (see Section~\ref{sssec:bench}).

\section{On the PnP convergence}
\label{sect:mon_tech}

Several works have recently focused on restoring the convergence of PnP algorithms using DNN denoisers. The majority of these works assume a nonexpansiveness constraint on the denoiser \citep{romano2017little, ryu2019plug, terris2020building, cohen2021regularization, hertrich2021convolutional}, but this constraint often comes with restrictive assumptions either on the algorithm, on the DNN architecture or on the operator $\nabla f$.

In \citet{pesquet2020learning}, casting the problem in the framework of monotone operator theory, we have shown that the convergence of a PnP algorithm can be ensured by introducing a well-chosen regularization in the training loss of the denoiser $\operatorname{D}$ enforcing its firm nonexpansiveness, without any form of limitation on the DNN architecture itself. Importantly, the approach also offers  a generalized notion of characterization of its limit as the solution to monotone inclusion problems, more general than pure optimization problems, hence opening a path towards generalizing the usual Bayesian MAP interpretation of optimization solutions. We summarize those results in the remainder of this section.

Once again concentrating on a differentiable $f$ such as \eqref{eq:def_fl2}, minimization problems like \eqref{eq:min_pb} can be reformulated (through their optimality conditions) as
\begin{equation}
\text{find $\widehat{\bm{x}}\in \RR^n$ s.t. }\qquad
    0 \in \nabla f(\widehat{\bm{x}}) + \lambda \partial r(\widehat{\bm{x}}),
\label{eq:varinc_prox}
\end{equation}
where $\partial r$ denotes the subdifferential of $r$, the subdifferential of $r\in\Gamma_0(\RR^n)$ being the set-valued operator defined as $\partial r(\bm{x}) = \{\bm{u}\in\RR^n | (\forall \bm{z} \in\RR^n) \langle \bm{z}-\bm{x}, \bm{u} \rangle + r(\bm{x}) \leq r(\bm{z})\}$. When $r$ is also differentiable, the subdifferential coincides with the gradient (\emph{i.e.} $\partial r = \nabla r$). Yet, the subdifferential is also defined for non-differentiable functions such as $\iota_{\mathcal{B}(\bm{y},\epsilon)}$; we refer the reader to \citet{bauschke2017convex} for more information. Problems as \eqref{eq:varinc_prox} are part of the class of so-called monotone inclusion problems and encompass unconstrained minimization problems such as uSARA \eqref{eq:uSARA}. They can however take a more general flavour and write as
\begin{equation}
\text{find $\widehat{\bm{x}}\in \RR^n$ s.t. }\qquad
    0 \in \nabla f(\widehat{\bm{x}}) + \operatorname{A}(\widehat{\bm{x}}),
\label{eq:varinc_monop}
\end{equation}
where $\operatorname{A}$ is an operator satisfying some maximal monotonicity assumptions \citep{bauschke2017convex}. It is important to stress that this class of operators includes, but is not restricted to, subdifferentials. Problem \eqref{eq:varinc_monop} is thus in general not equivalent to a minimization problem. 

Solving such inclusion problems can be done using a generalized version of the proximal FB algorithm \eqref{eq:prox_fb} with
\begin{equation}
(\forall k \in \NN),\qquad   \bm{x}_{k+1} = \operatorname{J}_{\gamma\!\operatorname{A}}(\bm{x}_k-\gamma \nabla f(\bm{x}_k)),
\label{eq:monotone_fb}
\end{equation}
where $\operatorname{J}_{\gamma\!\operatorname{A}}$ is the so-called resolvent of $\gamma\!\operatorname{A}$, yielding the convergence of the sequence $(\bm{x}_k)_{k\in\NN}$ to $\widehat{\bm{x}}$ satisfying \eqref{eq:varinc_monop}. As previously, the stepsize $\gamma$ must satisfy $0<\gamma<2/L$ where $L$ is the Lipschitz constant of $\nabla f$.

The resolvent operator can be seen as a generalized version of the proximal operator \eqref{eq:prox_def} \citep{bauschke2017convex}. In fact, \eqref{eq:prox_fb} is a special case of \eqref{eq:monotone_fb}, since when $\operatorname{A} = \lambda \partial r$ for $r\in\Gamma_0(\RR^n)$, one has $\operatorname{J}_{\gamma \lambda \partial r} = \operatorname{prox}_{\gamma \lambda r}$. A key characterization of resolvent operators is the following: the operator $\operatorname{J}$ is the resolvent of a maximally monotone operator $\operatorname{A}$ if and only if it is firmly nonexpansive. Technically, by definition, an operator $\operatorname{J}$ is firmly nonexpansive if there exists an operator $\operatorname{Q}$ with Lipschitz constant smaller or equal to 1 such that $\operatorname{J} = \left(\operatorname{I}+\operatorname{Q}\right)/2$, where $\operatorname{I}$ is the identity operator. Given the identical structure of PnP-FB and \eqref{eq:monotone_fb}, it follows that $\operatorname{D}$ in \eqref{eq:pnp_fb} should be firmly nonexpansive for the convergence of the PnP algorithm to be ensured, with the PnP solutions characterized as solutions of a monotone inclusion problem \eqref{eq:varinc_monop}. 

Thus, in order to impose the firm nonexpansiveness of $\operatorname{D}$, it is sufficient to impose that the Lipschitz constant of $\operatorname{Q}=2\operatorname{D}-\operatorname{I}$ is less than 1, or equivalently, that its Jacobian satisfies $\|\bm{\nabla}\operatorname{Q}(\bm{x})\|_{\rm{S}}\leq 1$ for all $\bm{x}$. Since most deep learning frameworks rely on Jacobian-vector products \citep{paszke2017automatic, balestriero2021fast, maddox2021fast}, one can, given some $\bm{x}\in\RR^n$, compute $\|\bm{\nabla}\operatorname{Q}(\bm{x})\|_{\rm{S}}$ via the power method. Thus, $\operatorname{D}$ would ideally be trained as a denoiser with a standard training loss, but under the constraint that $\|\bm{\nabla}\operatorname{Q}(\bm{x})\|_{\rm{S}}\leq 1$ for all $\bm{x}$. Such a constraint is not trivial to impose in practice and, as proposed in \citet{pesquet2020learning}, we relax it and introduce a regularization term in the training loss that penalizes non firmly nonexpansive networks. 

Given a database of noisy / groundtruth images $(\bm{z}_s,\bm{u}_s)_{1\leq s\leq S}$, and denoting $\bm{\theta} \in \RR^c$ the learnable parameters of $\operatorname{D}$, the training loss ensuring the firm nonexpansiveness of $\operatorname{D}$ reads: 
\begin{equation}
\begin{aligned}
    \underset{\bm{\theta} \in \RR^c}{\text{minimize}}\, \frac{1}{S}\sum_{s=1}^S \bigg(\mathcal{L}(\operatorname{D}_{\bm{\theta}}(\bm{z}_s)-\bm{u}_s)+\kappa\operatorname{max}\{\| \boldsymbol{\nabla} \operatorname{Q}_{\bm{\theta}}(\bm{z}_s)\|_{\rm{S}}, 1-\varepsilon\}\bigg),
\end{aligned}
\label{eq:training_loss_appendix}
\end{equation}
where $\mathcal{L}$ is a loss function and $\varepsilon>0$ is a safety margin parameter. As underlined in \citet{pesquet2020learning}, this penalty does not explicitly guarantee that the solution of this problem satisfies the constraint $\|\boldsymbol{\nabla}\operatorname{Q}(\bm{z})\|_{\rm{S}}\leq 1$ on the database, where it is effectively evaluated, let alone for all $\bm{z}\in\RR^n$. Various countermeasures are however deployed, which have been demonstrated to be effective at ensuring PnP convergence in practice (see \citet{pesquet2020learning}). Firstly, the safety margin parameter $\varepsilon>0$ is a mean to strengthen the penalty. Secondly, the term $\|\bm{\nabla}\operatorname{Q}\|_{\rm{S}}$ is not computed exactly at $\bm{z}_s$, but instead at a point sampled uniformly at random on the segment $[\bm{z}_s, \bm{u}_s]$, with the aim to explore a richer set of images than those of the database only. Last but not least, the value of $\kappa$ can, and should, be fine-tuned manually to optimize the balance between the regularization term and the standard loss $\mathcal{L}$.

We conclude this section by underlining that, contrary to other works \citep{ryu2019plug, terris2020building}, the proposed approach does not require to impose layer-wise Lipschitz constraints nor architectural restrictions on the DNN.

\bsp	
\label{lastpage}
\end{document}